\documentclass[%
 reprint,
 amsmath,amssymb,
 aps,
 nofootinbib,
]{revtex4-2}

\usepackage{mathrsfs}
\usepackage{graphicx}
\usepackage{dcolumn}
\usepackage{bm}

\usepackage{pstricks}
\usepackage{tikz}\label{key}
\usetikzlibrary{shapes,snakes}
\usepackage{float}

\begin{document}

\preprint{APS/123-QED}

\title{Study of ideal gases in curved spacetimes}

\author{Luis Arag\'on-Muñoz}
\affiliation{Instituto de Ciencias Nucleares, Universidad Nacional Aut\'onoma de M\'exico, 
	04510 Ciudad de M\'exico, M\'exico }
 \email{luis.aragon@correo.nucleares.unam.mx}
\author{Hernando Quevedo}%
 \email{quevedo@nucleares.unam.mx}
\affiliation{Instituto de Ciencias Nucleares, Universidad Nacional Aut\'onoma de M\'exico, 
	04510 Ciudad de M\'exico, M\'exico }
\affiliation{Dipartimento di Fisica and ICRA,  Universit\`a di Roma ``La Sapienza", I-00185 Roma, Italy}

\date{\today}

\begin{abstract}

The influence of a curved spacetime $M$ on the physical behavior of an ideal gas of $N$ particles is analyzed by considering  the phase space of the system as a region of the cotangent bundle $T^{*}M^{N}$ and using Souriau's Lie group {thermodynamics} to define the corresponding  probability distribution function. While the construction of the phase space respects the separability of the system, by forcing each particle to satisfy the so-called mass-shell constraint, the probability distribution is constructed by mixing the natural symplectic structure of the cotangent bundle with a Hamiltonian description of the system. In this way, the spacetime is introduced into the statistics and its isometries turn out to be of special interest because the distributions are parameterized by the elements of the Lie algebra of the isometry group, through the momentum map of the action of the isometries in $T^{*}M^{N}$. We find the Gibbs distribution that, in the simplest case of a flat spacetime, reduces to the so-called modified J\"{u}ttner distribution, used to describe ideal gases in the regime of special relativity. We also define a temperature-like function using the norm of a Killing vector, which allows us to recover the so-called Tolman-Ehrenfest effect. As a particular example, we study the outer region of a Schwarzschild black hole, for which a power series expansion of the Schwarzschild radius allows us to represent the partition function and the Gibbs distribution in terms of the corresponding quantities of the  Minkowski spacetime.
	
\begin{description}
\item[Keywords]
Statistical physics, General Relativity, Symplectic geometry, Momentum map, Souriau's Lie group statistics, Schwarzschild black hole
\end{description}
\end{abstract}

\maketitle


\section{Introduction} 
\label{sec:int}
The study of many-particle systems in situations far from laboratory conditions, such as high temperatures or intense gravitational fields, has interested physicists from various areas for many years, mainly for theoretical reasons to reinforce the foundations of physics itself \cite{tolman1934,bergmann1951,vankampen1968,horvath1968,werner1976,lipparini2008}. Recently, it has gained more interest due to the new advances in astrophysics that motivate the statistical discussion of systems such as hot plasmas \cite{pomeau2007,livadiotis2018,mati2020,bauche2015} or accretion disks around black holes \cite{uzdensky2008,bar2014,herpich2017,ftq2022}. Even on the large scale of the universe, statistical models for galaxies have been proposed \cite{	ogorodnikov1957,binney1987,pietronero2005,hameeda2021,ourabah2022}, which are similar to those that model dark matter in particle physics \cite{pietronero2007,feron2008,patwardhan2008,chavanis2015,hkv2022}.
In this regard, despite the great interest in the subject, there is no a well-established theory for statistical mechanics in curved backgrounds\footnote{See \cite{sato2021}, especially the introduction, for a more complete discussion of the problems that arise in the general relativity frame}.

In this work, we address the problem of the intersection between statistical mechanics and general relativity, and we investigate how the curvature of a background  {static} spacetime  affects the properties of a many-particle system (specifically, an ideal gas). Our analysis is oriented from the perspective of symplectic geometry and Hamiltonian systems\footnote{Different approaches to study the effect of curved spacetimes on many-particle systems can be found in the articles \cite{sato2021,kolekar2011,sarbach2022}.}. This perspective is based on the precursor ideas of Souriau et al. \cite{souriau1997structure, barbaresco2014, marle2016, barbaresco2019, marle2020gibbs}.  {In particular, we will focus on the model called Lie groups thermodynamics proposed by Souriau in 1969. The main idea of this model is simple. In classical statistical thermodynamics, a system in equilibrium with a heat bath is usually described by the Gibbs canonical ensemble, which can be defined on the phase space of the corresponding system. Souriau generalized the Gibbs construction to the case of a symplectic manifold equipped with a dynamical group, i.e., a group associated with the symmetries of the system.  This generalization allows us to investigate systems that usually cannot be considered in classical statistical thermodynamics. Although this approach  
implies the application of a rigorous mathematical formalism,  	
 the results obtained by using it are as physical as those of  classical statistical thermodynamics. To be more specific, in  
Lie groups thermodynamics the states of a many-particle system are described through probability distribution functions (on a symplectic manifold with Liouville's natural measure), which are parametrized by the infinitesimal generators of a symmetry of the system, though the exponential of the constant of motion associated with the symmetry. This means that the probability distribution function (pdf) is parametrized by a vector in the Lie algebra of the Lie group of the symmetries of a Hamiltonian system, by means of the exponential of the (negative) momentum map evaluated on that vector. This framework allows us to work with non-standard Hamiltonian systems in which the evolution is not generated by a function interpreted as the energy of the system, and the evolution parameter is not Newton's absolute and external time. In our case, this is essential because the Hamiltonian system depends on a geometrical-pure-nature function constructed directly by the metric tensor, and time loses the special treatment it receives in non-relativistic physics (in addition, the notion of energy of the system is a bit tricky, which is why we demand that the spacetime be static).} Also, this framework allows us to describe physical effects generated by gravitational fields such as the Tolman-Ehrenfest effect  {that claims the non-constancy of the temperature of a body in a thermal equilibrium state, when is affected by a gravitational field (see \cite{rovelli2011} for a complete description)}. In addition, 
in the limiting case of the Minkowski spacetime, our formulation  reduces to the one well known from the works of J\"{u}ttner et al \cite{cubero2007,chacon2009,aragon2018modified}, which is used to describe systems in the framework of special relativity.

 {This work is organized as follows. 
In Sec.(\ref{sec:lie_groups}), we review the formalism of Lie group thermodynamics. We first formulate the definition of a statistical state and then  show how a Lie group acts on a Hamiltonian system to generate the corresponding symmetries.  Finally, we define the generalized statistical states we will use later to study an ideal gas in a static spacetime. In addition, we use this section to introduce most of the mathematical formalism we will use throughout the work, like the momentum map and Souriau's vectors in the Lie algebra.}

 {The results of our work are presented in Secs.(\ref{sec:hamiltonian_description})--(\ref{sec:con}) and  organized as follows. }
In Sec.(\ref{sec:hamiltonian_description}), we begin by analyzing the Hamiltonian model that describes a free particle of mass $m$ in a spacetime  {$(M,g)$}, using its cotangent bundle $T^{*}M$ endowed with its natural symplectic form $\Omega$, a Hamiltonian function $H$ generated by the (inverse of the) metric tensor $g$ and a constraint hypersurface in $T^{*}M$, arising from the mass-shell condition that satisfies any massive particle. As part of this section, in subsection (\ref{subsec:mom_map}) the isometry group $G$ of the spacetime $(M,g)$ is presented, and we define its action on the cotangent bundle of the particle, showing that the isometries group generates symmetries of the proposed Hamiltonian system, {following Sec. (\ref{sec:lie_groups})}.

In Sec.(\ref{sec:ideal_gas}), we formulate the model of a dilute ideal gas as a set of $N$ identical particles enclosed in a box that travels in a spatially bounded geodesic trajectory. We impose the condition that the particles interact only with the internal walls of the box in very brief moments and in a perfectly elastic manner, in such a way that the Hamiltonian model presented in Sec.(\ref{sec:hamiltonian_description}), including the action from the isometry group,  can be generalized to describe this ideal gas by means of a) the separability of the system and b) the spatial restriction of the geodesics to the world tube that surrounds the geodesic of the box.  {
Finally, in Sec. (\ref{subsec:static_spacetime}), we introduce the concepts of static spacetimes and static states}.

 {In Sec. (\ref{sec:statiscal_distribution}), we define the static states of a diluted ideal gas through the Gibbs pdf introduced in Sec. (\ref{sub:generalized_statistical_states}). We prove the existence of static states, meaning that in the isometry Lie algebra, there are Souriau vectors that parameterize the states. We also show that the definition of static equilibrium allows us to always work at constant time from the perspective of  static observers.}

A decomposition of the Gibbs pdf of the ideal gas into one-particle distributions is done in Sec.(\ref{sec:one_particle_distribution}). This decomposition is based on the separability of the ideal gas and on an adequate proposal of phase space as a submanifold within the cotangent bundle of the ideal gas. Related to this phase space, the induced measure is deduced from the Liouville measure, giving rise to formal expressions both for the one-particle partition function and for the Gibbs one-particle distribution. As part of this section,  {in  subsections (\ref{sec:modified_juttner_like_pdf}) and (\ref{subsec:sch_spacetime}), we study a particular case of distributions parametrized by elements $(\beta >0)$ of the Lie subalgebra of the time translation subgroup (subalgebra of the Lie algebra of the whole isometry group) in the Minkowski spacetime and in the Schwarzschild spacetime, respectively. We find a distribution, which is reminiscent of the modified J\"{u}ttner distribution, used to describe ideal gases in the framework of special relativity, and a generalization to the case of the curved Schwarzschild spacetime}. In the limiting case of the Minkowski spacetime, we interpret the role of $\beta$ as the inverse of the temperature of a thermodynamic system with statistics given exactly by the modified J\"{u}ttner distribution.  {
For the  Schwarzschild spacetime outside the event horizon, we perform a power series expansion  of the Schwarzschild radius,  which allows us to express the statistical elements of the generalized J\"{u}ttner distribution  in terms of quantities of the flat Minkowski spacetime. In particular, we define a temperature-like function  that reduces to the Tolman-Ehrenfest temperature in the asymptotic flat limit}.

Finally, in Sec.(\ref{sec:equiv_temperatures}), we use the equivariance property of the momentum map to show the invariance of the total partition function with respect to a certain class of special isometries that map the phase space onto itself. This leads us to consider the possibility of an equivalence in the measurements of different static observers connected through the aforementioned special isometries, opening a possible new argument in the well-known debate on the law of transformation of temperature between different observers.

For complementarity, we include in   {four appendices specific calculations used in the development of the sections described above. In Appendix (\ref{ap:equivariant}), we deduce the so-called equivariance property of (some) momentum maps. In Appendix (\ref{ap:induced_measures}), we calculate the induced measure that synthesizes the physical and geometric properties of our statistical analysis}. Appendix (\ref{ap:partition_function}) deals with the calculation of an integral and the development of the expression of the one-particle partition function, and Appendix  (\ref{ap:series}) develops the series expansion of the one-particle partition function for the Schwarzschild case.

\section{Lie groups thermodynamics}
\label{sec:lie_groups}
  
In physics, the use of statistical distributions to model the state of a system, which   is computationally impossible to obtain by analytical methods due to the immense number of degrees of freedom and constituents, has turned out to be a powerful tool. Under the correct assumptions, the statistical approach connects with thermodynamics to such a degree that, in the physics community, it is widely accepted  that behind a thermodynamic system there should be a microscopic model of statistical nature \cite{aleksandr1949,callen1985,tuckerman2010}.

More specifically, knowing the statistical model behind a system is reduced to finding a pdf $\rho$, defined on the space of the microscopic states of the system, which allows us to obtain macroscopic information by averaging over the large number of microscopic degrees of freedom. In particular, after a precise identification of these macroscopic properties, which emerge from the possibly very complex individual dynamics of the particles, the same distribution $\rho$ allows us to find the fundamental equation of a thermodynamic system through the so-called partition function \cite{aleksandr1949,callen1985,tuckerman2010}.

In this way, we will use the Lie group thermodynamics formalism \cite{souriau1997structure, barbaresco2014, marle2016, barbaresco2019, marle2020gibbs}, where probability density functions are parameterized by vectors in the Lie algebra of the Lie group of symmetries of the physical system. In other words, and very roughly, statistics à la Souriau\footnote{Souriau's work in the field of statistical physics is only a small part of the program he set out to understand physics as a geometric theory , more specifically, as a symplectic theory \cite{souriau1997structure}. For more information on this, we highly recommend consulting \cite{barbaresco2021geometric}.} finds statistical models that are invariant with respect to  the symmetries of the underlying Hamiltonian model. This approach generalizes the usual statistical model that is only invariant under the Hamiltonian flow of the system itself, generally interpreted as translations in Newtonian time with the Hamiltonian function being the energy function.

We emphasize the fact that this formalism offers a generalization, rather than just a mathematical reformulation, since by its very nature, it can be applied to completely general Hamiltonian systems\footnote{In this regard, Souriau explores in \cite{souriau1997structure} several important cases as, for example, gases in centrifuges that describe processes such as the enrichment of Uranium 235 or the creation of ribonucleic acids. Souriau also analyzes relativistic systems, obtaining  the J\"uttner distribution for relativistic ideal gases \cite{souriau1997structure}. On the other hand, the Lie Group thermodynamics is also used in information theory and machine learning. For a more detailed discussion of applications, see \cite{barbaresco2021geometric}.}, where there is no need for a physical interpretation of the Hamiltonian and the evolution parameter in terms of the lab-measurable  energy and time. For this reason, we apply in this work the formalism of Lie group thermodynamics to the Hamiltonian of a 
many-particle system in a curved spacetime, taking the spacetime isometries as the symmetries of the system, and parameterizing the $\rho$ distribution of the statistical model with the infinitesimal generators of those isometries. To see exactly how this works, first we have to setup some statistical tools.

\subsection{Classical statistical states}
\label{sub:clasic_statistical_states}
Let us consider 
a physical system modeled as a Hamiltonian system $(U,\Omega,H)$, where the pair $( U,\Omega)$ is a symplectic manifold of dimension $2D$ and $H:U\rightarrow \mathbb{R}$ is the Hamiltonian function, which guides the dynamics of the system through the Hamiltonian flow $\varphi^{H}:\mathbb{R}\times U\rightarrow U$ generated by the Hamiltonian vector field $X^{H}\in \mathfrak{X}(U)$, which, in turn, is defined via the relation \cite{marsden1999}
\begin{equation}
\label{eq:hamiltonian_vector}
dH=-i_{X^{H}}\Omega.
\end{equation}
In other words, if the state of the system at $\tau=0$ is given by $u\in U$, at any other instant\footnote{The $\tau$ parameter is not really/needly related to time, so here "instant" is just and expression.} $\tau\in \mathbb{R}$ the state of the system will be given by $\varphi^{H}(\tau,u)\in U$. In this sense, for each $\tau\in \mathbb{R}$ the Hamiltonian flow defines a diffeomorphism on $U$ given by $\varphi^{H}_{\tau}:u\mapsto \varphi^{H }(\tau,u)$, such that it leaves the Hamiltonian system invariant \cite{marsden1999}:
\begin{eqnarray}
\label{eq:hamiltonian_flux_1}
\varphi^{H}_{\tau}(U)&=&U,\\
\label{eq:hamiltonian_flux_2}
H\circ\varphi_{\tau}^{H}&=&H,\\
\label{eq:hamiltonian_flux_3}
\,[\varphi^{H}_{\tau}]^{*}\Omega\circ \varphi^{H}_{\tau}&=&\Omega,
\end{eqnarray}
where $[\varphi_{\tau}^{H}]^{*}:T^{*}U\rightarrow T^{*}U$ is the pullback of $\varphi^{H}$.

At the local coordinate level, Eq.(\ref{eq:hamiltonian_vector}) is displayed as a set of first-order differential equations in the parameter $\tau$, which are known as Hamilton's equations \cite{marsden1999}, and whose solutions are the integral curves of $X^{H} $, that is, curves $\hat{\gamma}:\mathbb{R}\rightarrow U$ defined by the conditions
\begin{eqnarray}
\label{eq:ham_solution_1}
\hat{\gamma}(\tau)&=&\varphi_{\tau}^{H}(\hat{\gamma}(0)),\\
\label{eq:ham_solution_2}
X^{H}\circ \hat{\gamma}(\tau)&=&\displaystyle\frac{d\hat{\gamma}}{d\tau}.
\end{eqnarray}

The manifold $U$ is oriented by virtue of the existence of the Liouville top-form \cite{marsden1999,da2008lectures}
\begin{equation}
\label{eq:liouville_top}
\omega_{\Omega}=\displaystyle\frac{1}{D!}\Omega^{\wedge D}=\displaystyle\frac{1}{D!}\underbrace{\Omega\wedge \Omega\wedge...\wedge \Omega}_{D\textup{ times}}.
\end{equation}
Such a top-form induces, on the measurable space $(U,\mathfrak{B}(U))$ with the Borel $\sigma$-algebra\footnote{A $\sigma$-algebra of a set is the family of its subsets which, containing the void and the set itself, is closed under the complement and under the countable union \cite{guilleminmeasure,athreyameasure}. This definition formalizes the idea of how to divide a set to measure its parts. In particular, the Borel algebra $\mathfrak{B}(U)$ of a smooth manifold $U$ is the smallest $\sigma$-algebra of $U$ that contains its topology. As such, it is the set of all open sets of $U$ and its complements (the closed sets), as well as any countable unions and intersections of them \cite{guilleminmeasure,athreyameasure}.}, the so-called Liouville measure $\lambda_{\omega}:\mathfrak{B}(U)\rightarrow [0,+\infty)$ defined by \cite{da2008lectures}
\begin{equation}
\label{eq:liouville_measure}
\lambda_{\omega}(A)=\displaystyle\int_{A}\omega_{\Omega}(u),
\end{equation}
whose principal feature is that it assigns  volumes invariant under symplectomorphisms, that is, if $f\in \mathcal{C}^{\infty}(U)$ is any symplectomorphism, then for all $ A\in \mathfrak{B}(U)$ holds
\begin{equation}
\label{eq:invariance_omega}
\lambda_{\omega}(f(A))=\lambda_{\omega}(A).
\end{equation}
Indeed, given the invariance of the top-form $\omega_{\Omega}$ under the pullback $f^{*}$, inherited from the invariance of the symplectic 2-form (see Eq.(\ref{eq:hamiltonian_flux_3})), we have
\[\lambda_{\omega}(f(A))\!=\!\displaystyle\int_{f(A)}\!\!\!\!\!\!\!\!\omega_{\Omega}(u)\!=\!\displaystyle\int_{A}\!\!\!f^{*}\omega_{\Omega}\circ f(v)\!=\!\displaystyle\int_{A}\!\!\!\omega_{\Omega}(v)=\lambda_{\omega}(A).\]
In particular, the Liouville measure is invariant under the Hamiltonian flow $\varphi^{H}$ of the system, a fact that is known as Liouville's theorem \cite{marsden1999}.

With this brief background of measure theory, we can define a statistical state of a physical system as a probability density function (pdf) $\rho:U\rightarrow [0,+\infty)$, related to (\ref{eq:liouville_measure}) in the form \cite{marle2020gibbs}
\begin{equation}
\label{eq:statistical_state}
\lambda_{\rho}(A)=\displaystyle\int_{A}d\lambda_{\omega}(u)\rho(u)=\displaystyle\int_{A}\omega_{\Omega}(u)\rho(u),
\end{equation}
such that it satisfies the normalization condition:
\begin{equation}
\label{eq:normalization}
\displaystyle\int_{U}d\lambda_{\omega}(u)\rho(u)=1.
\end{equation}

The physical interpretation of $\rho$ is that of the ensemble theory: when the number of particles of the physical system is large, and the correlations between them are few, the dimension of the phase space $U$ must be even larger in order to capture all the degrees of freedom of the system \cite{callen1985}. For example, in a mechanical system of $N$ particles, the dimension of $U$ is around the order $6N$ \cite{callen1985}. In any case, the system of differential equations (\ref{eq:hamiltonian_vector}) that must be solved to find the Hamiltonian flow $\varphi^{H}$, coupled to the set of initial conditions, to fully specify the state of the system, ends up being computationally overwhelming. For this reason it is necessary to use statistical methods. Instead of trying to get exactly the dynamic state $u$ at all times, we can generate enough copies (an infinite number, if necessary) of the system and, simultaneously, perform the measurement of the dynamic state in each of these copies, generating with this a big data set at all times. In this sense, $u$ becomes a random variable and $\rho$ describes the way in which $u$ is distributed over the entire manifold $U$. Thus, physically we interpret $\lambda_{\rho}(A)$ as the probability of finding the state of the system in the measurable subset $A\subset U$.

It is common that the dynamic state of a system does not completely visit all $U$, but for dynamic reasons it is constrained to remain only in a certain region $\mathcal{A}\subset U$, which must be invariant under the Hamiltonian flow:
\begin{equation}
\label{eq:invarianza_soporte}
\varphi^{H}(\mathcal{A})=\mathcal{A}.
\end{equation}
In such a case, we look for the statistical state to have support over $\mathcal{A}$, i.e.,  the statistical model constraints the system to remain in $\mathcal{A}$, requiring that there be no distributed probability in the complement $\mathcal{A}^{c}=U-\mathcal{A}$, that is\footnote{Formally, the support of a statistical state is the smallest of all closed sets with measure 1 \cite{guilleminmeasure,athreyameasure}.}
\begin{equation}
\label{eq:soporte}
\lambda_{\omega}(\mathcal{A})=\displaystyle\int_{\mathcal{A}}d\lambda_{\omega}(u)\rho(u)=1.
\end{equation}
In this regard, it is a well-known fact that the probability distribution that best describes the state of a system, for which there is no prior information, is the constant and uniform probability distribution \cite{jaynes1957information}
\begin{equation}
\label{eq:constant_m}
\rho_{0}(u)=\displaystyle\frac{\Theta_{\mathcal{A}}(u)}{\textup{Vol}(\mathcal{A})},
\end{equation}
where $\textup{Vol}(\mathcal{A})=\lambda_{\omega}(\mathcal{A})$ is the volume assigned to $\mathcal{A}$ by the Liouville measure, and 
\begin{equation}
\label{eq:characteristics}
\Theta_{W}(u)=\left\{\begin{array}{lll}
1,&&u\in W,\\
0,&&u\not\in W
\end{array}\right.,  
\end{equation}
a notation we will use throughout this work. 

Opposite to this situation, we have the case where we are not completely unaware of the properties of the state of the system, because either from an experimental or a theoretical point of view, we know that some macroscopic property comes from a symmetry of the microscopic system \cite{callen1985,callensym}. This is the case, precisely, when the evolution of the system is Hamiltonian and the conservation of $H$ is taken into account (see Eq.(\ref{eq:hamiltonian_flux_2})), in which case the best distribution available to describe the statistics of the system is the Gibbs distribution \cite{marle2020gibbs}
\begin{equation}
\label{eq:gibbs}
\rho(\beta;u)=\displaystyle\frac{\Theta_{\mathcal{A}}(u)}{Z(\beta)}e^{-\beta H(u)},
\end{equation}
where $Z(\beta)$ is the so-called partition function,
\begin{equation}
\label{eq:partition_normal}
Z(\beta)=\displaystyle\int_{\mathcal{A}}d\lambda_{\omega}(u)e^{-\beta H(u)},
\end{equation}
and $\beta>0$ is a real parameter that can be identified with the inverse of the temperature of a system in thermodynamic equilibrium from the statistical point of view \cite{marle2020gibbs}. We say that a statistical state is an equilibrium statistical state if it is invariant under the Hamiltonian flow of the system \cite{tuckerman2010},
\begin{equation}
\label{eq:estado_equilibrio}
\mathscr{L}_{X^{H}}\rho(u)=0.
\end{equation}
Naturally,  (\ref{eq:gibbs}) is an equilibrium state because its only dependence on $u$ is through $\Theta_{\mathcal{A}}$ and $H$, and both functions are invariant under the Hamiltonian flow of the system (see Eqs.(\ref{eq:hamiltonian_flux_2}) and (\ref{eq:invarianza_soporte})).

\subsection{Symmetry groups and momentum map}
\label{sub:momentum_map}
Consider the more general case in which we do not have only a monoparametric group of symmetries of the system, as is the case of the family $\{\varphi^{H}_{\tau}\}_{\tau\in \mathbb{R}}$ generated by $\varphi^{H}$, instead we have a $G$-parametric group of symplectomorphisms, where $G$ is a Lie group of dimension $\textup{dim}(G)=k$. Concretely, given a Lie group $G$ with the group product represented by concatenation,
\[\forall\Lambda_{1},\Lambda_{2}\in G,\quad \Lambda_{1}\Lambda_{2}\in G,\] 
and identity element  $e\in G$ such that
\[\forall \Lambda\in G,\quad \Lambda e=e \Lambda =\Lambda.\]
Furthermore, we define a smooth left action on $U$ as a map
\begin{equation}
\label{eq:action}
\Psi:G\times U\rightarrow U\ ,
\end{equation}
which, via the diffeomorphisms $\Psi_{\Lambda}:u\mapsto \Psi(\Lambda,u)$, satisfies the properties \cite{marsden1999,nakahara2018,izquierdogroups}
\begin{eqnarray}
\label{eq:action_1}
\Psi_{e}&=&\textup{Id}_{U}:u\mapsto u,\\
\label{eq:action_2}
\Psi_{\Lambda_{1}}\circ \Psi_{\Lambda_{2}}&=&\Psi_{\Lambda_{1}\Lambda_{2}}.
\end{eqnarray}

If for each $\Lambda\in G$ the diffeomorphism $\Psi_{\Lambda}$ is a symplectomorphism of the symplectic structure $\Omega$, then we say that the action is symplectic and defines a $G$-parametric $\{\Psi_{\Lambda}\}_{\Lambda\in G}$ group of symplectomorphisms. If, in addition, for all $\Lambda\in G$ the Hamiltonian remains unchanged under the symplectomorphism $\Psi_{\Lambda}$, 
\begin{equation}
\label{eq:hamiltonian_invariance}
H\circ \Psi_{\Lambda}=H,
\end{equation}
then we say that the action $\Psi$ leaves the Hamiltonian system $(U,\Omega,H)$ invariant and, as such, each $\Lambda\in G$ defines a symmetry of the system through $\Psi_{\Lambda}$. In this way, we can precisely understand $\varphi^{H}$ as a particular case of the symplectic action of the Abelian group $\mathbb{R}=(\mathbb{R},+)$ on $U$ that leaves the same Hamiltonian system $(U,\Omega,H)$ invariant.

Associated with the concept of a smooth left action we have the concept of  fundamental fields of the Lie algebra of the group, which we will denote as $\mathfrak{g}$. The fundamental vector field $\psi_{\chi}\in \mathfrak{X}(T^{*}U)$ of the element $\chi \in \mathfrak{g}$, is the vector field over $U$ generated by the action of the one-parameter subgroup $\{\textup{exp}(s\chi)\}_{s\in \mathbb{R}}$ \cite{marsden1999}, 
\begin{equation}
\label{eq:fun_fields}
\psi_{\chi}(u)=\displaystyle\frac{d}{ds}\Psi_{e^{s\chi}}(u)\Bigr|_{s=0},
\end{equation}
where $\textup{exp}:\mathbb{R}\times \mathfrak{g}\rightarrow G$ is the exponential map that relates group elements near the identity with displacement/tangent vectors in the tangent space of $G$, under the identification $\mathfrak{g}\simeq T_{e}G$ \cite{nakahara2018}.

\begin{figure}[H]
	\centering
	\tikzset{every picture/.style={line width=0.75pt}} 

\begin{tikzpicture}[x=0.75pt,y=0.75pt,yscale=-0.85,xscale=0.85]

\draw   (301.83,73.28) .. controls (312.83,55.28) and (345,60) .. (389.83,75.28) .. controls (434.66,90.56) and (434.83,165.28) .. (406.83,193.28) .. controls (378.83,221.28) and (265.83,244.28) .. (282,161) .. controls (298.17,77.72) and (290.83,91.28) .. (301.83,73.28) -- cycle ;
\draw   (82,154.41) .. controls (82,118.84) and (110.84,90) .. (146.41,90) .. controls (181.99,90) and (210.83,118.84) .. (210.83,154.41) .. controls (210.83,189.99) and (181.99,218.83) .. (146.41,218.83) .. controls (110.84,218.83) and (82,189.99) .. (82,154.41) -- cycle ;
\draw   (122.4,86.64) -- (195.09,86.64) -- (143.6,137.36) -- (70.91,137.36) -- cycle ;
\draw [color={rgb, 255:red, 208; green, 2; blue, 27 }  ,draw opacity=1 ]   (133,112) .. controls (208.68,125.08) and (163.24,142.74) .. (197.22,164.29) ;
\draw [shift={(198.83,165.28)}, rotate = 30.74] [color={rgb, 255:red, 208; green, 2; blue, 27 }  ,draw opacity=1 ][line width=0.75]      (0, 0) circle [x radius= 3.35, y radius= 3.35]   ;
\draw [shift={(133,112)}, rotate = 9.81] [color={rgb, 255:red, 208; green, 2; blue, 27 }  ,draw opacity=1 ][fill={rgb, 255:red, 208; green, 2; blue, 27 }  ,fill opacity=1 ][line width=0.75]      (0, 0) circle [x radius= 3.35, y radius= 3.35]   ;
\draw [color={rgb, 255:red, 65; green, 117; blue, 5 }  ,draw opacity=1 ]   (306,109) .. controls (357.05,157.54) and (400.51,140.8) .. (392.24,182.34) ;
\draw [shift={(391.83,184.28)}, rotate = 102.8] [color={rgb, 255:red, 65; green, 117; blue, 5 }  ,draw opacity=1 ][line width=0.75]      (0, 0) circle [x radius= 3.35, y radius= 3.35]   ;
\draw [shift={(306,109)}, rotate = 43.55] [color={rgb, 255:red, 65; green, 117; blue, 5 }  ,draw opacity=1 ][fill={rgb, 255:red, 65; green, 117; blue, 5 }  ,fill opacity=1 ][line width=0.75]      (0, 0) circle [x radius= 3.35, y radius= 3.35]   ;
\draw [color={rgb, 255:red, 208; green, 2; blue, 27 }  ,draw opacity=1 ][line width=1.5]    (133,112) -- (192.93,125.41) ;
\draw [shift={(196.83,126.28)}, rotate = 192.61] [fill={rgb, 255:red, 208; green, 2; blue, 27 }  ,fill opacity=1 ][line width=0.08]  [draw opacity=0] (11.61,-5.58) -- (0,0) -- (11.61,5.58) -- cycle    ;
\draw [color={rgb, 255:red, 65; green, 117; blue, 5 }  ,draw opacity=1 ][line width=1.5]    (306,109) -- (352.93,153.53) ;
\draw [shift={(355.83,156.28)}, rotate = 223.5] [fill={rgb, 255:red, 65; green, 117; blue, 5 }  ,fill opacity=1 ][line width=0.08]  [draw opacity=0] (11.61,-5.58) -- (0,0) -- (11.61,5.58) -- cycle    ;
\draw    (171.83,221.28) .. controls (187.51,247.74) and (259.85,253.07) .. (296.62,225.98) ;
\draw [shift={(298.83,224.28)}, rotate = 141.15] [fill={rgb, 255:red, 0; green, 0; blue, 0 }  ][line width=0.08]  [draw opacity=0] (8.93,-4.29) -- (0,0) -- (8.93,4.29) -- cycle    ;
\draw   (274.49,83.64) -- (347.17,83.64) -- (337.51,134.36) -- (264.83,134.36) -- cycle ;
\draw    (165.83,80.28) .. controls (189.47,65.5) and (233.48,43.94) .. (287.36,74.83) ;
\draw [shift={(289.83,76.28)}, rotate = 210.96] [fill={rgb, 255:red, 0; green, 0; blue, 0 }  ][line width=0.08]  [draw opacity=0] (8.93,-4.29) -- (0,0) -- (8.93,4.29) -- cycle    ;
\draw   (151.83,275) -- (332.83,275) -- (332.83,407.28) -- (151.83,407.28) -- cycle ;
\draw [color={rgb, 255:red, 208; green, 2; blue, 27 }  ,draw opacity=1 ]   (165.83,298.28) .. controls (192.99,280.82) and (185.34,310.4) .. (204.02,300.33) ;
\draw [shift={(205.83,299.28)}, rotate = 328.24] [color={rgb, 255:red, 208; green, 2; blue, 27 }  ,draw opacity=1 ][line width=0.75]      (0, 0) circle [x radius= 3.35, y radius= 3.35]   ;
\draw [shift={(165.83,298.28)}, rotate = 327.26] [color={rgb, 255:red, 208; green, 2; blue, 27 }  ,draw opacity=1 ][fill={rgb, 255:red, 208; green, 2; blue, 27 }  ,fill opacity=1 ][line width=0.75]      (0, 0) circle [x radius= 3.35, y radius= 3.35]   ;
\draw [color={rgb, 255:red, 65; green, 117; blue, 5 }  ,draw opacity=1 ]   (165.83,326.28) .. controls (192.99,308.82) and (185.34,338.4) .. (204.02,328.33) ;
\draw [shift={(205.83,327.28)}, rotate = 328.24] [color={rgb, 255:red, 65; green, 117; blue, 5 }  ,draw opacity=1 ][line width=0.75]      (0, 0) circle [x radius= 3.35, y radius= 3.35]   ;
\draw [shift={(165.83,326.28)}, rotate = 327.26] [color={rgb, 255:red, 65; green, 117; blue, 5 }  ,draw opacity=1 ][fill={rgb, 255:red, 65; green, 117; blue, 5 }  ,fill opacity=1 ][line width=0.75]      (0, 0) circle [x radius= 3.35, y radius= 3.35]   ;
\draw [color={rgb, 255:red, 208; green, 2; blue, 27 }  ,draw opacity=1 ][line width=1.5]    (165,355) -- (206.83,355.25) ;
\draw [shift={(210.83,355.28)}, rotate = 180.35] [fill={rgb, 255:red, 208; green, 2; blue, 27 }  ,fill opacity=1 ][line width=0.08]  [draw opacity=0] (11.61,-5.58) -- (0,0) -- (11.61,5.58) -- cycle    ;
\draw [color={rgb, 255:red, 65; green, 117; blue, 5 }  ,draw opacity=1 ][line width=1.5]    (166,383) -- (207.83,383.25) ;
\draw [shift={(211.83,383.28)}, rotate = 180.35] [fill={rgb, 255:red, 65; green, 117; blue, 5 }  ,fill opacity=1 ][line width=0.08]  [draw opacity=0] (11.61,-5.58) -- (0,0) -- (11.61,5.58) -- cycle    ;

\draw (70,182) node [anchor=north west][inner sep=0.75pt]    {$G$};
\draw (414,180) node [anchor=north west][inner sep=0.75pt]    {$U$};
\draw (204,216) node [anchor=north west][inner sep=0.75pt]    {$\Psi(\bullet,u)$};
\draw (195,31) node [anchor=north west][inner sep=0.75pt]    {$[ \Psi(\bullet,u)]_{*}$};
\draw (221,286) node [anchor=north west][inner sep=0.75pt]    {$\textup{exp}(s\chi)$};
\draw (221,317) node [anchor=north west][inner sep=0.75pt]    {$\Psi_{\textup{exp}(s\chi)}( u)$};
\draw (224,344) node [anchor=north west][inner sep=0.75pt]    {$\chi\in T_{e}G$};
\draw (224,369) node [anchor=north west][inner sep=0.75pt]    {$\psi_{\chi}(u)\in T_{u}U$};
\draw (85,66) node [anchor=north west][inner sep=0.75pt]    {$T_{e}G$};
\draw (347,69) node [anchor=north west][inner sep=0.75pt]    {$T_{u}U$};

\end{tikzpicture}
	\caption{Fundamental fields as pushforwards.}
	\label{fig:fun_fields}
\end{figure}
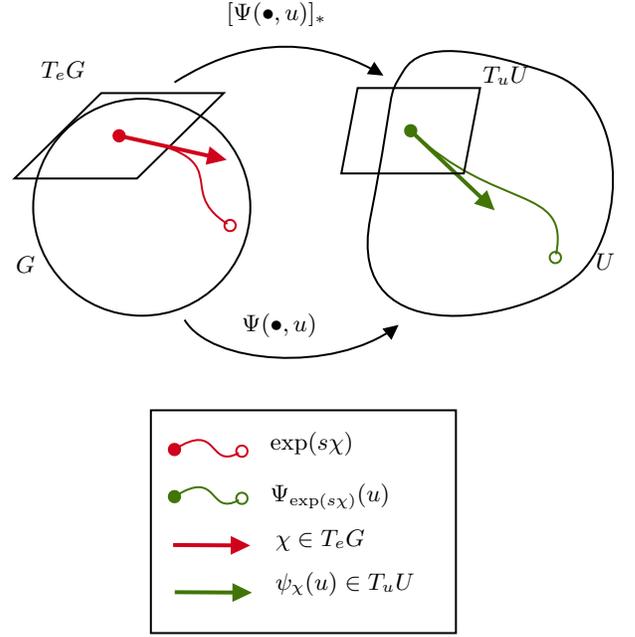

Another way of understanding the concept of the fundamental field $\psi_{\chi}$ of $\chi\in \mathfrak{g}$, is as the pushforward (see Fig.(\ref{fig:fun_fields}))
\begin{equation}
\label{eq:fun_fields_push}
\psi_{\chi}(u)=[\Psi_{e}(u)]_{*}\chi.
\end{equation}

\noindent From this idea we can deduce that the assignment of fundamental fields to elements in the Lie algebra is a linear assignment, that is, for any finite number of elements $\chi_{i}\in \mathfrak{g}$ and for any equal finite number of real constants $\alpha_{i}\in \mathbb{R}$, it is fulfilled
\begin{equation}
\label{eq:fun_fields_linearity}
\psi_{\alpha^{i}\chi_{i}}(u)=\alpha^{i}\psi_{\chi_{i}}(u).
\end{equation}

When the action of $G$ on $U$ is symplectic, the fundamental fields (\ref{eq:fun_fields}) are infinitesimal generators of symplectomorphisms, i. e.,
\[\mathscr{L}_{\psi_{\chi}}\Omega=\frac{d}{ds}[\Psi_{e^{s\chi}}]^{*}\Omega\circ \Psi_{e^{s\chi}}\Bigr|_{s=0}=\displaystyle\frac{d}{ds}\Omega\Bigr|_{s=0}=0,\]
and, at least locally, are Hamiltonian vector fields for some real functions $\mu_{\chi}:U\rightarrow \mathbb{R}$, that is, it is satisfied
\begin{equation}
\label{eq:hamiltonians}
i_{\psi_{\chi}}\Omega=-d\mu_{\chi}.
\end{equation}
Indeed, this result is obtained using Cartan's magic formula considering that all closed forms are, at least locally, exacts forms \cite{nakahara2018}:
\[\begin{array}{c}
\mathscr{L}_{\psi_{\chi}}\Omega=i_{\psi_{\chi}}d\Omega+d(i_{\psi_{\chi}}\Omega)=d(i_{\psi_{\chi}}\Omega)=0\\
\\
\Downarrow\\
\\
i_{\psi_{\chi}}\Omega=-d\mu_{\chi}.
\end{array}\]
When the fundamental fields (\ref{eq:fun_fields}) are globally Hamiltonian, Eq.(\ref{eq:hamiltonians}) is valid on all $U$. A particular scenario where this happens is when the symplectic form is exact for a 1-form $\vartheta\in \Lambda^{1}(U)$, $\Omega=d\vartheta$, and the action $\Psi$ leaves this symplectic potential invariant,
\begin{equation}
\label{eq:potential}
\,[\Psi_{\Lambda}]^{*}\vartheta\circ \Psi_{\Lambda}=\vartheta.
\end{equation}
Consequently, the action is symplectic,
\[\begin{array}{ll}
\,[\Psi_{\Lambda}]^{*}\Omega\circ \Psi_{\Lambda}\!\!\!&=[\Psi_{\Lambda}]^{*}(d\vartheta)\circ \Psi_{\Lambda}\\
\\
&=d\left([\Psi_{\Lambda}]^{*}\vartheta\circ\Psi_{\Lambda}\right)=d\vartheta=\Omega,
\end{array}\]
and the fundamental fields are globally Hamiltonian for the functions:
\begin{equation}
\label{eq:hamiltonianas_chi}
\mu_{\chi}(u)=i_{\psi_{\chi}}\vartheta(u).
\end{equation}
Indeed, taking into account that Eq.(\ref{eq:potential}) implies $\mathscr{L}_{\psi_{\chi}}\vartheta=0$, and again using Cartan's magic formula, we get
\[\begin{array}{c}
\mathscr{L}_{\psi_{\chi}}\vartheta=i_{\psi_{\chi}}d\vartheta+d(i_{\psi_{\chi}}\vartheta)=i_{\psi_{\chi}}\Omega+d\mu_{\chi}=0\\
\\
\Downarrow\\
\\
d\mu_{\chi}=-i_{\psi_{\chi}}\Omega.
\end{array}\]
 
Given the linearity in the RHS of (\ref{eq:hamiltonianas_chi}) in the vectors $\chi$ of the Lie algebra (remember Eq.(\ref{eq:fun_fields_linearity})), the function $\mu_{\chi}$ in the LHS must also be linear in $\chi$. This observation allows us to define, associated with the symplectic action $\Psi$ of $G$ over $U$, the map $\mu:U\rightarrow \mathfrak{g} ^{*}$ called the momentum map given by \cite{marsden1999}
\begin{equation}
\label{eq:momentum_map}
i_{\chi}\mu(u)=\mu_{\chi}(u).
\end{equation}
An important property of the momentum map is that, when $G$ acts on $(U,\Omega,H)$ by symmetries, then the functions $\mu_{\chi}$ are integral of motion for any element $\chi \in \mathfrak{g}$;
this means that they remain constant along the curves $\hat{\gamma}$, which  satisfy both Eq.(\ref{eq:ham_solution_1}) and Eq.(\ref{eq:ham_solution_2}). This can be proved as follows:
\[\begin{array}{ll}
(\mathscr{L}_{X^{H}}\mu_{\chi})(u)\!\!\!&=(i_{X^{H}}d\mu_{X})(u)\\
\\
&=-\Omega(u)(\psi_{\chi},X^{H})\\
\\
&=\Omega(u)(X^{H},\psi_{\chi})=-i_{\psi_{\chi}}(-i_{X^{H}}\Omega)(u)\\
\\
&=-(i_{\psi_{\chi}}dH)(u)=-(\mathscr{L}_{\psi_{\chi}}H)(u)=0.
\end{array}\]
In this way, we can interpret the momentum map $\mu$ as a geometric object that stores all the constants of motion of the Hamiltonian system $(U,\Omega,H)$, such that to extract the conserved quantity associated with the symmetry generated by the monoparametric group $\{\Psi_{e^{s\chi}}\}_{s\in \mathbb{R}}$, it is enough to evaluate $\mu$ in $\chi$.

In the example we are considering with the Abelian group $\mathbb{R}$, acting on $(U,\Omega=d\vartheta,H)$ through the Hamiltonian flow $\varphi^{H}$, as the Lie algebra associated with $\mathbb{R}$ is just $\mathbb{R}$, which is trivially identified with its dual vector space $\mathbb{R}^{*}$,  the momentum map of the action is trivially the same as the Hamiltonian function,
\begin{equation}
\label{eq:hamiltonian_mom_map}
\mu=Hd\tau\in \mathbb{R}^{*}.
\end{equation}
Then, the fundamental fields $aX^{H}=\frac{d\varphi^{H}_{a\tau}}{d\tau }\bigr|_{\tau=0}$ of the vectors $a\frac{\partial}{\partial \tau}\in \mathbb{R}$ are associated with the Hamiltonians $aH=i_{a\frac{ \partial}{\partial \tau}}Hd\tau$, as we can see from 
\[-i_{aX^{H}}\Omega=-ai_{X^{H}}\Omega=a dH=d(aH).\]

Given the importance of the momentum map in the formalism of Lie groups thermodynamics, we emphasize here the manner it transforms under the action of the group $G$ itself. We will say that the momentum map $\mu$ is equivariant if, for any elements $\Lambda\in G$ and $x\in U$, the next relation is fulfilled \cite{souriau1997structure,marsden1999}:
\begin{equation}
\label{eq:equivariant}
\mu\circ \Psi_{\Lambda}(x)=\textup{Ad}_{\Lambda}^{*}(\mu(x)),
\end{equation}
where $\textup{Ad}^{*}:G\times \mathfrak{g}^{*}\rightarrow \mathfrak{g}^{*}$ is the co-Adjoint action (\ref{eq:coAdjoint}) of the group $G$ on the dual of its Lie algebra. See 
Appendix(\ref{ap:equivariant}) for a more detailed explanation of this result.

\subsection{Generalized statistical states}
\label{sub:generalized_statistical_states}
Let us summarize the concepts described in the last two subsections. Let $\rho$ be a pdf of a statistical state of a Hamiltonian system $(U,\Omega=d\vartheta,H)$ with a Lie group that implements the symmetries of the system by means of a symplectic action $\Psi:G\times U\rightarrow U$. As a special requirement, suppose that for each $\Lambda\in G$, $\Psi_{\Lambda}$ maps the support $\mathcal{A}$ of $\rho$ to itself,
\begin{equation}
\label{eq:soporte_invariance}
\Psi_{\Lambda}(\mathcal{A})=\mathcal{A}.
\end{equation}
The claim that the Gibbs distribution defined in Eq.(\ref{eq:gibbs}) best fits the statistics of a Hamiltonian system, where the average of the Hamiltonian itself is a macroscopic conserved quantity along the state of the system, can be generalized considering that, for all $\chi\in \mathfrak{g}$, the averages of the functions (\ref{eq:hamiltonianas_chi}) also must be preserved because they are integrals of motions of the system \cite{marle2020gibbs}. As a result of this generalization, we can define the generalized Gibbs states \cite{marle2020gibbs}
\begin{equation}
\label{eq:gibbs_generalized}
\rho(\chi;u)=\displaystyle\frac{\Theta_{\mathcal{A}}(u)}{Z(\chi)}e^{-\mu_{\chi}(u)},
\end{equation}
where $Z(\chi)$ is the generalized partition function
\begin{equation}
\label{eq:partition_generalized}
Z(\chi)=\displaystyle\int_{\mathcal{A}}d\lambda_{\omega}(u)e^{-\mu_{\chi}(u)},
\end{equation}
and the element $\chi\in \mathfrak{g}$ is a Souriau vector\footnote{In the Lie group thermodynamics literature, these vectors are called ``generalized temperatures" \cite{marle2020gibbs}; however, we will use this name in a more physical context to designate a different concept  that will appear later in this work.}, which is, roughly speaking, a vector of the Lie algebra that ensures the existence of the integrals $\int_{\mathcal{A}}d\lambda_{\omega}(u)e^{-\mu_{\chi}(u)}$ and $\int_{\mathcal{A}}d\lambda_{\omega}(u)\mu(u)e^{-\mu_{\chi}(u)}$ \cite{marle2020gibbs}.

Given the importance of  Souriau vectors, we formulate a rigorous definition as follows. A vector $\chi\in \mathfrak{g}$ is a Souriau vector if and only if there is a neighborhood
\begin{equation}
\label{eq:neigh}
V_{\chi}\subset \mathfrak{g}
\end{equation}
and an integrable function $f_{\chi}:U\rightarrow \mathbb{R}$ over all $\mathcal{A}$, such that for every point $u\in \mathcal{A}$ and every vector $\chi'\in V_{\chi}$, the function $f_{\chi}$ is an upper bound of the function $e^{-\mu_{\chi'}(u)}$ \cite{marle2020gibbs}:
\begin{equation}
\label{eq:souriau_vector}
e^{-\mu_{\chi'}(u)}\leq f_{\chi}(u),\qquad \forall u\in \mathcal{A},\quad \chi'\in V_{\chi}.
\end{equation}

We will represent the set of Souriau vectors for a given momentum map as the subset $\mathfrak{g}_{\star}\subset \mathfrak{g}$, i. e.,
\begin{equation}
\label{eq:souriau_vector_set}
\mathfrak{g}_{\star}=\{\chi\in \mathfrak{g}:\chi\textup{ is a Souriau vector}\}.
\end{equation}
The set $\mathfrak{g}_{\star}$ has interesting properties. For example, if it is not empty, it is an open convex subset \cite{marle2020gibbs}. Furthermore, if the momentum map is equivariant (see Eq.(\ref{eq:equivariant})), then $\mathfrak{g}_{\star}$ is invariant under the adjoint action of the whole group $G$ \cite{marle2020gibbs}
\begin{equation}
\label{eq:invariance_gstar}
\textup{Ad}_{\Lambda}(\mathfrak{g}_{\star})=\mathfrak{g}_{\star}.
\end{equation}
This is true because if $\chi\in \mathfrak{g}$ is a Souriau vector with neighborhood $V_{\chi}$ and function $f_{\chi}$ such that both satisfy the Souriau's condition (\ref{eq:souriau_vector}), then $\textup{Ad}_{\Lambda}(\chi)$ also has a neighborhood
\[V_{\textup{Ad}_{\Lambda}(\chi)}=\{\textup{Ad}_{\Lambda}(\chi')\}_{\chi'\in V_{\chi}},\]
such that Eq.(\ref{eq:souriau_vector}) is satisfied by the function $f_{\chi}$:
\[\begin{array}{ll}
e^{-\mu_{\textup{Ad}_{\Lambda}(\chi')}(u)}\!\!\!&=e^{-(\textup{Ad}_{\Lambda^{-1}}^{*}\mu)_{\chi'}(u)}\\
\\
&=e^{-\mu_{\chi'}\circ \Psi_{\Lambda^{-1}}(u)}=e^{-\mu_{\chi'}(v)}\leq f(v).
\end{array}\]

In the case of the Abelian group $\mathbb{R}$, which acts as a group of symmetry in $(U,\Omega=d\vartheta,H)$ through $\varphi^{H}$, note that $\beta\frac{\partial}{\partial \tau}\in \mathbb{R}$ is a Souriau vector for $\beta>0$ if $H$ is interpreted as the positive energy of the system. Indeed,  in this case, we can define around $\beta\frac{\partial}{\partial \tau}$ the neighborhood
\[V_{\beta\frac{\partial}{\partial \tau}}=\left\{\beta'\frac{\partial}{\partial \tau}\in \mathbb{R}:\beta-\epsilon \leq\beta'\leq \beta+\epsilon\right\}\]
for $\epsilon \ll 1$ such that $\beta>\epsilon$. Then, for any $\beta'\frac{\partial}{\partial \tau}\in V_{\beta}$, the function $e^{-i_{\beta'\frac{\partial}{\partial \tau}}H(u) d\tau}=e^{-\beta'H(u)}$ is a positive-definite and decreasing function in all $\mathcal{A}$, and thus it is bounded from above by any constant function with value $c\geq 1$, satisfying the defining condition (\ref{eq:souriau_vector}) for a Souriau vector. 

With this  preamble of Lie group thermodynamics, we can now build the Hamiltonian system that models a dilute ideal gas in a static spacetime, and determine the corresponding generalized Gibbs statistical states.
\color{black}

\section{Hamiltonian description of a massive free particle}
\label{sec:hamiltonian_description}
Consider a particle of mass $m$ in a spacetime $(M,g)$, where $M$ is a smooth $D=1+d$-dimensional manifold ($d\geq 3$) and $g$ is a Lorentzian metric with signature $(-1,+d)$. We know that, if the particle is free, its world line in $M$ will be a geodesic curve with respect to the torsionless Levi-Civita connection \cite{moore}. 
  This configuration can be described by means of a Hamiltonian system as follows. Let $(T^{*}M,\Omega=d\vartheta)$ be a symplectic manifold, where $T^{*}M$ is the cotangent bundle of $M$, whose points can be taken as ordered pairs $u=(x,p)\in T^{*}M$ with $x\in M$ and $p\in T_{x}^{*}M$, which induce a vector bundle structure over $M$ with natural projection $\pi:(x,p)\mapsto x$ \cite{nakahara2018}. The symplectic structure $\Omega=d\vartheta$ emerges naturally in these manifolds from the 1-form $\vartheta$ defined as \cite{da2008lectures}
\begin{equation}
\label{eq:canonical_1_form}
\vartheta(u)=\vartheta(x,p)=[\pi(u)]^{*}p(x),
\end{equation}
or equivalently for any vector field $X\in \mathfrak{X}(T^{*}M)$,
\begin{equation}
\label{eq:canonical_1_form_2}
i_{X}\vartheta(x,p)=i_{[\pi(u)]_{*}X}p(x).
\end{equation}

On this symplectic manifold, the Hamiltonian function is given by \cite{ehlers1973}
\begin{equation}
\label{eq:hamiltonian}
H(x,p)=\displaystyle\frac{g^{-1}(x)(p,p)}{2m},
\end{equation}
where $g(x)^{-1}$ is the inverse metric of $g(x)$. In this way, the curves $\hat{\gamma}:\mathbb{R}\rightarrow T^{*}M$ that satisfy  Eqs.(\ref{eq:ham_solution_1}) and (\ref{eq:ham_solution_2}), can be expressed as the pair
\begin{equation}
\label{eq:solutions}
\tau\mapsto \hat{\gamma}(\tau)=(\gamma(\tau),p\circ \gamma(\tau)),
\end{equation}
where $\pi\circ \widehat{\gamma}=\gamma$ is a geodesic curve of the metric tensor $g$, while the vertical part $p\circ \gamma(\tau)$ is related to the velocity vector field $U\circ \gamma( \tau)=\frac{d\gamma}{d\tau}$ of the geodesic. Thus,  $m\,U\in \mathfrak{X}(M)$ is the metric gradient of $p$ over the geodesics $\gamma$ \cite{ehlers1973}:
\begin{equation}
\label{eq:gradient_p}
m\,U\circ \gamma(\tau)=\textup{grad}\,p\circ \gamma(\tau).
\end{equation}
This means that, for any vector field $X\in \mathfrak{X}(M)$, we have
\begin{equation}
\label{eq:gradient_p_2}
m\,g(\gamma(\tau))(U,X)=i_{X}p\circ \gamma(\tau).
\end{equation}
\color{black}
As a consequence of the semi-Riemannian nature of the pair $(M,g)$, there are three different types of distances that can be defined in $M$ through $g$ and, therefore, three different types of geodesics (time-like, space-like and null-like), each with its unique physical and geometric properties \cite{moore}. In the case of free massive particles, the trajectories described by means of Eq.(\ref{eq:hamiltonian}) are time-like geodesics. Geometrically, this means that the velocity vector field $U$ is a time-like vector field,
\[g(\gamma(\tau))(U,U)<0.\]
In particular, without losing generality, we can parameterize the geodesic by the arc length to obtain \cite{moore}
\begin{equation}
\label{eq:time_like_U}
g(\gamma(\tau))(U,U)=-1,
\end{equation}
and using the relation between $p$ and $U$ (see Eq.(\ref{eq:gradient_p_2})) we can \textit{lift} Eq.(\ref{eq:time_like_U}) to the cotangent bundle $T^{*}M$ as
\begin{equation}
\label{eq:time_like_p}
-m^{2}=g(\gamma(\tau))^{-1}(p,p).
\end{equation}
In other words, once the initial conditions of the particle are given, the generalized positions and momenta of the particle must satisfy \cite{moore}
\begin{equation}
\label{eq:mass_shell}
-m^{2}=g(x)^{-1}(p,p).
\end{equation}

Eq.(\ref{eq:mass_shell}) fixes, in the cotangent bundle $T^{*}M$, a hyperboloid where the trajectory $\widehat{\gamma}$ of the particle is constrained to be. Such a hyperboloid can be seen as the zero-level hypersurface of the so-called mass-shell function $\mathcal{P}:T^{*}M\rightarrow \mathbb{R}$, defined by
\begin{equation}
\label{eq:mass_shell_function}
\mathcal{P}(x,p)=m^{2}+g(x)^{-1}(p,p).
\end{equation}
The fact that the trajectory of the particle is constrained to the zero-level hypersurface of Eq.(\ref{eq:mass_shell_function}), is based on the fact that the function $\mathcal{P}$ is proportional to the Hamiltonian (\ref{eq:hamiltonian}), i. e., $\mathcal{P}=2 m H+m^{2}$, which allows us to verify that $\mathcal{P}$ remains constant throughout the Hamiltonian flow of the system, just as the Hamiltonian function $H$ itself does (see Eq.(\ref{eq:hamiltonian_flux_2})).

\subsection{Action and momentum map of the group of isometries of the spacetime}
\label{subsec:mom_map}
Let $G$ be the $k$-dimensional Lie group generated by all continuous and smooth isometries of the spacetime $(M,g)$, that means, the group generated by compositions of all diffeomorphisms $\Lambda\in \textup{Diff}(M)$ that satisfies \cite{nakahara2018}
\begin{equation}
\label{eq:isometries_pull}
\Lambda^{*}g\circ\Lambda=g.
\end{equation}
  
Note that $G$ acts in a smooth-left way through $\boldsymbol{\Psi}:G\times M\rightarrow M$ as
\begin{equation}
\label{eq:action_bold}
\boldsymbol{\Psi}_{\Lambda}(x)=\Lambda(x).
\end{equation}
Thus, the fundamental fields (\ref{eq:fun_fields}) of $\chi\in \mathfrak{g}$, 
\begin{equation}
\label{eq:killing_vector}
\xi_{\chi}(x)=\displaystyle\frac{d}{ds}\boldsymbol{\Psi}_{e^{s\chi}}(x)\Bigr|_{s=0},
\end{equation}
turn out to be  Killing vector fields of the spacetime $(M,g)$ \cite{nakahara2018}:
\begin{equation}
\label{eq:killing_equation}
\mathscr{L}_{\xi_{\chi}}g=0.
\end{equation}  
\color{black}
Geometrically, each isometry $\Lambda\in G$ is a diffeomorphism that preserves both the distance between points in spacetime $(M,g)$, as well as the norms of their vector fields \cite{nakahara2018}, therefore, each $\Lambda\in G$ can be thought of as a mapping that takes time-like geodesics and sends them to time-like geodesics. In this way, we will hope that in some sense the isometry group $G$ will be a group of symmetries of the Hamiltonian that we create in the previous section. Let's find out exactly how this works.

The fact that the isometries map solutions of the Hamilton equations (\ref{eq:hamiltonian_vector}) into solutions of the same equations, on the same zero-level hypersurface $\mathcal{P}^{-1}(0)$, allows us to define the action $\Psi:G\times T^{*}M\rightarrow T^{*}M$ of the Lie group $G$ on the cotangent bundle $T^{*}M$ as a cotangent lift of the action defined in Eq.(\ref{eq:action_bold}), i. e., by \cite{marsden1999}
\begin{eqnarray}
\nonumber
\Psi_{\Lambda}(x,p)&=&\bigl(\boldsymbol{\Psi}_{\Lambda}(x),[\boldsymbol{\Psi}_{\Lambda^{-1}}]^{*}p\bigr) \\
\label{eq:action_iso}
&=&\bigl(\Lambda(x),[\Lambda^{-1}]^{*}p\bigr).
\end{eqnarray}
Notice that the map $\Psi$ is a smooth left action, because trivially satisfies Eq.(\ref{eq:action_1}) as well as Eq.(\ref{eq:action_2}):
\[\begin{array}{ll}
\Psi_{\Lambda}\circ \Psi_{\Lambda'}(x,p)\!\!\!&=\Psi_{\Lambda}\bigl(\Lambda'(x),[\Lambda^{'-1}]^{*}p\bigr)\\
\\
&=\bigl(\Lambda \Lambda'(x),[\Lambda^{-1}]^{*}\circ [\Lambda^{'-1}]^{*}p\bigr)\\
\\
&=\bigl(\Lambda \Lambda'(x),[\Lambda^{'-1} \Lambda^{-1}]^{*}p\bigr)\\
\\
&=\bigl(\Lambda \Lambda'(x),[(\Lambda \Lambda')^{-1}]^{*}p\bigr)=\Psi_{\Lambda\Lambda'}(x,p).
\end{array}\]
  
Besides, the action $\Psi$ is $\pi$-related to the action $\boldsymbol{\Psi}$,
\begin{equation}
\label{eq:pi_relation}
\pi\circ \Psi_{\Lambda}=\boldsymbol{\Psi}_{\Lambda}\circ \pi.
\end{equation}
As a consequence of Eq.(\ref{eq:pi_relation}), the fundamental fields associated with the action $\Psi$ are projected by the pushforward $[\pi]_{*}:T(T^{*}M)\rightarrow TM$ into the Killing fields 
\begin{equation}
\label{eq:fun_fields_iso}
\,[\pi(u)]_{*}\psi_{\chi}(u)=\xi_{\chi}\circ \pi(u).
\end{equation}
Indeed,
\[\begin{array}{ll}
\,[\pi(u)]_{*}\psi_{\chi}(u)\!\!\!&=\displaystyle\frac{d}{ds}\pi\circ \Psi_{e^{s\chi}}(u)\Bigr|_{s=0}\\
\\
&=\displaystyle\frac{d}{ds}\boldsymbol{\Psi}_{e^{s\chi}}\circ \pi(u)\Bigr|_{s=0}=\xi_{\chi}\circ \pi(u).
\end{array}\]
\color{black}
Once the action of the isometry group on the cotangent bundle has been defined, as well as the fundamental fields of the isometry algebra, we can see that for each isometry $\Lambda\in G$ the map $\Psi_{\Lambda}$ leaves invariant the Hamiltonian system that we introduced in the previous section to describe a free massive particle; this means that it leaves invariant both the symplectic form and the Hamiltonian function defined in Eq.(\ref{eq:hamiltonian}), as well as the hyperboloid $\mathcal{P}^{-1}(0)$ generated by the mass-shell function defined in Eq.(\ref{eq:mass_shell_function}), i. e.,  
\begin{eqnarray}
\label{eq:ham_sym_3}
\mathcal{P}\circ \Psi_{\Lambda}(x,p)&=&\mathcal{P}(x,p).
\end{eqnarray}
Indeed, first we show that the 1-form $\vartheta$ is conserved allong $\Psi_{\Lambda}$ and thus $\Psi_{\Lambda}$ is a symplectomorphism. To do this we use an arbitrary point $u=(x,p)\in T^{*}M$ and an arbitrary vector field $X\in \mathfrak{X}(T^{*}M)$, as well as Eq.(\ref{eq:canonical_1_form_2}):
  
\[\begin{array}{ll}
\,i_{X}\bigl([\Psi_{\Lambda}]^{*}\vartheta\circ \Psi_{\Lambda}(u)\bigr)\!\!\!&=i_{[\Psi_{\Lambda}]_{*}}\vartheta\circ \Psi_{\Lambda}(u)\\
\\
&=([\boldsymbol{\Psi}_{\Lambda^{-1}}]^{*}p)(\boldsymbol{\Psi}_{\Lambda}(x))(\pi_{*}\circ [\Psi_{\Lambda}]_{*}X)\\
\\
&=p(x)([\boldsymbol{\Psi}_{\Lambda^{-1}}]_{*}\circ \pi_{*}\circ [\Psi_{\Lambda}]_{*}X)\\
\\
&=p(x)([\boldsymbol{\Psi}_{\Lambda^{-1}}\circ \pi\circ \Psi_{\Lambda}]_{*}X)\\
\\
&=p(x)([\boldsymbol{\Psi}_{\Lambda^{-1}}\circ \boldsymbol{\Psi}_{\Lambda}\circ \pi]_{*}X)\\
\\
&=p(x)(\pi_{*}X)=i_{X}\vartheta(u).
\end{array}\]
\color{black}
Then, let us prove Eq.(\ref{eq:hamiltonian_invariance}) by taking into account that the isometries of the metric $g$ are also isometries of its inverse metric tensor, i. e., $[\Lambda^{-1}]_{*}(g\circ \Lambda)^{-1}=g^{-1}$:
  
\[\begin{array}{ll}
H\circ \Psi_{\Lambda}(u)\!\!\!&=\displaystyle\frac{1}{2m}g(\boldsymbol{\Psi}_{\Lambda}(x))^{-1}([\boldsymbol{\Psi}_{\Lambda^{-1}}]^{*}p,[\boldsymbol{\Psi}_{\Lambda^{-1}}]^{*}p)\\
\\
&=\displaystyle\frac{1}{2m}\Bigl([\boldsymbol{\Psi}_{\Lambda^{-1}}]^{*}(g\circ \boldsymbol{\Psi}_{\Lambda}(x))^{-1}\Bigr)(p,p)\\
\\
&=\displaystyle\frac{1}{2m}\Bigl([\Lambda^{-1}]^{*}(g\circ \Lambda(x))^{-1}\Bigr)(p,p)\\
\\
&=\displaystyle\frac{g(x)^{-1}(p,p)}{2m}=H(u).
\end{array}\]
\color{black}
Finally, using both Eq.(\ref{eq:hamiltonian_invariance}) and the identification $\mathcal{P}=2mH+m^{2}$, the validity of Eq.(\ref{eq:ham_sym_3}) becomes quite clear.

A consequence of the fact that action $\Psi$ leaves invariant the presymplectic potential $\vartheta$ is that we can identify the fundamental fields $\psi_{\chi}$, $[\pi]_{*}$-related with $\xi_{\chi}$ v\'ia Eq.(\ref{eq:fun_fields_iso}), as Hamiltonian fields of the functions (see Eq.(\ref{eq:hamiltonianas_chi}))
\begin{equation}
\label{eq:mom_map_ham}
\mu_{\chi}(x,p)=i_{\psi_{\chi}}\vartheta(x,p)=i_{\xi_{\chi}}p(x).
\end{equation}
Using the relation between $p$ and $U$ along the trajectory of the particle (see. Eq(\ref{eq:gradient_p_2}), we can re-express the integrals of motion $\mu_{\chi}\circ\hat{\gamma}$ as projections of the velocity vector of the particle with the infinitesimal generator of isometries, i.e., Killing vector field $\xi_{\chi}$:
\begin{equation}
\label{eq:mom_map_velocity}
\mu_{\chi}\circ \hat{\gamma}(\tau)=m g\bigl(\gamma(\tau)\bigr)(U,\xi_{\chi}).
\end{equation}
  
This momentum map   
will be central in the next sections, where we will generalize this one-particle model to an $N$-particle model. To this end, we first show that indeed it is an equivariant momentum map (see the definition in Eq.(\ref{eq:equivariant})) by using the property (see Appendix  (\ref{ap:equivariant}))
\begin{equation}
\label{eq:property_killing}
\,\Lambda_{*}\xi_{\chi}=\xi_{\textup{Ad}_{\Lambda}(\chi)}.
\end{equation}
Indeed, for any $\Lambda\in G$ and $(x,p)\in T^{*}M$, we have
\[\begin{array}{ll}
\mu_{\chi}\circ \Psi_{\Lambda}(x,p)\!\!\!&=i_{\xi_{\chi}}\Bigl([\Lambda^{-1}]^{*}p\circ \Lambda(x)\Bigr)\\
\\
&=i_{[\Lambda^{-1}]_{*}\xi_{\chi}}p(x)=i_{\xi_{\textup{Ad}_{\Lambda^{-1}}(\chi)}}p(x)\\
\\
&=\mu_{\textup{Ad}_{\Lambda^{-1}}(\chi)}(x,p)=(\textup{Ad}_{\Lambda}^{*}\mu)_{\chi}(x,p).
\end{array}\]
\color{black}
\section{Ideal gas made up of massive free particles}
\label{sec:ideal_gas}
On the spacetime $(M,g)$, we now consider a system formed by $N$ identical particles of mass $m$, which do not interact with each other. Given the independence between the particles of the system, it is not difficult to see that the world line of each of them will be a time-like geodesic, which can be modeled with the Hamiltonian system that we introduced in the previous section. Thus, due to the separability of the system, we have a Hamiltonian description formed by $N$ copies of the one-particle Hamiltonian system. That is, the Hamiltonian description of the $N$-particle system works now on the cotangent bundle $T^{*}M^{N}$, with symplectic form
\begin{equation}
\label{eq:symplectic_N}
 \Omega^{(N)}=\sum_{A=1}^{N}\varpi^{*}_{A}\Omega_{A}
\end{equation}
and Hamiltonian 
\begin{equation}
\label{eq:hamiltonian_N}
  H^{(N)}=\sum_{A=1}^{N}H_{A}\circ \varpi_{A},
\end{equation}
where the subscript $A$ is a label to differentiate the $A$-th particle from the rest of the  $N$ particles. In this way, $\Omega_{A}$ is the symplectic form defined in the cotangent bundle $T^{*}M_{A}$ of the $A$-th particle and $H_{A}$ is its Hamiltonian defined, also, in $T^{*}M_{A}$ by means of Eq.(\ref{eq:hamiltonian}). Moreover, the naturals projections from $T^{*}M^{N}$ to any $T^{*}M_{A}$ are denoted by
\begin{equation}
\label{eq:natural_projection}
\varpi_{A}:T^{*}M^{N}\rightarrow T^{*}M_{A},
\end{equation}
while $\pi_{A}:T^{*}M_{A}\rightarrow M_{A}$ still stands for the natural projection in the one-particle cotangent bundle.

The trajectory of the $A$-th particle of the $N$-particle system is constrained to the $\mathcal{P}^{-1}_{A}(0)$ hyperboloid because such  a trajectory is a time-like geodesic on $M_{A}$. In the description of the complete system of $N$ particles, this means that the trajectory of the total system is constrained to the N-codimension region of the total cotangent bundle $T^{*}M^{N}$, which is defined as
\begin{equation}
\label{eq:A_region}
  A=\mathcal{P}_{1}^{-1}(0)\times \mathcal{P}^{-1}_{2}(0)\times...\times \mathcal{P}^{-1}_{N}(0).
\end{equation}
  
It is worth mentioning that the action $\Psi$ of the isometry group $G$ on the cotangent bundle $T^{*}M_{A}$ (see Eq.(\ref{eq:action_iso})) can also be directly generalized to an action of the isometry group on the cotangent bundle $T^{*}M^{N}$ when acting with $\Psi$ on each space individually,
\begin{equation}
\label{eq:action_N}
\Psi^{(N)}_{\Lambda}(u)=\Bigl(\Psi_{\Lambda}\circ \varpi_{1}(u),...,\Psi_{\Lambda}\circ \varpi_{N}(u)\Bigr).
\end{equation}
This generalized action has the same symplectic properties as the one-particle action and gives rise to a momentum map $\mu^{(N)}:T^{*}M^{N}\rightarrow \mathfrak{g}^{*}$ defined by the sum
\begin{equation}
\label{eq:momentum_map_total}
\mu^{(N)}(u)=\displaystyle\sum\limits_{A=1}^{N}\mu\circ \varpi_{A}(u),
\end{equation}
such that for any $\chi\in \mathfrak{g}$ we have
\begin{equation}
\label{eq:mom_map_iso_N}
i_{\chi}\mu^{(N)}(u)=\mu_{\chi}^{(N)}(u)=\displaystyle\sum\limits_{A=1}^{N}\mu_{\chi}\circ \varpi_{A}(u).
\end{equation}
\color{black}
Let us assume that the aforementioned $N$ particles are found inside a box with impenetrable, undeformable, and perfectly elastic walls. Suppose that this box is large enough for the particles to continue without interacting with each other. Moreover, the particles interact only in very brief moments (and in a perfectly elastic way) with the internal walls of the box so that the $N$ particles model a dilute ideal gas. Furthermore, let us assume that the box is small enough to also move along a time-like geodesic\footnote{On the constraints for material objects to move along time-like geodesics, just as massive point particles do, see \cite{gerloch}.}. We will use $\tau \mapsto \gamma_{\textup{box}}(\tau)$ for the time-like geodesic that follows the box, where from now on the $\tau$ parameter will exclusively represent the arc-length parameter of $\gamma_{\textup{box}}$. 

\begin{figure}[H]
	\begin{center}
		\tikzset{every picture/.style={line width=0.75pt}} 

\begin{tikzpicture}[x=0.75pt,y=0.75pt,yscale=-1,xscale=1]

\draw  [fill={rgb, 255:red, 193; green, 187; blue, 187 }  ,fill opacity=1 ][dash pattern={on 4.5pt off 4.5pt}] (100,313.2) .. controls (100,305.74) and (120.59,299.7) .. (146,299.7) .. controls (171.41,299.7) and (192,305.74) .. (192,313.2) .. controls (192,320.66) and (171.41,326.7) .. (146,326.7) .. controls (120.59,326.7) and (100,320.66) .. (100,313.2) -- cycle ;
\draw    (146,313.2) -- (174.53,287.02) ;
\draw [shift={(176,285.67)}, rotate = 137.45] [color={rgb, 255:red, 0; green, 0; blue, 0 }  ][line width=0.75]    (10.93,-3.29) .. controls (6.95,-1.4) and (3.31,-0.3) .. (0,0) .. controls (3.31,0.3) and (6.95,1.4) .. (10.93,3.29)   ;
\draw    (154,208.67) -- (132.74,154.86) ;
\draw [shift={(132,153)}, rotate = 68.44] [color={rgb, 255:red, 0; green, 0; blue, 0 }  ][line width=0.75]    (10.93,-3.29) .. controls (6.95,-1.4) and (3.31,-0.3) .. (0,0) .. controls (3.31,0.3) and (6.95,1.4) .. (10.93,3.29)   ;
\draw    (100,313.2) .. controls (114,241) and (87,204) .. (75,114.8) ;
\draw    (192,313.2) .. controls (190,251) and (221,213) .. (219,114.8) ;
\draw  [fill={rgb, 255:red, 255; green, 255; blue, 255 }  ,fill opacity=1 ][dash pattern={on 4.5pt off 4.5pt}] (75,114.8) .. controls (75,107.34) and (107.24,101.3) .. (147,101.3) .. controls (186.76,101.3) and (219,107.34) .. (219,114.8) .. controls (219,122.26) and (186.76,128.3) .. (147,128.3) .. controls (107.24,128.3) and (75,122.26) .. (75,114.8) -- cycle ;
\draw    (146,313.2) .. controls (216,258) and (96,161) .. (148,86) ;

\draw (198,305) node [anchor=north west][inner sep=0.75pt]    {$\mathcal{S}_{\textup{box}}$};
\draw (153,67) node [anchor=north west][inner sep=0.75pt]    {$\gamma _{box}$};
\draw (102,162) node [anchor=north west][inner sep=0.75pt]    {$U_{box}$};
\draw (176,262.7) node [anchor=north west][inner sep=0.75pt]    {$U_{box}$};
\end{tikzpicture}
	\end{center}
	\caption{Piece of the world tube $\Sigma$  generated in the proper time interval $\tau=0$ and $\tau=\tau_{0}$.}
	\label{fig:world_tube}
\end{figure}
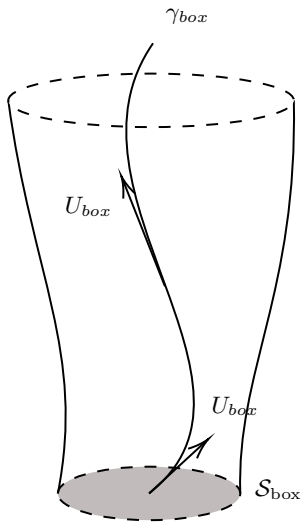

To model the diluted ideal gas described above, we can consider that each of the $N$ particles  has a geodesic trajectory,  but is bounded to remain within the  world tube $\Sigma\subset M$ that surrounds the geodesic $\gamma_{\textup{box}}$. More specifically, if $\mathcal{S}_{\textup{box}}$ is the region of spacetime that the box occupies at proper time $\tau=0$, $\Sigma$ is the region obtained by letting each point $u\in \mathcal{S}_{\textup{box}}$ evolve through the $\gamma_{\textup{box}}$ time-like geodesic (see Fig.(\ref{fig:world_tube})).

In this way, the Hamiltonian description of the $N$-particle system  must be modified to take into account the effect of confinement in the positions imposed by the box. This confinement effect can be achieved by defining a new Hamiltonian by multiplying each Hamiltonian function $H_{A}$ in $H^{(N)}$ by the composition $\Theta_{\Sigma}\circ \varpi_{A}$,
  
\begin{equation}
\label{eq:new_hamiltonian}
\mathcal{H}=\displaystyle\sum\limits_{A=1}^{N}\Theta_{\Sigma}\circ \varpi_{A}\,H_{A}\circ \varpi_{A}.
\end{equation}
Although the presence of the characteristic function $\Theta_{\Sigma}$ (cf. Eq.(\ref{eq:characteristics})) does not affect the Hamiltonian dynamics inside the box, it does limit the states in the cotangent bundle $T^{*}M^{N}$ that the ideal gas can reach. Taking this into account, we change the former region defined in  Eq.(\ref{eq:A_region}) by the new one
\begin{equation}
\label{eq:new_region}
\begin{array}{ll}
\mathcal{A}\!\!\!&=\Bigl((\pi_{1}\circ \varpi_{1})^{-1}(\Sigma)\cap \mathcal{P}_{1}^{-1}(0)\Bigr)\\
&\phantom{=\Bigl(\Bigr)}\times...\times\Bigl((\pi_{N}\circ \varpi_{N})^{-1}(\Sigma)\cap \mathcal{P}_{N}^{-1}(0)\Bigr).
\end{array}
\end{equation}

To formulate a statistical approach to the problem, we will consider that the limitation of the accessible states is such that the system adopts a \textit{static state}, in the sense that we will define in the next section. But first, since we have put the world tube $\Sigma$ as a boundary for our analysis, it will be convenient to reduce the group of isometries $G$ to the subgroup $G_{\Sigma}\subset G$ that particularly fixes $\Sigma $, that is, the subgroup defined by
\begin{equation}
\label{eq:subgroup_G_Sigma}
G_{\Sigma}=\{\Lambda\in G: \Lambda(\Sigma)=\Sigma\}.
\end{equation}
Since the isometries map  time-like geodesics onto other time-like geodesics \cite{nakahara2018}, limiting ourselves to isometries in $G_{\Sigma}$ means that we are only interested in symmetries of the Hamiltonian system $( T^{ *}M^{N},\Omega^{(N)},\mathcal{H})$ that preserve the geodesics inside the world tube, i.e., that do not put or take out particles from the box. Naturally, each $\Lambda\in G_{\Sigma}$ also leaves invariant the region $\mathcal{A}$,
\begin{equation}
\label{eq:invariance_A}
\Psi^{(N)}_{\Lambda}(\mathcal{A})=\mathcal{A},
\end{equation}
since all isometries leave intact $\mathcal{P}^{-1}_{A}(0)$, as we already said before.
\color{black}

\subsection{The ideal gas in static spacetimes}
\label{subsec:static_spacetime}
Before defining what we will understand in this article by a static state, we must consider certain aspects of the nature of the spacetime where the  ideal gas lives. Although so far we have not imposed any special conditions on $M$, from now on we will limit ourselves to working exclusively with static spacetimes, that are spacetimes where almost everywhere there exists a time-like Killing vector field $\xi_{0}\in \mathfrak{X}(M)$, 
\begin{eqnarray}
\label{eq:stationary_1}
g(x)(\xi_{0},\xi_{0})&<&0,\\
\label{eq:stationary_2}
\mathscr{L}_{\xi_{0}}g(x)&=&0,
\end{eqnarray}
whose $g$-orthogonal distribution is involutive \cite{kriele1999,wald}. This means that for any point $x\in M$ and any pair of tangent vectors $X,Y\in \Pi_{x}$ the Lie bracket between $X$ and $Y$ belongs to $\Pi_{x}$,
\begin{equation}
\label{eq:involutive}
\,[X,Y]\in \Pi_{x},
\end{equation}
where $\Pi_{x}\subset T_{x}M$ is the vector sub-space of $T_{x}M$ generated by all vectors, which are orthogonal to $\xi_{0}(x)$:
\begin{equation}
\label{eq:g_orthogonal}
\Pi_{x}=\{X\in T_{x}M:\,g(x)(X,\xi_{0})=0\}.
\end{equation}

Since the involution property in Eq.(\ref{eq:involutive}) is a necessary and sufficient condition of integrability \cite{marsden1999}, an important property of static spacetimes is that the Killing vector field $\xi_{0 }$ generates a foliation whose leaves are tangent to the orthogonal distribution defined in Eq.(\ref{eq:g_orthogonal}) \cite{kriele1999, wald}, and it allows us to express, at least locally for an open subset $W\subset M$, the $W$ subset as the product $W=\mathbb{R}\times S$, where $S$ is a $d$-dimensional Riemannian manifold (the leaves of the foliation).

Geometrically, the foliation generated by time-like vector field $\xi_{0}$ allows us to find a kind of coordinate system $(x^{\mu})=(t,x^{1},x^{2},...,x^{d})=(t,\vec{x})$, where $x^{0}=t$ is identified as a temporal coordinate generated by the flow of $\xi_{0}$, i. e., $\xi_{0}=\frac{\partial}{\partial t}$, while $(\vec{x})=(x^{1},...,x^{d})$ are spatial coordinates in the leaves of the foliation. In this way, the leaves of the foliation are level-zero hypersurfaces of the coordinate function $t$. In such a coordinate system, and in the induced cotangent basis, the components of the metric tensor are independent of the $t$ coordinate, as can be deduced directly from the Killing condition (see Eq.(\ref{eq:stationary_1})), i. e., $\frac{\partial g_{\alpha\beta}}{\partial t}=0$. This allows us to rewrite the metric tensor $g$ as 
\begin{equation}
\label{eq:static_g}
g(\vec{x})=-|g_{00}(\vec{x})|dt\otimes dt+g_{S}(\vec{x}),
\end{equation}
where $g_{S}$ is a Riemmanian metric tensor on the foliation leaves \cite{kriele1999}.

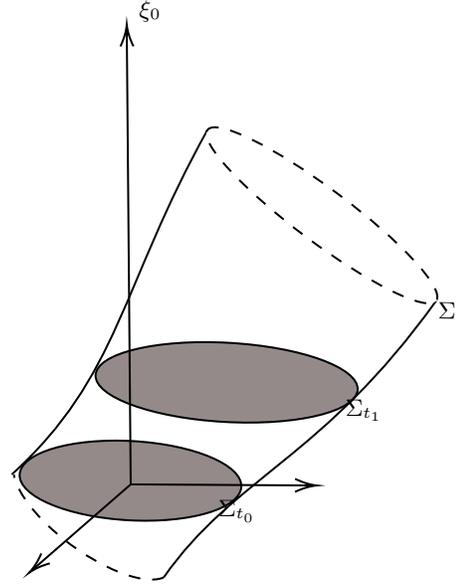
\begin{figure}[]
	\centering
		\tikzset{every picture/.style={line width=0.75pt}} 
	
	\begin{tikzpicture}[x=0.75pt,y=0.75pt,yscale=-1,xscale=1]
	
	\draw  [fill={rgb, 255:red, 148; green, 138; blue, 138 }  ,fill opacity=1 ] (170.18,250.77) .. controls (170.8,239.74) and (196.32,232.2) .. (227.19,233.93) .. controls (258.06,235.66) and (282.58,246.01) .. (281.96,257.03) .. controls (281.34,268.06) and (255.82,275.6) .. (224.95,273.87) .. controls (194.08,272.14) and (169.56,261.79) .. (170.18,250.77) -- cycle ;
	\draw  [fill={rgb, 255:red, 148; green, 138; blue, 138 }  ,fill opacity=1 ] (208.47,200.58) .. controls (209.09,189.55) and (239.18,182.27) .. (275.68,184.32) .. controls (312.17,186.36) and (341.26,196.96) .. (340.64,207.99) .. controls (340.02,219.02) and (309.93,226.3) .. (273.44,224.25) .. controls (236.94,222.21) and (207.85,211.61) .. (208.47,200.58) -- cycle ;
	\draw    (166.18,250.77) .. controls (220.61,201.31) and (221.13,155.51) .. (264.91,76.87) ;
	\draw    (242.85,302.71) .. controls (278.48,251.68) and (323.08,242.93) .. (380.26,163.07) ;
	\draw  [fill={rgb, 255:red, 255; green, 255; blue, 255 }  ,fill opacity=1 ][dash pattern={on 4.5pt off 4.5pt}] (264.91,76.87) .. controls (269.38,70.9) and (298.82,85.36) .. (330.67,109.16) .. controls (362.52,132.96) and (384.73,157.1) .. (380.26,163.07) .. controls (375.8,169.04) and (346.36,154.59) .. (314.51,130.79) .. controls (282.65,106.98) and (260.45,82.85) .. (264.91,76.87) -- cycle ;
	\draw    (226.18,255.77) -- (224.19,25.77) ;
	\draw [shift={(224.18,23.77)}, rotate = 89.51] [color={rgb, 255:red, 0; green, 0; blue, 0 }  ][line width=0.75]    (10.93,-3.29) .. controls (6.95,-1.4) and (3.31,-0.3) .. (0,0) .. controls (3.31,0.3) and (6.95,1.4) .. (10.93,3.29)   ;
	\draw  [draw opacity=0][dash pattern={on 4.5pt off 4.5pt}] (242.85,302.71) .. controls (242.2,303) and (241.48,303.24) .. (240.68,303.42) .. controls (228.35,306.22) and (203.39,294.39) .. (184.92,276.98) .. controls (174.55,267.2) and (168.44,257.75) .. (167.25,250.84) -- (207.24,271.89) -- cycle ; \draw  [dash pattern={on 4.5pt off 4.5pt}] (242.85,302.71) .. controls (242.2,303) and (241.48,303.24) .. (240.68,303.42) .. controls (228.35,306.22) and (203.39,294.39) .. (184.92,276.98) .. controls (174.55,267.2) and (168.44,257.75) .. (167.25,250.84) ;  
	\draw    (226.18,255.77) -- (318,256.32) ;
	\draw [shift={(320,256.33)}, rotate = 180.35] [color={rgb, 255:red, 0; green, 0; blue, 0 }  ][line width=0.75]    (10.93,-3.29) .. controls (6.95,-1.4) and (3.31,-0.3) .. (0,0) .. controls (3.31,0.3) and (6.95,1.4) .. (10.93,3.29)   ;
	\draw    (226.18,255.77) -- (176.51,299.02) ;
	\draw [shift={(175,300.33)}, rotate = 318.95] [color={rgb, 255:red, 0; green, 0; blue, 0 }  ][line width=0.75]    (10.93,-3.29) .. controls (6.95,-1.4) and (3.31,-0.3) .. (0,0) .. controls (3.31,0.3) and (6.95,1.4) .. (10.93,3.29)   ;
	
	\draw (379.9,162.48) node [anchor=north west][inner sep=0.75pt]  [rotate=-1.09]  {$\Sigma $};
	\draw (333,211.33) node [anchor=north west][inner sep=0.75pt]    {$\Sigma _{t_{1}}$};
	\draw (268.96,262.03) node [anchor=north west][inner sep=0.75pt]    {$\Sigma _{t_{0}}$};
	\draw (229,10) node [anchor=north west][inner sep=0.75pt]    {$\xi _{0}$};

	\end{tikzpicture}	
	\caption{Slices obtained from the intersection of the world tube with constant-time hypersurfaces.}
	\label{fig:slices_world_tube}
\end{figure}

We will say that the observers that move with the flow of the Killing vector field $\xi_{0}$ are static observers. In this sense, we define a static state of the ideal gas of $N$ particles enclosed in a box as a state in which the macroscopic characteristics do not change according to measurements made by any static observer, i. e., the macroscopic characteristics do not change with the coordinate time $t$. This means that the macroscopic state of the system appears as static in the slices $\Sigma_{t_{0}}$ defined as the intersections between the world tube $\Sigma$ and the level hypersurface $t^{-1}(t_{0})$ of the coordinate time function $t=t(x)$ (see  Fig.(\ref{fig:slices_world_tube})):
\begin{equation}
\label{eq:slices_world_tube}
\Sigma_{t_{0}}=t^{-1}(t_{0})\cap \Sigma.
\end{equation}

Note that the full world tube can be seen as the union of all the slices,
\begin{equation}
\label{eq:world_tube_complete}
\Sigma=\bigcup\limits_{t_{0}\in \mathbb{R}}\Sigma_{t_{0}}.
\end{equation}
In this way, the world tube is invariant under translations in time $t$ generated by $\xi_{0}$, identifying its flow as a monoparametric subgroup of $G_{\Sigma}$,
\begin{equation}
\label{eq:xi0_G_sigma}
\{e^{t\chi_{0}}\}_{t\in \mathbb{R}}\subset G_{\Sigma}.
\end{equation}
Here, we have identified the vector $\chi_{0}$ as the element of $\mathfrak{g}$ whose fundamental field associated with the action $\boldsymbol{\Psi}$ of $G$ in $M$ is $\xi_{0}$, $\boldsymbol{\psi}_{\chi_{0}}=\xi_{0}$, such that the flux generated by $\varphi_{t}^{\epsilon_{0} }$ is simply the action of the monopametric group generated by $\chi_{0}$, i. e.,
\begin{equation}
\label{eq:flow_relationship}
\varphi_{t}^{\xi_{0}}=\boldsymbol{\Psi}_{e^{t\chi_{0}}}\simeq e^{t\chi_{0}}.
\end{equation}

Returning to the isometry group, over Hamilton's solutions $\hat{\gamma}$, the Hamiltonian functions of the fundamental fields $\alpha \xi_{0}=\psi_{\alpha\chi_{0}}$ are those given by (see Eq.(\ref{eq:mom_map_ham}))
\begin{equation}
\label{eq:momentum_map_static}
\mu_{\alpha\chi_{0}}\circ\hat{\gamma}=-\alpha E,
\end{equation}
where $E$ can be interpreted physically, using Noether's theorem \cite{marle2020gibbs}, as the energy of the particle measured by a static observer, with the understanding that it is the conserved quantity associated with the translations generated by $\xi_{0}$ \cite{sachs2012general}:
\begin{equation}
\label{eq:static_energy}
E=-g(\gamma)(mU,\xi_{0}).
\end{equation}
It is important to remember that the existence of a time-like vector field in a whole spacetime defines a time orientation \cite{sachs2012general}. In this sense, we will say that a vector field $X\in \mathfrak{X}(M)$ is future pointing, with respect to the temporal orientation imposed by the Killing vector field $\xi_{0 }$, if, being a time-like vector field, its projection on $\xi_{0}$ is negative at all points $x\in M$ \cite{sachs2012general}:
\begin{eqnarray}
\label{eq:future_1}
g(x)(X,X)&<&0,\\
\label{eq:future_2}
g(x)(X,\xi_{0})&<&0.
\end{eqnarray}
If, instead, the vector field $-X$ is future pointing, then $X$ points to the past. In this sense, the velocities of the particles would point to the future or to the past defined by $\xi_{0}$, depending on whether we consider that the static observers register the history of the ideal gas in the same direction in which their clocks move or \textit{review a reverse succession of photographs of the system}. In other words, choosing whether the particles travel to the future defined by $\xi_{0}$ or to the past, is equivalent to choosing an orientation of the arrow of time of the  system. We will choose that the particles travel into the future, in such a way that the energy (\ref{eq:static_energy}) is always positive, 
$E>0$.
\color{black}
\section{Statistical states of the ideal gas in a static spacetime}
\label{sec:statiscal_distribution}
  
Let $\rho^{(N)}:\mathcal{A}\rightarrow [0,+\infty)$ be a statistical state with support $\mathcal{A}$ given by Eq.(\ref{eq:new_region}), over the measured space $(T^{*}M^{N}, \Omega^{(N)},\lambda_{\omega^{(N)}})$, where $\lambda_{\omega^{(N)}}$ is the Liouville measure generated from $\Omega^{(N)}$ via the Liouville top-form (see Eq.(\ref{eq:liouville_measure}))
\begin{equation}
\label{eq:liouville_top_form_N}
\omega^{(N)}=\displaystyle\frac{1}{(ND)!}(\Omega^{(N)})^{\wedge ND}.
\end{equation}	
To be more specific, let $\rho^{(N)}$ be a generalized Gibbs statistical state (see Eq.(\ref{eq:gibbs_generalized})), which is the kind of statistical state that best describe the status of a Hamiltonian system with symmetries in a Lie group $G_{\Sigma}$ \cite{marle2020gibbs}, i. e.,
\begin{equation}
\label{eq:gibbs_iso}
\rho^{(N)}(\chi;u)=\displaystyle\frac{\Theta_{\mathcal{A}}(u)}{Z(\chi)}e^{-\mu^{(N)}_{\chi}(u)},
\end{equation}
where $Z:\mathfrak{g}_{\star}\rightarrow \mathbb{R}$ is the partition function
\begin{equation}
\label{eq:partition_iso}
Z(\chi)=\displaystyle\int_{\mathcal{A}}d\lambda_{\omega^{(N)}}(u)\,e^{-\mu^{(N)}_{\chi}(u)}.
\end{equation}

First, we show that  in the subalgebra $\mathfrak{g}_{\Sigma}\simeq T_{e}G_{\Sigma}$, there exist Souriau vector, so that the definition of $\rho^{(N)}$ is consistent. Indeed, let us check that for any $\beta>0$, the vector $-\beta\chi_{0}$, which is already in $\mathfrak{g}_{\Sigma}$ by virtue of the definition of $\chi_{0}$, is a Soriau vector:
\begin{equation}
\label{eq:souriau_vector_beta}
-\beta \chi_{0}\in \mathfrak{g}_{\star}.
\end{equation}
To do this, we have to show that there is a neighborhood $V_{-\beta\chi_{0}}\subset \mathfrak{g}_{\Sigma}$ and a function $f_{-\beta\chi_{0}}:T^{*}M^{N}\rightarrow \mathbb{R}$ such that the condition of Souriau (expressed in the Eq.(\ref{eq:souriau_vector})) is fulfilled. The neighborhood is
\begin{equation}
\label{eq:neighborhood}
V_{-\beta\chi_{0}}=\left\{-\beta \chi_{0}+\epsilon \chi: \chi\in \mathfrak{g}_{\Sigma},\quad \xi_{\chi}^{0}=dt(\xi_{\chi})= \frac{\beta}{2}\right\}
\end{equation}
where $0\leq \epsilon \ll 1$, and the function is $f_{-\beta\chi_{0}}=c$ for any constant $c\geq 1$. Indeed, the condition $0\leq \epsilon \ll 1$ assures us that, at least up to the linear order, the Killing vector field associated with any vector $\chi'\in V_{-\beta\xi_{0}}$ is time-like (see Eq.(\ref{eq:future_1})),
\[\begin{array}{ll}
g(x)(\xi_{\chi'},\xi_{\chi'})\!\!\!&\simeq -\beta^{2}|g_{00}(\vec{x})|+2\beta \epsilon |g_{00}(\vec{x})|\xi_{\chi}^{0}\\
\\
&=-\beta^{2} |g_{00}(\vec{x})|(1-\epsilon)<0,
\end{array}\]
and is past pointing (see Eq.(\ref{eq:future_2})): 
\[\begin{array}{ll}
g(x)(\xi_{0},\xi_{\chi'})\!\!\!&\simeq \beta |g_{00}(\vec{x})|-\epsilon |g_{00}(\vec{x})|\xi_{\chi}^{0}\\
\\
&=\beta |g_{00}(\vec{x})|\left(1-\displaystyle\frac{\epsilon}{2}\right)>0.
\end{array}\]
So, according to Eq.(\ref{eq:static_energy}) and with our $U$-future pointing convention, the function $i_{\chi'}\mu^{(N)}$ is always positive and the function $\exp(-\mu_{\chi'}^{(N)})$ is a positive-definite and decreasing function everywhere in  $\mathcal{A}$, and thus it is bounded from above by any constant function with value $c\geq 1$. Of course, having $\{-\beta \xi_{0}\}_{\beta>0}\in \mathfrak{g}_{\star}$ is enough to prove that $g_{\star}$ is an open convex subset and there are many other Souriau vectors. 
 
Although we can give $\rho^{(N)}$ the same physical interpretation through ensemble theory, let us note that, as we already anticipated at the beginning of this work, the interpretation of the Hamiltonian flow of the system is not that of the action of a group of translations in time, nor is the Hamiltonian itself the energy of the particles in the ideal gas. The Hamiltonian model is entirely geometric and, therefore, the concept of equilibrium statistical state must be interpreted geometrically. In this way, we say that the state $\rho$ is of static equilibrium, if it is invariant under the action of the subgroup $\{e^{s\chi_{0}}\}_{s\in \mathbb{R}}\subset G_{ \Sigma}$:
\begin{equation}
\label{eq:static_equilibrium}
\rho\circ \Psi_{e^{s\chi_{0}}}=\rho.
\end{equation}

The definition of static statistical states (static states, from now on) given in Eq.(\ref{eq:static_equilibrium}) allows us to show that not all generalized Gibbs states are static, but only those that are parameterized by Souriau vectors that commute with $\chi_{0}$. We will call these Souriau vectors static, and we will denote the subset that contains them as
\begin{equation}
\label{eq:souriau_static_vectors}
\mathfrak{g}_{\odot}=\{\chi\in \mathfrak{g}_{\star}: [\chi,\chi_{0}]=0\}.
\end{equation}
Indeed, this is a consequence of the equivariance property (see Eq.(\ref{eq:equivariant}))  satisfied by $\mu^{(N)}$, inherited from the equivariance of the momentum map defined in Eq.(\ref{eq:mom_map_ham}). To see this, let us first see that equality 
\begin{equation}
\label{eq:gibbs_static}
\rho^{(N)}(\chi;\Psi_{e^{s\chi_{0}}}(u))=\rho^{(N)}(\chi ;u)\qquad \forall u\in \mathcal{A},
\end{equation}
is equivalent to asking for $\textup{Ad}_{e^{-s\chi_{0}}}(\chi)=\chi$,
\[\begin{array}{ll}
\rho^{(N)}(\chi;\Psi_{e^{s\chi_{0}}}(u))\!\!\!&=\displaystyle\frac{\Theta_{\mathcal{A}}\circ \Psi_{e^{s\chi_{0}}}(u)}{Z(\chi)}e^{-\mu_{\chi}^{(N)}\circ \Psi_{e^{s\chi_{0}}}(u)}\\
\\
&=\displaystyle\frac{\Theta_{\mathcal{A}}(u)}{Z(\chi)}e^{-(\textup{Ad}_{e^{s\chi_{0}}}^{*}\mu^{(N)})_{\chi}(u)}\\
\\
&=\displaystyle\frac{\Theta_{\mathcal{A}}(u)}{Z(\chi)}e^{-\mu_{\textup{Ad}_{e^{-s\chi_{0}}}(\chi)}(u)}=\rho^{(N)}(\chi;u).
\end{array}\]
Then, it is enough to use the infinitesimal version of the condition $\textup{Ad}_{e^{-s\chi_{0}}}(\chi)=\chi$. Indeed, considering that  $\frac{d}{ds}\textup{Ad}_{e^{-s\chi_{0}}}(\chi)|_{s=0}=\frac{d}{ds}\chi|_{s=0}=0$ and $\frac{d}{ds}\textup{Ad}_{e^{-s\chi_{0}}}(\chi)|_{s=0}=-[\chi_{0},\chi]$, we obtain 
\[\rho^{(N)}(\chi;\Psi_{e^{s\chi_{0}}}(u))=\rho^{(N)}(\chi;u)\quad\Longleftrightarrow\quad [\chi_{0},\chi]=0.\]
\color{black}
An important feature of static states  is that we can distribute identically the random variable of the dynamic state of the system in the slices $\Sigma_ {t}$ of the world tube, so that all these slices can be identified with the  equivalence relation $\Sigma_{t_{0}}\sim \Sigma_{t_{1}}$, based on the distribution of a static state. Choosing a representative $\Sigma_{0}$, we can move from the support $\mathcal{A}$ to the new (and final) support
  
\begin{equation}
\label{eq:new_support_2}
\begin{array}{ll}
\mathcal{M}^{(N)}\!\!\!&=\Bigl((\pi_{1}\circ \varpi_{1})^{-1}(\Sigma_{0})\cap \mathcal{P }_{1}^{-1}(0)\Bigr)\times...\times\\
\\
&\phantom{=\Bigl(\Bigr)}\Bigl((\pi_{N}\circ \varpi_{N})^{-1}(\Sigma_{0 })\cap \mathcal{P}_{N}^{-1}(0)\Bigr).
\end{array}
\end{equation}
For these static states, we identify the phase space of the system as the  $2N(D-1)=2Nd$-dimensional submanifold defined in Eq.(\ref{eq:new_support_2}).

The fact that we have an equivalence relationship between the slices at different times, allows us to reduce the group of isometries once again to only those that belong to the subgroup
\begin{equation}
\label{eq:subG0}
G_{0}=\{\Lambda\in G_{\Sigma}: \Lambda(\Sigma_{0})=\Sigma_{0}\},
\end{equation}
where the equivalence relation is taken into account in the expression $\Lambda (\Sigma_ {0})=\Sigma_{0}$. Naturally, $\chi_{0}\in \mathfrak{g}_{0}\simeq T_{e}G_{0}$ by virtue of its definition. Also, we keep the nested family of sets
\begin{equation}
\label{eq:nested_sets}
\mathfrak{g}_{\odot}\subset \mathfrak{g}_{\star}\subset \mathfrak{g}_{0}\subset \mathfrak{g}_{\Sigma}\subset \mathfrak{g}.
\end{equation}

Redefining the support $\mathcal{A}$ as $\mathcal{M}^{(N)}$, and taking into account the new structure given in Eq.(\ref{eq:nested_sets}), we will work from now on with the generalized-Gibbs-static states given by
\begin{equation}
\label{eq:gibbs_iso_final}
\rho^{(N)}(\chi;u)=\displaystyle\frac{\Theta_{\mathcal{M}^{(N)}}(u)}{Z(\chi)}e^{-\mu^{(N)}_{\chi}(u)},
\end{equation}
where $Z:\mathfrak{g}_{\odot}\rightarrow \mathbb{R}$ is the partition function
\begin{equation}
\label{eq:partition_iso_final}
Z(\chi)=\displaystyle\int_{\mathcal{M}^{(N)}}d\lambda_{\omega^{(N)}}(u)\,e^{-\mu^{(N)}_{\chi}(u)}.
\end{equation}
\color{black}
\section{The one-particle distribution}
\label{sec:one_particle_distribution}
  
Note that the top-form $\omega^{(N)}$ (see Eq.(\ref{eq:liouville_top_form_N})) can be expressed as a product of the one-particle Liouville top-forms:
\begin{equation}
\label{eq:liouville_top_form_product}
\omega^{(N)}=\varpi_{1}^{*}\omega_{\Omega_{1}}\wedge \varpi_{2}^{*}\omega_{\Omega_{2}}\wedge...\wedge \varpi_{N}^{*}\omega_{\Omega_{N}}.
\end{equation}
The easiest way to prove this statement is using a version of the multinomial theorem, with differential 2-forms instead of  real numbers, taking into account that the wedge product between differential forms of even degree is commutative \cite{nakahara2018}:
\[\begin{array}{ll}
\omega^{(N)}\!\!\!&=\displaystyle\frac{1}{(ND)!}\left(\displaystyle\sum\limits_{A=1}^{N}\varpi_{A}^{*}\Omega_{A}\right)^{\wedge ND}\\
\\
&=\displaystyle\sum\limits_{\stackrel{k_{1},k_{2},...,k_{N}\geq0}{k_{1}+...+k_{N}=ND}}\displaystyle\frac{(\varpi_{1}^{*}\Omega_{1})^{\wedge k_{1}}}{k_{1}!}\wedge...\wedge \displaystyle\frac{(\varpi_{N}^{*}\Omega_{N})^{\wedge k_{N}}}{k_{N}!}\\
\\
&=\displaystyle\sum\limits_{\stackrel{k_{1},k_{2},...,k_{N}\geq0}{k_{1}+...+k_{N}=ND}}\varpi_{1}^{*}\left(\displaystyle\frac{\Omega_{1}^{\wedge k_{1}}}{k_{1}!}\right)\wedge...\wedge \varpi_{N}^{*}\left(\displaystyle\frac{\Omega_{N}^{\wedge k_{N}}}{k_{N}!}\right).
\end{array}\]
Now, taking into account that no index $k_{i}$ can be greater than $D$, since that would give rise to a $2k>2D$-form on a manifold of dimension $2D$, then necessarily $k_{1}= k_{2}=....=k_{N}=D$, and so we find the desired result:
\[\begin{array}{ll}
\omega^{(N)}\!\!\!&=\varpi_{1}^{*}\left(\displaystyle\frac{\Omega_{1}^{\wedge D}}{D!}\right)\wedge...\wedge \varpi_{N}^{*}\left(\displaystyle\frac{\Omega_{N}^{\wedge D}}{D!}\right)\\
\\
&=\varpi_{1}^{*}\omega_{\Omega_{1}}\wedge...\wedge \varpi_{N}^{*}\omega_{\Omega_{N}}.
\end{array}\]
\color{black}
The separability of the  Liouville top-form of the Hamiltonian system $(T^{*}M^{N},\Omega^{(N)},\mathcal{H})$ is the result of 
the construction of the system itself.  In the same way, note that the simplest way to construct $\mathcal{M}^{(N)}$ is by using the Cartesian product
\begin{equation}
\label{eq:separability}
\mathcal{M}^{(N)}=\mathcal{M}_{1}\times \mathcal{M}_{2}\times...\times\mathcal{M}_{N },
\end{equation}
where $\mathcal{M}_{A}\simeq \mathcal{M}$ is the submanifold of codimension 2 in $T^{*}M$, defined as \textit{the portion of the mass-shell hyperboloid} $\mathcal{P}^{-1}(0)$ \textit{above} $\Sigma_{0}$,
\begin{equation}
\label{eq:end_support}
\mathcal{M}=\bigcup_{x\in \Sigma_{0}}\mathcal{P}^{-1}_{x}(0),
\end{equation}
where $\mathcal{P}_{x}:T_{x}^{*}M\rightarrow \mathbb{R}$ is the constraint on the mass-shell function (see Eq.(\ref{eq:mass_shell_function})) to the cotangent space to $M$ in $x$:
\begin{equation}
\label{eq:mass_shell_constriction}
\mathcal{P}_{x}(p)=\mathcal{P}(x,p).
\end{equation}
  
The Cartesian product (\ref{eq:separability}), which determines the  phase space of the Hamiltonian system $(T^{*}M^{N},\Omega^{(N)},\mathcal{H})$, exploits the separability of the total Liouville measure into the individual measures of each particle, in such a way that the integrals $\int_{\mathcal{M}^{(N)}}d\lambda_{\omega^{(N)}}( ...)$ can be separated as\footnote{Basically, we are taking into account that $\lambda_{\omega^{(N)}}$ is a measure product of the product of (finite) measured spaces $(T^ {*}M,\mathfrak{B}(T^{*}M),\omega_{\Omega})$ with measures $\lambda_{\omega}$, such that we can use Fubini's theorem in the support $\mathcal{M}^{(N)}=\mathcal{M}_{1}\times...\times \mathcal{M}_{N}$ \cite{athreyameasure}.}
\begin{eqnarray}
\nonumber
\displaystyle\int_{\mathcal{M}^{(N)}}\!\!\!\!\!\!\!\!d\lambda_{\omega^{(N)}}(u )(...)\!\!\!&=&\!\!\!\!\displaystyle\int_{\mathcal{M}_{1}}\!\!\!\!\!d\lambda_{\omega_{1}}|_ {\mathcal{M}_{1}}\circ \varpi_{1}(u)...\\
\label{eq:separability_integrals}
&\phantom{=}&\displaystyle\int_{\mathcal{M}_{N}}\!\!\!\!\!\!\!\!d\lambda_{\omega_{N}}|_{\mathcal{M}_{N}}\circ \varpi_{N}(u)(...),
\end{eqnarray}
where $\lambda_{\omega_{A}}|_{\mathcal{M}_{A}}\simeq \lambda_{\omega}|_{\mathcal{M}}$ is the induced measure, via $\lambda_{\omega}$, in the submanifold $\mathcal{M}$ seen as an embedding:
\begin{equation}
\label{eq:induced_measure}
\lambda_{\omega}|_{\mathcal{M}}(A)=\displaystyle\int_{A}\omega_{\Omega}|_{\mathcal{M}}(u).
\end{equation}
Precisely, this induced measure incorporates the restrictions of the phase space to be projected onto slices of the world tube and takes into account the time-like nature of the geodesics of each particle. 

In Appendix (\ref{ap:induced_measures}) we show that in a coordinate system $(t,\vec{x})$, where the metric tensor $g$ splits into the sum (\ref{eq:static_g}), the induced volume top-form $\omega_{\Omega}|_{\mathcal{M}}$ reduces to
\begin{equation}
\label{eq:medida_inducida_local}
\omega_{\Omega}|_{\mathcal{M}}(\vec{x},\vec{P})=  \displaystyle\frac{m|g_{S}(\vec{x})|^{\frac{1}{2}}d^{d}x\wedge d^{d}P}{(m^{2}+\Vert\vec{P}\Vert^{2})^{\frac{1}{2}}},
\end{equation}
where $|g_{S}|$ stands for the absolute value of the determinant of $g_{S}$, and $P=(P^{0},...,P^{d}) =(P^{0},\vec{P})$ are the components of the covectors $p\in T_{x}M$ in the orthonormal frame of a static observer, satisfying 
\begin{equation}
\label{eq:def_Psquare}
\Vert\vec{P}\Vert^{2}=\displaystyle\sum\limits_{i=1}^{d}[P^{i}]^{2}.
\end{equation}

Taking into account the results (\ref{eq:separability_integrals}) and (\ref{eq:medida_inducida_local}), we can express the integrals $\int_{\mathcal{M}^{(N )}}d\lambda_{\omega^{(N)}}(u)(...)$ as
\begin{eqnarray}
\nonumber
\displaystyle\int_{\mathcal{M}^{(N)}}\!\!\!\!\!\!\!\!d\lambda_{\omega^{(N)}}(u)(...)\!\!\!&=&\!\!\!m^{N}\!\!\!\!\displaystyle\int_{\Sigma_{0}}\!\!\!\!\textup{vol}_{g_{S}}(\vec{x}_{1})\!\displaystyle\int_{\mathbb{R}^{d}}\!\!\displaystyle\frac{d^{d}P_{1}}{P_{1}^{0}}...\\
\label{eq:integrals_M_N}
&\phantom{=}& \displaystyle\int_{\Sigma_{0}}\!\!\!\!\textup{vol}_{g_{S}}(\vec{x}_{N})\!\displaystyle\int_{\mathbb{R}^{d}}\!\!\displaystyle\frac{d^{d}P_{N}}{P_{N}^{0}}(...),
\end{eqnarray}
where $\vec{x}_{A}$ are the coordinates of the projection $\pi_{A}\circ \varpi_{A}(u)$ on $\Sigma_{0}$ for a point $u \in \mathcal{M}^{(N)}$,  $\vec{P}_{A}$ are Cartesian coordinates in $\mathbb{R}^{d}$, and $ P_{A}^{0}=(m^{2}+\Vert \vec{P}_{A}\Vert^{2})^{\frac{1}{2}}$.
\color{black}
Using the separation of integrals stated in Eq.(\ref{eq:integrals_M_N}), we can re-express the generalized Gibbs distribution $\rho^{(N)}$ as the product 
\begin{equation}
\label{eq:gibbs_separable}
\rho^{(N)}(\chi;u)=N!\prod\limits_{A=1}^{N}\rho(\chi:\varpi_{A}(u)),
\end{equation}
where  $N!$ appears as a corrective factor associated with the no-distinguishability of the particles  \cite{callen1985}. Moreover,  $\rho:\mathfrak{g}_{\odot}\times T^{*}M\rightarrow [0,+\infty)$ is the one-particle Gibbs distribution defined by
\begin{equation}
\label{eq:gibbs_one_particle}
\rho(\chi;u)=\displaystyle\frac{\Theta_{\mathcal{M}}(u)}{z(\chi)}e^{-\mu_{\chi}(u)},
\end{equation}
where $z:\mathfrak{g}_{\odot}\rightarrow \mathbb{R}$ is the one-particle partition function $Z(\chi)=\frac{z(\chi)^{N}}{N!}$ with
\begin{equation}
\label{eq:partition_function_one_particle}
z(\chi)=m\displaystyle\int_{\Sigma_{0}}\!\!\!\textup{vol}_{g_{S}}(\vec{x})\displaystyle\int_{\mathbb{R}^{d}}\!\! \displaystyle\frac{d^{d}P}{P^{0}}e^{-P^{0}\xi_{\chi}^{0}-\vec{P}\cdot \vec{\xi}_{\chi}}.
\end{equation}
Here, $P^{0}=(m^{2}+\Vert \vec{P}\Vert^{2})^{\frac{1}{2}}$,   $(\xi_{\chi}^{0},\vec{\xi}_{\chi})$ are the components of the Killing vector field in the orthonormal basis generated by the Cartesian coordinates $(P^{0},\vec{P})$,  and 
\begin{equation}
\label{eq:product}
\vec{P}\cdot \vec{\xi}_{\chi}=\displaystyle\sum\limits_{i=1}^{d}P^{i}\xi_{\chi}^{i}.
\end{equation}

 As a particular case of a one-particle Gibbs function,
 in Appendix (\ref{ap:partition_function}), we show that when the Killing vector field $\xi_{\chi}$ is time-like, the corresponding one-particle partition function simplifies to 
\begin{equation}
\label{eq:partition_function_time_like}
z_{g}(\chi)=2m^{d}(2\pi)^{\frac{d-1}{2}}\displaystyle\int_{\Sigma_{0}}\textup{vol}_{g_{S}}(\vec{x}) k_{\frac{d-1}{2}}(m\Vert\xi_{\chi}(\vec{x})\Vert),
\end{equation}
where $k_{\nu}(\alpha)$ is defined as the quotient
\begin{equation}
\label{eq:k_nu}
k_{\nu}(\alpha)=\displaystyle\frac{K_{\nu}(\alpha)}{\alpha^{\nu}},
\end{equation}
with $K_{\nu}$ as the modified Bessel function of the second kind and order $\nu$, given by \cite{abramowitz}
\begin{equation}
\label{eq:besse_second_kind}
K_{\nu}(\alpha)=\left(\displaystyle\frac{\alpha}{2}\right)^{\nu}\displaystyle\frac{\Gamma(\frac{1}{2})}{\Gamma(\nu+\frac{1}{2})}\displaystyle\int_{1}^{\infty}\!\!d\zeta e^{-\alpha \zeta} (\zeta^{2}-1)^{\nu-\frac{1}{2}}.
\end{equation}  
Here $\Gamma$ is the gamma-function defined by \cite{abramowitz}
\begin{equation}
\label{eq:gamma_function}
\Gamma(\alpha)=\displaystyle\int_{0}^{\infty}\!\!d\zeta \zeta^{\alpha-1}e^{-\zeta}.
\end{equation}

In this scenario, inside the integral given in  Eq.(\ref{eq:partition_function_time_like}) the dependency on the Killing fields is given through their norms. An interesting case, as we will see later, is when we work with the static Souriau vector $-\beta \chi_{0}\in \mathfrak{g}_{\odot}$ for $\beta>0$, because then the Killing field is proportional to $-\xi_{0}$ through a real positive constant $\beta>0$,
\begin{equation}
\label{eq:special_killing}
\xi_{-\beta\chi_{0}}(x)\equiv\xi_{\beta}(x)=-\beta \displaystyle\frac{\partial}{\partial t},
\end{equation}
then $\rho(-\beta\chi_{0};u)\equiv \rho_{\beta}(u)$, and (\ref{eq:gibbs_one_particle}) reduces to
\begin{equation}
\label{eq:gibbs_special}
\rho_{\beta}(u)=\displaystyle\frac{\Theta_{\mathcal{M}}(u)}{z(\chi)}e^{-\beta E(u)},
\end{equation}
with partition function $z(-\beta\chi_{0})=z(\beta)$ given by
\begin{equation}
\label{eq:partition_function_special}
z(\beta)=2m^{d}(2\pi)^{\frac{d-1}{2}}\displaystyle\int_{\Sigma_{0}}\textup{vol}_{g_{S}}(\vec{x}) k_{\frac{d-1}{2}}(m\beta |g_{00}(\vec{x})|^{\frac{1}{2}}).
\end{equation}

To understand  the role of the constant $\beta>0$ in the statistical $\rho_{\beta}$-description of the dilute ideal gas, we will describe in detail two examples. First, we will consider the trivial case of the flat Minkowski spacetime, modeled by the Cartesian space $\mathbb{R}^{D}$ and endowed with the flat Lorentzian metric $\eta$. Then, we will study  the Schwarzchild spacetime $(M,g_{\textup{Sch}})$, which describes a spacetime with a static, finite, and spherically symmetric distribution of matter.
\color{black}
\subsection{Modified  J\"{u}ttner one-particle distributions in Minkowski spacetime}
\label{sec:modified_juttner_like_pdf}

In the Minkowski spacetime $(\mathbb{R}^{D},\eta)$, the isometry group is the Poincar\'e group, defined as the semi-direct product
\begin{equation}
\label{eq:poincare_group}
 GP=\textup{SO}^{+}(1,d)\ltimes \mathbb{R}^{D}
 \end{equation}
between the restricted Lorentz group, formed by spatial rotations and boosts, and the Abelian group of spacetime translations $(\mathbb{R}^{D},+)\sim \mathbb{R}^{D}$ \cite{kriele1999}. In this sense, it is clear that this spacetime is static. In fact, because it is flat, the Cartesian coordinate system $(t,\vec{x})$ extends to all spacetime and allows us to write $\eta$ as
\begin{equation}
\label{eq:eta}\eta=-dt\otimes dt+\sum_{i=1}^{d}dx^{i}\otimes dx^{i},
\end{equation}
and the norm of Killing vectors of the type $\xi_{\beta}$ is $\Vert\xi_ {\beta}\Vert=\beta$. This simplifies the calculation of the partition function from Eq.(\ref{eq:partition_function_special}) to
\begin{equation}
\label{eq:partition_function_minkowski}
z(\beta)=2 m^{d} V \Bigl(\displaystyle\frac{2\pi}{m\beta}\Bigr)^{\frac{d-1}{2}} K_{\frac{d-1}{2}}(m\beta),
\end{equation}
where $V$ is the volume of the slice $\Sigma_{0}$ of the world tube $\Sigma$,
\begin{equation}
\label{eq:volume_slice}
V=\displaystyle\int_{\Sigma_{0}}\!\!\!\textup{vol}_{\eta_{S}}(\vec{x})=\displaystyle\int_{\Sigma_{0}} \!\!\!\!d^{d}x.
\end{equation}

With respect to the world tube $\Sigma$ of the box containing the gas, we know that all Minkowski spacetime geodesics are expressed in coordinates $(t,\vec{x})$ as straight lines  \cite{moore,kriele1999,wald}. Thus, without losing generality, we can work in the co-moving frame of the box with these coordinates, which is a valid frame for a static observer because in this geometry the integral curves of $\xi_{0}$ are time-like geodesics. This means that the world tube of the box is generated as $\mathbb{R}\times S_{\textup{box}}$, where  $S_{\textup{box}}$ is the region occupied by the box  $\mathbb{R}^{d}$. Then,
\begin{equation}
\label{eq:slice}
\Sigma_{0}=S_{\textup{box}},
\end{equation}
and $V$ is actually the real volume of the box.

   In this way, translations in time $t$ leave $S_{\textup{box}}$ invariant because the slices of $\Sigma$ at different times are all the same. The remaining isometries present in $GP_{0}$ (see Eq.\ref{eq:subG0})) depend on the geometry of the box and, therefore, on the region $S_{\textup{box}}$. For example, if the box is a sphere in $\mathbb{R}^{d}$, then the subgroup of spatial rotations leaves the slice $S_{\textup{box}}$ invariant. For more general cases, there can be continuous or discrete rotational symmetries only around certain directions or axes of symmetry of the box.

The boosts are not elements of $\mathfrak{g}_{\odot}$ because, although they could leave $S_{\textup{box}}$ invariant in some cases, and thus be placed in $\mathfrak {gp}_{0}\simeq T_{e}GP_{0}$ an even in $\mathfrak{gp}_{\star}\simeq T_{e}GP_{\star}$, the infinitesimal boost generators do not commute with the time translations generator $\chi_{0}$ \cite{kriele1999}.
\color{black}
What is important in this example is that by identifying $\beta>0$ with the inverse of the ideal gas temperature $T$, the partition function (\ref{eq:partition_function_minkowski}) leads to the partition function $z_{\eta}(\beta)$ of the so-called modified J\"uttner  {distribution}\footnote{Technically, the modified J\"uttner distribution described in references \cite{chacon2009,aragon2018modified,cubero2007} is
	\[\rho_{\textup{MJ}}=\frac{\Theta_{\mathcal{M}}(u)}{2m^{d}V}\left(\displaystyle\frac{m\beta}{2 \pi}\right)^{\frac{d-1}{2}}\frac{e^{-\beta m \gamma}}{\gamma K_{\frac{d-1}{2}}( m\beta)},\]
	where $\gamma=(1+\frac{\Vert \vec{p}\Vert^{2}}{m^{2}})^{\frac{1}{2}}=\frac{p^ {0}}{m}$. The modified J\"uttner distribution presented here differs by a factor $\gamma^{-1}$,
	\[\rho_{\textup{MJ}}=\frac{\rho}{\gamma}.\]
	The reason for this discrepancy is that this factor $\frac{1}{\gamma}=\frac{m}{p^{0}}$, which appears in our integral of moments with the top-form $\sigma_{ x}=d^{d}p \frac{m}{p^{0}}$, is absorbed into the definition of $\rho_{\textup{MJ}}$ in order to integrate over the Liouville/Lebesgue measures $d^{d}x d^{d}p$ and not over the measure $d^{d}x d^{d}p \frac{m}{p^{0}}$ we use here, i. e.,
	\[\int_{\mathcal{M}}d^{d}x d^{d}p \rho_{\textup{MJ}}=\int_{\mathcal{M}} d^{d}x d^{d }p \frac{\rho}{\gamma}= \int_{\mathcal{M}} d^{d}x d^{d}p \frac{m}{p^{0}} \rho_{\beta }.\]} \cite{chacon2009,aragon2018modified,cubero2007}
\begin{equation}
\label{eq:mod_juttner}
\rho_{\eta}(\beta)=\displaystyle\frac{\Theta_{\mathcal{M}}}{2m^{d} V}\left(\displaystyle\frac{m\beta}{2\pi}\right)^{\frac {d-1}{2}}\displaystyle\frac{e^{-\beta E_{\eta}}}{K_{\frac{d-1}{2}}(m\beta)},
\end{equation}
where $E_{\eta}=mU^{t}$ especially stands for the particle energy in the Minkowski scenario.

The $\rho_{\eta}$ distribution models with high precision the data obtained from  simulations of gases in one, two, and three dimensions, when the counting of microstates is performed on a hypersurface of constant time \cite{chacon2009,cubero2007}.

Of course, it is interesting to identify the parameter $\beta>0$ with the inverse of the temperature of the relativistic ideal gas, since geometrically $\beta$ is the norm of the Killing vector fields $\xi_{\beta}=-\beta\xi_{0}$ and, therefore, the inverse of the temperature measured by the static observer $\mathcal{O}$, co-moving with the box, satisfies the relation
\begin{equation}
\label{eq:ehrenfest_tolman_minkowski}
\displaystyle\frac{1}{T}\Bigr|_{\mathcal{O}}=\Vert \xi_{\beta}\Vert.
\end{equation}
 {In the framework of relativistic kinetic theory, this relationship has already been reported in \cite{chacon2009}, when  working with the (unmodified) J\"uttner distribution.}

\subsection{Modified J\"{u}ttner-like one-particle distributions in Schwarzschild spacetime}
\label{subsec:sch_spacetime}
As a second non-trivial example of a statistical system,  consider a diluted ideal gas inside a box that travels along a (spatially bounded) time-like geodesic in a $D=4$-dimensional spacetime, generated by a static and  spherically symmetric distribution of matter with mass $ M_{0}$ and radius $R_{0}$. Specifically, let us study only the region outside the matter distribution described  via the Schwarzschild spacetime, which is  static, spherically symmetric, and asymptotically flat \cite{moore}. The metric tensor is
\begin{equation}
\label{eq:sch_metric}
g_{\textup{Sch}}=-Rdt\otimes dt+\displaystyle\frac{dr\otimes dr}{R}+r^{2}d\theta\otimes d\theta+r^{2}\sin^{2 }\theta\, d\phi\otimes d\phi,
\end{equation}
where $R$ is the shorthand notation for $R=1-\frac{r_{s}}{r}$ and $r_{s}=2M$ is the so-called Schwarzschild radius \cite{moore}.

Physically, in the region defined by the two inequalities $r>r_{s}$ and $r>R_{0}$, the coordinates $(x^{\mu})=(t,r,\theta,\phi)$ used in the last expression have the following interpretations: $t$ is the time measured by a static observer at a spatial distance that is infinitely large (with respect to the center of the matter distribution), $\theta$ and $\phi$ are the azimuthal and polar angles, respectively, and $r$ is the inverse square root of the Gaussian curvature of a family of nested spheres centered on the matter distribution \cite{moore}.

Unlike the Minkowski spacetime, in Schwarzschild spacetime, the isometry group is not the full Poincar\'e group composed of rotations, boosts, and translations (see Eq.(\ref{eq:poincare_group})). The Schwarzschild group of isometries is, in fact, only the subgroup formed by the spatial rotations and the temporal translations (hence the properties of being spherically symmetric and static \cite{moore,kriele1999,wald}). We will denote this group of isometries as
\begin{equation}
\label{eq:sch_group}
G_{\textup{Sch}}=\mathbb{R}\times \textup{SO}(3),
\end{equation}
with the associated Lie algebra
\begin{equation}
\label{eq:sch_algebra}
\mathfrak{g}_{\textup{Sch }}=\mathbb{R}\times \mathfrak{so}(3).
\end{equation}
  
In Schwarzschild spacetime, unlike Minkowski spacetime, geodesics are not, in general, as simple as straight lines. However, from the time and spherical symmetries, together with the asymptotic flatness property, we can derive the following properties for geodesics. Firstly, we can have particles falling radially towards the central distribution, and particles orbiting the distribution. Secondly, a geodesic curve remains in the same hyperplane that contains the center of the distribution, which without losing generality we can take as $\theta=\frac{\pi}{2}$.

Furthermore, in the Schwarzschild spacetime, circular geodesics describe the motion of test particles orbiting the mass distribution on the equatorial hyperplane\footnote{This geodesic, as a curve on $M$, is perfectly described without any ambiguity other than that associated with spherical symmetry and the personal identification that we made of the plane of the geodesic as the hyperplane $\theta=\frac{\pi}{2 }$. This is because, even by fixing the coordinates $\theta$ and $r$, three of the four coordinate equations of the geodesic do not vanish. Furthermore, it is necessary to take into account the mass-shell condition and the fact that the radius $r_{0}$ of the orbit itself is not arbitrary, but depends on the parameters of the system: the Schwarzschild radius and the angular momentum of the particle. For a more complete explanation see \cite{moore}.} $\theta=\frac{\pi}{2}$. Note that if we choose this circular geodesic for our box, then a static observer would identify the world tube as a cylindrical helix (like a coil spring). In any case, the slices at some time are, again, 3-dimensional regions $S_{\textup{box}}$ defined by the shape of the box. In this way, all slices of the world tube are inside a 	
toroid $T_{S_{\textup{box}}}^{2}$, generated by the rotation of the box on the equatorial hyperplane (see Fig.(\ref{fig:slice_sch})). In this case, the subgroup $G_{\textup{Sch}_{0}}$ of isometries that leaves invariant (considering the equivalence relation between slices) $\Sigma_{0}$ is the subgroup that maps the toroid in itself, i.e., the rotations on the $\theta=\frac{\pi}{2}$-hyperplane:
\begin{equation}
\label{eq:Gsch0}
G_{\textup{Sch}_{0}}=\mathbb{R}\times S^{1}.
\end{equation}
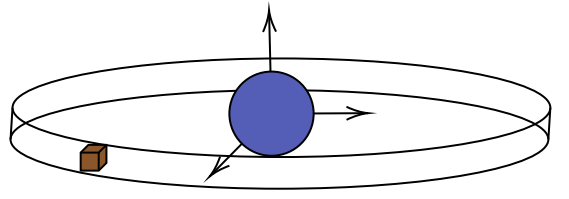
\begin{figure}[]
	\begin{center}
			\tikzset{every picture/.style={line width=0.75pt}} 
	
	\begin{tikzpicture}[x=0.75pt,y=0.75pt,yscale=-1,xscale=1]
	
	\draw    (280.25,126.25) -- (279.05,76) ;
	\draw [shift={(279,74)}, rotate = 88.63] [color={rgb, 255:red, 0; green, 0; blue, 0 }  ][line width=0.75]    (10.93,-3.29) .. controls (6.95,-1.4) and (3.31,-0.3) .. (0,0) .. controls (3.31,0.3) and (6.95,1.4) .. (10.93,3.29)   ;
	\draw    (280.25,126.25) -- (327,126.01) ;
	\draw [shift={(329,126)}, rotate = 179.71] [color={rgb, 255:red, 0; green, 0; blue, 0 }  ][line width=0.75]    (10.93,-3.29) .. controls (6.95,-1.4) and (3.31,-0.3) .. (0,0) .. controls (3.31,0.3) and (6.95,1.4) .. (10.93,3.29)   ;
	\draw    (280.25,126.25) -- (250.4,156.57) ;
	\draw [shift={(249,158)}, rotate = 314.55] [color={rgb, 255:red, 0; green, 0; blue, 0 }  ][line width=0.75]    (10.93,-3.29) .. controls (6.95,-1.4) and (3.31,-0.3) .. (0,0) .. controls (3.31,0.3) and (6.95,1.4) .. (10.93,3.29)   ;
	\draw   (149.63,123.25) .. controls (149.63,109.51) and (210.35,98.38) .. (285.25,98.38) .. controls (360.15,98.38) and (420.88,109.51) .. (420.88,123.25) .. controls (420.88,136.99) and (360.15,148.13) .. (285.25,148.13) .. controls (210.35,148.13) and (149.63,136.99) .. (149.63,123.25) -- cycle ;
	\draw   (148.63,139.25) .. controls (148.63,125.51) and (209.35,114.38) .. (284.25,114.38) .. controls (359.15,114.38) and (419.88,125.51) .. (419.88,139.25) .. controls (419.88,152.99) and (359.15,164.13) .. (284.25,164.13) .. controls (209.35,164.13) and (148.63,152.99) .. (148.63,139.25) -- cycle ;
	\draw    (419.88,139.25) -- (420.88,123.25) ;
	\draw    (148.63,139.25) -- (149.63,123.25) ;
	\draw  [fill={rgb, 255:red, 139; green, 87; blue, 42 }  ,fill opacity=1 ] (184,145.9) -- (187.9,142) -- (197,142) -- (197,151.1) -- (193.1,155) -- (184,155) -- cycle ; \draw   (197,142) -- (193.1,145.9) -- (184,145.9) ; \draw   (193.1,145.9) -- (193.1,155) ;
	\draw  [fill={rgb, 255:red, 84; green, 94; blue, 182 }  ,fill opacity=1 ] (259,126.25) .. controls (259,114.51) and (268.51,105) .. (280.25,105) .. controls (291.99,105) and (301.5,114.51) .. (301.5,126.25) .. controls (301.5,137.99) and (291.99,147.5) .. (280.25,147.5) .. controls (268.51,147.5) and (259,137.99) .. (259,126.25) -- cycle ;
	\end{tikzpicture}
		\caption{Slice $\Sigma_{0}$ of the box world tube as a toroid $T_{S_{\textup{box}}}^{2}$.}
		\label{fig:slice_sch}
	\end{center}
\end{figure}
\noindent In this case, the infinitesimal generator $\chi_{\phi}$ of the subgroup $S^{1}$ of rotations commutes with the time-translation generator $\chi_{0}$ \cite{moore}. Then, if $\chi_{\phi}$ is a Souriau vector, then it is also a static Souriau vector.
 \color{black}
Regarding the ideal gas states described by one-particle Gibbs distributions parameterized by elements of the type $-\beta\chi_{0} \in \mathfrak{g}_{\odot}$, we note that in this case the norm of the Killing vector fields $\xi_{\beta}$ is given by
\begin{equation}
\label{eq:killing_sch_specials}
\Vert\xi_{\beta}\Vert=\beta R^{\frac{1}{2}}=\beta \left(1-\displaystyle\frac{r_{s}}{r}\right)^{\frac{1}{2}},
\end{equation}
while the volume $\textup{vol}_{g_{\textup{Sch}_{S}}}$ induced on $S_{\textup{box}}$ is expressed as
\begin{equation}
\label{eq:gamma_sch}
\textup{vol}_{g_{\textup{Sch}_{S}}}(r,\theta,\phi)=\displaystyle\frac{r^{2}\sin\theta}{R^{\frac{1}{2}}}dr\wedge d\theta\wedge d\phi,
\end{equation}
resulting in the one-particle partition function 
\begin{equation}
\label{eq:partition_function_sch}
z(\beta)=4\pi m^{3}\displaystyle\int_{S_{\textup{box}}}\!\!\!\!\!\!\!\!drd\theta d\phi r^{2}\sin(\theta) \displaystyle\frac{k_{1}(m\beta R^{\frac{1}{2}})}{R^{\frac{1}{2}}} .
\end{equation}

To study the properties of the partition function  $z(\beta)$, it is convenient to perform an expansion in power series of the Schwarzschild radius. As we will see at the end of this subsection, this will allow us to express $z(\beta)$ in terms of the aforementioned powers in $r_{s}$ and the partition function of the Minkowski flat spacetime, i. e., in terms of $z_{\eta}(\beta)$ defined in Eq.(\ref{eq:partition_function_minkowski}) for $\beta=\frac{1}{T}$.

Let us start by expanding the integrand of Eq.(\ref{eq:partition_function_sch}) into a power series of $\frac{r_{s}}{r}$, in the case in which $\frac{r_{s}}{r}<1$, 
\begin{equation}
\label{eq:integrand_series}
r^{2}\sin(\theta)\displaystyle\frac{k_{1}(m\beta R^{\frac{1}{2}})}{R^{\frac{1}{2}}}=r^{2}\sin(\theta) k_{1}(m\beta)\displaystyle\sum\limits_{n=0}^{\infty}A_{n}(m\beta)\left(\displaystyle\frac{r_{s}}{r}\right)^{n},
\end{equation}
where the coefficients $A_{n}$ are defined as
\begin{equation}
\label{eq:A_n}
A_{n}(x)=\displaystyle\sum\limits_{\ell\geq k\geq 0}^{n}b_{\ell,k}\left[P_{0}(k ;x)-P_{1}(k;x)\!\left(1-\displaystyle\frac{\mathcal{E}(x)}{2}\right)\right],
\end{equation}
for the quantities $\mathcal{E}$, $b_{n,k}$, $P_{\kappa}$ and $a_{\kappa}$ given by
\begin{eqnarray}
\label{eq:mathcal_E}
\mathcal{E}&=&x\displaystyle\frac{K_{2}(x)}{K_{1}(x)},\\
\nonumber\\
\label{eq:b_n_k}
b_{n,k}&=&\displaystyle\sum\limits_{j=0}^{k}(-1)^{j+n}\binom{k}{j}\binom{\frac{j}{2 }}{n},\\
\nonumber\\
\label{eq:pol_P_coef}
P_{\kappa}(k;x)&=&\displaystyle\sum\limits_{j_{2}=0}^{k}\displaystyle\sum\limits_{j_{1}=0}^{j_{ 2}}\displaystyle\sum\limits_{j_{0}=0}^{j_{1}-\kappa}a_{\kappa}(k,j_{2},j_{1},j_{ 0}) x^{2j_{0}},\\
\nonumber\\
\label{eq:a_k}
a_{\kappa}(k,j_{2},j_{1},j_{0})\!\!\!&=&\displaystyle\frac{(-1)^{j_{2}+j_{1}+j_{0}}(-j_{2})_{2(j_{2 }-j_{1})}}{2^{j_{2}-2j_{1}+2j_{0} }(k-j_ {2})!(j_{2}-j_{1})!}\times\\
\nonumber\\
\nonumber
&&\displaystyle\frac{(-1)_{k-j_{2}}(1)_{j_{1}}(j_{1}-j_{0}-\kappa)!}{j_{0}!j_{2}!(j_{1}-2j_{0}-\kappa)!(-j_{1})_{j_{0}} (1)_{j_{0}+\kappa}},
\end{eqnarray}
where the symbol $\binom{k}{j}$ stands for the binomial coefficient defined as $\binom{k}{j}=\frac{k(k-1)...(k-j+1)}{j!}$ and  $(x)_{k}$ is the Pochhammer symbol defined as \cite{abramowitz}
\begin{equation}
\label{eq:pochhammer}
(x)_{k}=\left\{\begin{array}{rll}
\displaystyle\frac{\Gamma(x+k)}{\Gamma(x)},&\qquad& x>0,\\
(-1)^{k}\displaystyle\frac{\Gamma(1-x)}{\Gamma(1-x-k)},&&x<0.
\end{array}\right.
\end{equation}
In Appendix (\ref{ap:series}), we include details of these calculations. Moreover, the first five coefficients $A_{n}$ are given in Table (\ref{tab:coeff_A}).
 It is only important to mention that the evaluation of $\mathcal{E}$ in $m\beta$ is physically identified as $\beta$ times the expected value of the Minkowski energy, calculated with respect to the Minkowski-Gibbs distribution $\rho_{\eta}$, i. e.,
\begin{equation}
\label{eq:mathcal_E_mean_energy}
\mathcal{E}(m\beta)=\beta \langle E_{\eta}\rangle_{\eta}=\displaystyle\frac{\beta}{V}\displaystyle\int_{S_{\textup{box}}}\!\!d^{3}x\,E_{\eta}(\vec{x}).
\end{equation}
Indeed, using the identity $\frac{dk_{\nu}}{d\alpha}=-\alpha^{\nu}K_{\nu+1}(\alpha)$ \cite{abramowitz}, first we verify the relation
\[\begin{array}{ll}
\mathcal{E}(m\beta)\!\!\!&=m\beta \displaystyle\frac{K_{2}(m\beta)}{K_{1}(m\beta)}=-\displaystyle\frac{1}{K_{1}(m\beta)}\displaystyle\frac{\partial}{\partial (m\beta)}\bigl(k_{1}(m\beta)\bigr)\\
\\
&=-\displaystyle\frac{m\beta}{4\pi m^{3} V k_{1}(m\beta)}\displaystyle\frac{\partial\bigl(4\pi m^{3} V k_{1}(m\beta)\bigr)}{\partial (m\beta)}\\
\\
&=-\displaystyle\frac{m\beta}{z_{\eta}(\beta)}\displaystyle\frac{\partial z_{\eta}}{\partial (m\beta)},
\end{array}\]
and then, employing the definition of the one-particle partition function (see Eq.(\ref{eq:partition_function_one_particle})), we find
\[\begin{array}{ll}
\mathcal{E}(m\beta)&=-\displaystyle\frac{\beta}{z_{\eta}}\displaystyle\frac{\partial}{\partial (m\beta)}\left(m\!\displaystyle\int_{S_{\textup{box}}}\!\!\!\!d^{3}x \displaystyle\int_{\mathbb{R}^{3}}\!\!\displaystyle\frac{d^{3}P}{P^{0}} e^{-m\beta E_{\eta}}\right)\\
\\
&=\displaystyle\frac{\beta}{z_{\eta}}\left(m\!\displaystyle\int_{S_{\textup{box}}}\!\!\!\!d^{3}x \displaystyle\int_{\mathbb{R}^{3}}\!\!\displaystyle\frac{d^{3}P}{P^{0}}e^{-m\beta E_{\eta}} E_{\eta}\right)\\
\\
&=\displaystyle\frac{\beta}{V}\displaystyle\int_{S_{\textup{box}}}d^{3}x E_{\eta}(\vec{x})=\beta \langle E_{\eta}\rangle_{\eta}.
\end{array}\]

In Eq.(\ref{eq:integrand_series}), we note that the integrand $r^{2}\sin(\theta) k_{1 }(m\beta)$ of the Minkowski partition function $z_{\eta}(\beta)$ multiplies the power series of $\frac{r_{s}}{r}$. Therefore, we can express the one-particle partition function of the Schwarzschild case as a power series of average values,
\begin{equation}
\label{eq:partition_function_series}
z(\beta)=z_{\eta}(\beta)\left[1+\displaystyle\sum\limits_{n=1}^{\infty} \left\langle \displaystyle\frac {A_{n}(m\beta)}{r^{n}}\right\rangle_{\!\!\eta}r_{s}^{n}\right],
\end{equation}
where the averages values are calculated with respect to the modified J\"{u}ttner distribution, that is, they can be represented as the integrals
\begin{equation}
\label{eq:average_minkowski}
\left\langle\displaystyle\frac {A_{n}(m\beta)}{r^{n}} \right\rangle_{\eta}=\displaystyle\frac{1}{V}\displaystyle\int_{S_{\textup{box}}}\!\!drd\theta d\phi r^{2}\sin(\theta) \displaystyle\frac {A_{n}(m\beta)}{r^{n}},
\end{equation}
where the volume $V$ of the box is now given by
\begin{equation}
\label{eq:volume_minkowski}
V=\int_{S_{\textup{box}}}\!\!\!\!dr d\theta d\phi r^{2}\sin(\theta).
\end{equation}

Similarly, we can expand the Gibbs factor $e^{-\mu_{\beta}}$ into power series of $r_{s}$, taking into account that the energy $E$, defined earlier in Eq.(\ref{eq:static_energy}), can be expressed in terms of $E_{\eta}$ as
\begin{equation}
\label{eq:energy_sch}
E=-g_{00}E_{\eta}=\left(1-\frac{r_{s}}{r}\right)E_{\eta}.
\end{equation}
Thus, for the Gibbs factor $e^{(1-\frac{r_{s}}{r})E_{\eta}}$ we find the series expansion
\begin{equation}
\label{eq:gibbs_factor_sch_series}
e^{-\mu_{\beta}}=e^{-\beta E_{\eta}}\displaystyle\sum\limits_{n=0}^{\infty}\displaystyle\frac{(\beta E_{\eta})^{n}}{n! r^{n}} r_{s}^{n}.
\end{equation}

Now, calculating the quotient of the series (\ref{eq:gibbs_factor_sch_series}) and the series (\ref{eq:partition_function_series}), as explained in reference \cite{grads2014}, we find an expression for the Schwarzschild one-particle Gibbs pdf in terms of the modified J\"{u}ttner distribution  as
\begin{equation}
\label{eq:gibs_sch_series}
\rho(\beta)=\rho_{\eta}(\beta)\left[1+\displaystyle\sum\limits_{n=1}^{\infty}\mathcal{A}_{n}(m\beta,r)r_{s}^{n}\right],
\end{equation}
where the coefficients $\mathcal{A}_{n}$ are defined by\footnote{This closed-form for the coefficients of a series quotient, instead of the recursive form described in \cite{grads2014}, is obtained by considering the inverse of the series in the denominator and multiplying it, using the Cauchy product \cite{grads2014,apostol1997}, with the numerator series.} 
\begin{equation}
\label{eq:mathcal_A_coeff}
\mathcal{A}_{n}(m\beta,r)=\displaystyle\sum\limits_{\ell=0}^{n} J\left[\left\langle \displaystyle\frac{A_{\ell}(1;m\beta)}{r^{\ell}}\right\rangle_{\eta}\right]\displaystyle\frac{(\beta E_{\eta})^{n-\ell}}{(n-\ell)! r^{n-\ell}},
\end{equation}
and the expression $J[\langle A_{\ell}(m\beta)/r^{\ell}\rangle_{ \eta}]$ stands for
\begin{equation}
\label{eq:JAl}
\begin{array}{l}
J\left[\langle \frac{A_{\ell}}{r^{\ell}}\rangle_{\eta}\right]=\\
\\
(-1)^{\ell}\textup{det}\left(\begin{array}{ccccc}
\langle \frac{A_{1}}{r}\rangle_{\eta}&1&0&...&0\\
\langle \frac{A_{2}}{r^{2}}\rangle_{\eta}&\langle \frac{A_{1}}{r}\rangle_{\eta}&1&...&0\\
\vdots&\vdots&\vdots&\ddots&\vdots\\
\langle \frac{A_{\ell-1}}{r^{\ell-1}}\rangle_{\eta}&\langle \frac{A_{\ell-2}}{r^{\ell-2}}\rangle_{\eta}&\langle \frac{A_{\ell-3}}{r^{\ell-3}}\rangle_{\eta}&...&1\\
\langle \frac{A_{\ell}}{r^{\ell}}\rangle_{\eta}&\langle \frac{A_{\ell-1}}{r^{\ell-1}}\rangle_{\eta}&\langle \frac{A_{\ell-2}}{r^{\ell-2}}\rangle_{\eta}&...&\langle \frac{A_{1}}{r}\rangle_{\eta}
\end{array}\right).
\end{array}
\end{equation}
The first three coefficients defined  of the series (\ref{eq:mathcal_A_coeff}) can be seen in Table (\ref{tab:mathcal_A}) of Appendix (\ref{ap:series}).

From Eqs.(\ref{eq:partition_function_series}), (\ref{eq:energy_sch}), and (\ref{eq:gibs_sch_series}), it is clear that the behavior of the statistical description of the ideal gas is similar to that of the modified J\"{u}ttner function  in the Minkowski spacetime, in the limit $r\rightarrow \infty$. As we move away from the distribution of matter, the  Schwarzschild spacetime \textit{begins to flatten}, i. e., when $r\gg r_{s}$ is satisfied and only the first few terms in the given series survive. This is also the case with the norm of the  Killing vector $\xi_{\beta}$ (see Eq.(\ref{eq:killing_sch_specials})), which tends (at infinity $r\rightarrow \infty$) to the inverse of the temperature defined for the Minkowski spacetime (see Eq.(\ref{eq:ehrenfest_tolman_minkowski})).
  
With this idea in mind, we can define a generalized temperature function that, at least in static and asymptotically flat spacetimes, tends to the thermodynamic concept of the temperature of a relativistic ideal gas:
\begin{equation}
\label{eq:Temperature}
\mathcal{T}(\vec{x})=\displaystyle\frac{1}{\Vert\xi_{\beta}(\vec{x})\Vert}=\displaystyle\frac{|g_{00}(\vec{x})|^{-\frac{1}{2}}}{\beta}=|g_{00}(\vec{x})|^{-\frac{1}{2}}T,
\end{equation}
which is (barring perhaps suitable constants of proportionality) how the Ehrenfest-Tolman effect shows up in a static spacetime \cite{rovelli2011}. As follows from Eq.(\ref{eq:Temperature}), the temperature is a 
positive definite quantity.  
In the particular case of Schwarzschild spacetime, (\ref{eq:Temperature}) looks like
\begin{equation}
\label{eq:Temperatura_sch}
\mathcal{T}(r,\theta,\phi)=\mathcal{T}(r)=\left(1-\displaystyle\frac{r_{s}}{r}\right)^{-\frac{1}{2}}T.
\end{equation}

Calculating the gradient of the function $\mathcal{T}$, we find the expression
\begin{equation}
\label{eq:gradient_T}
\displaystyle\frac{\textup{grad}\mathcal{T}}{\mathcal{T}}=-\displaystyle\frac{r_{s}}{2 r^{2}}\displaystyle\frac{\partial}{\partial r}=-\displaystyle\frac{M}{r^{2}}\displaystyle\frac{\partial}{\partial r},
\end{equation}
which tells us that the temperature increases as we go deeper and deeper into the family of nested spheres, i. e., when we get closer to the central distribution of matter. This result is in accordance with our physical expectations. 
\color{black}
\section{Equivalence of the temperature-like functions}
\label{sec:equiv_temperatures}
   Let us return to the one-particle function of the dilute ideal gas, $z:\mathfrak{g}_{\odot}\rightarrow \mathbb{R}$, which we defined in Eq.(\ref{eq:partition_function_special}), to discuss an important property concerning the momentum map $\mu$ and  the $\Vert\xi_{\chi}\Vert$-norms of the Killing fields, related to the $\chi\in \mathfrak{g}_{\odot}$ parameters of the $\rho$ distribution. Due to the equivariance of the momentum map $\mu$ (see Eq.(\ref{eq:equivariant}) to remember the concept), the partition function turns out to be invariant under the subgroup $\{e^{s\chi}\}_{\chi\in \mathfrak{g}_{\odot}}$, that is,
\begin{equation}
\label{eq:partition_function_invariance}
z(\textup{Ad}_{\Lambda}(\chi))=z(\chi).
\end{equation}

Before proving Eq.(\ref{eq:partition_function_invariance}), it is important to first establish that the subset $\mathfrak{g}_{\odot}$ of the family of nested sets in Eq.(\ref{eq:nested_sets}), i. e., the subset of all Souriau static vectors, is invariant under the adjoint action of the subgroup $\{e^{s\chi}\}_{\chi\in \mathfrak{g}_{\odot}}$:
\begin{equation}
\label{eq:godot_invariant}
\textup{Ad}_{e^{s\chi}}(\mathfrak{g}_{\odot})=\mathfrak{g}_{\odot}.
\end{equation}
This is due to the fact that $\mathfrak{g}_{\star}$ is Ad-invariant (see Eq.(\ref{eq:invariance_gstar})), and also because if $\textup{Ad}_{e^{-\tau\chi_{0}}}(\chi')=\chi'$, then 
\[\begin{array}{ll}
\textup{Ad}_{e^{-\tau\chi_{0}}}\bigl(\textup{Ad}_{e^{s\chi}}(\chi')\!\!\!&=\textup{Ad}_{e^{-\tau\chi_{0}}e^{s\chi}}(\chi')\\
\\
&=\textup{Ad}_{e^{-\tau\chi_{0}}e^{s\chi}e^{\tau\chi_{0}}e^{-\tau\chi_{0}}}(\chi')\\
\\
&=\textup{Ad}_{\textup{exp}(s\textup{Ad}_{e^{-\tau \chi_{0}}}(\chi))e^{-\tau\chi_{0}}}(\chi')\\
\\
&=\textup{Ad}_{e^{-\tau\chi_{0}}}(\chi')=0,
\end{array}\]
so that if $[\chi',\chi_{0}]=0$, then $[\textup{Ad}_{e^{s\chi}}(\chi'),\chi_{0}]=0$.

To prove Eq.(\ref{eq:partition_function_invariance}), we use 
\[\mu_{\textup{Ad}_{\Lambda}(\chi)}=(\textup{Ad}_{\Lambda^{-1}}^{*}\mu)_{\chi}= \mu_{\chi}\circ \Psi_{\Lambda^{-1}},\]
in the integral that defines $z$,
\[\begin{array}{ll}
z(\textup{Ad}_{\Lambda}(\chi))\!\!\!&=\displaystyle\int_{\mathcal{M}}d\lambda_{\omega}(u) e^{-\mu_{\textup{Ad}_{\Lambda}(\chi)}(u)}\\
\\
&=\displaystyle\int_{\mathcal{M}}d\lambda_{\omega}(u) e^{-(\textup{Ad}_{\Lambda^{-1}}^{*}\mu)_{\chi}(u)}\\
\\
&=\displaystyle\int_{\mathcal{M}}d\lambda_{\omega}(u) e^{-\mu_{\chi}\circ\Psi_{\Lambda^{-1}}(u)},
\end{array}\]
and then we apply the change of variable $v=\Psi_{\Lambda^{-1}}(u)$, taking into account that $\Psi_{\Lambda^{-1}}$ is a diffeomorphism that leaves invariant $\mathcal{M}$ and, more specifically, it is a symplectomorphism that also leaves  $\lambda_{\omega}$ invariant:
\[\begin{array}{ll}
z(\textup{Ad}_{\Lambda}(\chi))\!\!\!&=\displaystyle\int_{\mathcal{M}}d\lambda_{\omega}(u) e^{-\mu_{\chi}\circ\Psi_{\Lambda^{-1}}(u)}\\
\\
&=\displaystyle\int_{\mathcal{M}}d\lambda_{\omega}(v)\,e^{-\mu_{\chi}(v)}=z(\chi).
\end{array}\]
\color{black}
To study the physical implications of the invariance (\ref{eq:partition_function_invariance}), let us return to the special case in which the static Souriau vector $\chi=-\beta\chi_{0}\in \mathfrak{g}_{\odot}$ generates time-like Killing vector fields, whose norms $\Vert\xi_{\beta}\Vert$ are the inverse of the  temperature-like function (\ref{eq:Temperature}). We first notice that for any isometry if $\xi_{\beta}$ is a time-like vector field, then $\xi_{\textup{Ad}_{\Lambda}(\beta)}=\Lambda_{* }\xi_{\beta}$ is also a time-like vector field,
\[\begin{array}{ll}
g(\Lambda(x))(\Lambda_{*}\xi_{\beta},\Lambda_{*}\xi_{\beta})\!\!\!&=(\Lambda^{*}g\circ \Lambda)(x)(\xi_{\beta},\xi_{\beta})\\
\\
&=g(x)(\xi_{\beta},\xi_{\beta})<0.
\end{array}\]
In second place, if we interpret $z(\beta)$ as an average of quantity $e^{-\mu_{\beta}}$, where $\mu_{\beta}$ is $ \beta$-times the energy of the particle with geodesic $\gamma$, measured by the observer with velocity $\xi_{0}$,
\[\mu_{\beta}\circ \gamma=-g(\gamma)(mU,\xi_{\beta})=-\beta g(\gamma)(mU,\xi_{0})=\beta E,\]
then we can interpret $z(\textup{Ad}_{\Lambda}(\beta))$ as the same average but now from the quantities $e^{-\mu_{\textup{Ad}_{\Lambda}(\beta)}}$, where $\mu_{\textup{Ad}_{\Lambda}(\beta)}$ can be understood as $\beta$-times the energy measured by an observer with velocity $\Lambda_{*}\xi_{0}$ (see Eq.(\ref{eq:pushforward_fundamentals_fields})),
\[\begin{array}{ll}
\mu_{\textup{Ad}_{\Lambda}(\beta)}\!\!\!&=-g(mU, \xi_{\textup{Ad}_{\Lambda}(\beta)})\\
\\
&=-g(mU ,\Lambda_{*}\xi_{\beta})=-\beta g(mU,\Lambda_{*} \xi_{0}).
\end{array}\]
Thus, we can say that in the family of inverse temperatures-like functions  $\{\Vert \xi_{\textup{Ad}_{\Lambda}(\beta)}\Vert\}_{\Lambda\in \{e^{s\chi}\}_{\chi\in \mathfrak{g}_{\odot}}}$, the one-particle partition function of the ideal gas is the same. Also, we can say  that under measurements of the family of static observers with velocity in $\{\Lambda_{*}\xi_{0}\}_{\Lambda\in \{e^{s\chi}\}_{\chi\in \mathfrak{g}_{\odot}}}$, the ideal gas one-particle partition function is the same.

\section{Conclusions}
\label{sec:con}
In this work, we show that the influence of the curvature of a background spacetime on a simple many-particle system, such as a dilute ideal gas, can be introduced into the classical theory of statistical mechanics through the tools of symplectic geometry and through ideas found in the so-called Lie group thermodynamics developed by Souriau et al \cite{souriau1997structure,barbaresco2014,marle2016,barbaresco2019,marle2020gibbs}. In this sense, the connection with spacetime occurs through the nature of a free particle and its trajectory as a time-like geodesic, and it is formally introduced into the phase space formulation through the so-called mass-shell constraint. In this way, the metric tensor is encapsulated in the different expressions that appear in the theory. A second way in which spacetime is glimpsed in the ideal gas model is through its isometry group, which gives us the concept of the momentum map as a generalization to the idea of energy as a conserved quantity of a free particle.

The Souriau concept of statistical state to describe a Hamiltonian system of many particles, which in its generality can be applied to all kinds of Hamiltonian systems, is also crucial in the formulation presented here. In this sense, we consider that the present work generalizes the relativistic systems studied by Souriau, from the regime of special relativity to the regime of general relativity.

The example of Schwarzschild spacetime is simple enough to present these static statistical states without falling into a trivial case, which is recovered with the Minkowski spacetime. In this sense, the appearance of the modified J\"{u}ttner distributions is interesting in the context of the old debate on what is the correct distribution for a relativistic ideal gas \cite{cubero2007,chacon2009,aragon2018modified}, and should be analyzed in more detail in future works. Also, it would be interesting to consider more general cases of ideal gases orbiting black holes with angular momentum and both angular momentum and electric charge, i. e., the outer cases of Kerr and Kerr-Newman black holes. It is worth mentioning that these more general spacetimes raise the need to generalize the concept, presented here, of a static equilibrium state to a steady one since these spacetimes are stationary (not static) and there are frame-dragging effects in them \cite{moore,kriele1999,caroll2019}.

Finally, we highlight the result (\ref{eq:partition_function_invariance}) that tells us about the invariance of the partition function under the action of certain isometries in the parameter space, i. e., the Lie algebra of the group of isometries. In the particular case where these parameters are interpreted in terms of the inverse of a temperature, this property of the partition function indicates that all the static states of the system are equivalent for all the temperatures connected through the Adjunct action. In other words, the measurements made by static observers related by means of these special isometries could be macroscopically equivalent.
\begin{acknowledgements}

This work was partially supported  by PAPIIT-DGAPA-UNAM, Grant No. 114520, and Conacyt-Mexico, Grant No. A1-S-31269.

\end{acknowledgements}

\appendix
\section{Equivariance of the momentum map}
\label{ap:equivariant}
  Let $\mu$ be the momentum map of some left and symplectic action $\Psi:G\times U\rightarrow U$, in some symplectic manifold $(U,\Omega)$. Then, for any elements $\Lambda\in G$ and $x\in U$, the next relation is fulfilled \cite{souriau1997structure,marsden1999,beckett2022}:
\begin{equation}
\label{eq:equivariance}
\mu\circ \Psi_{\Lambda}(x)=\textup{Ad}_{\Lambda}^{*}(\mu(x))+\vartheta(\Lambda),
\end{equation}
where $\textup{Ad}^{*}:G\times \mathfrak{g}^{*}\rightarrow \mathfrak{g}^{*}$ is the co-Adjoint action of the group $G$ on the dual of its Lie algebra, defined for any $\alpha \in \mathfrak{g}^{*}$, $\beta\in \mathfrak{g}$ and $\Lambda\in G$ as
\begin{equation}
\label{eq:coAdjoint}
\langle \textup{Ad}_{\Lambda}^{*}(\alpha),\beta\rangle=\langle \alpha,\textup{Ad}_{ \Lambda^{-1}}(\beta)\rangle,
\end{equation}
and where $\vartheta:G\rightarrow \mathfrak{g}^{*}$ is a symplectic 1-cocycle, that is, a map that satisfies the property \cite{souriau1997structure,marsden1999,beckett2022}
\begin{equation}
\label{eq:2_cocycle}
\vartheta(\Lambda_{1}\Lambda_{2})=\textup{Ad}^{*}_{\Lambda_{1}}\bigl(\vartheta(\Lambda_{2})\bigr)+\vartheta (\Lambda_{1}).
\end{equation}

The result (\ref{eq:equivariance}) is a direct consequence of the definition of the momentum map and of the fact that the pushforward under the action $\Psi$ of a fundamental field is identified with another fundamental field, specifically, with the fundamental field of the same element but under the Adjoint action of $G$. Indeed, let us first prove \cite{marsden1999}
\begin{equation}
\label{eq:pushforward_fundamentals_fields}
\,[\Psi_{\Lambda}]_{*}\psi_{\chi}=\psi_{\textup{Ad}_{\Lambda}(\chi)}
\end{equation}
in a few lines:
\[\begin{array}{ll}
\,[\Psi_{\Lambda}]_{*}\psi_{\chi}\!\!\!&=\displaystyle\frac{d}{ds}\Psi_{\Lambda}\circ\Psi_{e^{s\chi}}\Bigr|_{s=0}=\displaystyle\frac{d}{ds}\Psi_{\Lambda e^{s\chi}}\Bigr|_{s=0}\\
\\
&=\displaystyle\frac{d}{ds}\Psi_{\Lambda e^{s\chi}\Lambda^{-1}\Lambda}\Bigr|_{s=0}=\displaystyle\frac{d}{ds}\Psi_{\Lambda e^{s\chi}\Lambda^{-1}}\circ \Psi_{\Lambda}\Bigr|_{s=0}\\
\\
&=\displaystyle\frac{d}{ds}\Psi_{e^{s\textup{Ad}_{\Lambda}(\chi)}}\circ \Psi_{\Lambda}\Bigr|_{s=0}=\psi_{\textup{Ad}_{\Lambda}(\chi)}\circ \Psi_{\Lambda}.
\end{array}\]
This result can be used to derive (\ref{eq:equivariance}). Indeed,  for any vector field $Y\in \mathfrak{X}(U)$, remembering that $\Psi$ acts through symplectomorphisms, we have
\[\begin{array}{ll}
i_{Y}d(\mu_{\chi}\circ \Psi_{\Lambda})\!\!\!&=i_{Y}\Psi_{\Lambda}^{*}(d\mu_{\chi})\\
\\
&=i_{Y}\Psi_{\Lambda}^{*}\left(-i_{\psi_{\chi}}\Omega\right)=\Omega([\Psi_{\Lambda}]_{*}Y,\psi_{\chi})\\
\\
&=(\Psi_{\Lambda}^{*}\Omega)(Y,[\Psi_{\Lambda^{-1}}]^{*}\psi_{\chi})=\Omega(Y,[\Psi_{\Lambda^{-1}}]^{*}\psi_{\chi})\\
\\
&=\Omega(Y,\psi_{\textup{Ad}_{\Lambda^{-1}}(\chi)})=i_{Y}(d\mu_{\textup{Ad}_{\Lambda^{-1}(\chi)}}).
\end{array}\]
Since the above is valid for all $Y\in \mathfrak{X}(U)$, then it is satisfied at the level $d(\mu_{\chi}\circ \Psi_{\Lambda}-\mu_{\textup{Ad}_{\Lambda}^{-1}(\chi)})=0$, i. e., \[
\langle \mu\circ \Psi_{\Lambda}-\textup{Ad}_{\Lambda}^{*}(\mu),\chi\rangle=f(\Lambda,\chi),\] 
where $f(\Lambda,\chi)$ is some constant over $U$. In fact, given the linearity in $\langle \mu\circ\Psi_{\Lambda}-\textup{Ad}_{\Lambda^{-1}}^{*}(\mu),\chi\rangle$ with respect to $\chi$, it is clear that $f(\Lambda,\chi)$ must be linear in $\chi$, so we can define $f(\Lambda,\chi)=\langle \vartheta(\Lambda) ,\chi\rangle$ and with it arrive at the expression  (\ref{eq:equivariance}). On the other hand, the property (\ref{eq:2_cocycle}) follows after a bit of algebra, taking into account that both $\Psi$ and $\textup{Ad}^{*}$ are left actions:
\[\begin{array}{ll}
\vartheta(\Lambda_{1}\Lambda_{2})\!\!\!&=\mu\circ \Psi_{\Lambda_{1}\Lambda_{2}}-\textup{Ad}_{\Lambda_{1}\Lambda_{2}}^{*}(\mu)\\
\\
&=\mu\circ \Psi_{\Lambda_{1}}\circ \Psi_{\Lambda_{2}}-\textup{Ad}_{\Lambda_{1}}^{*}\circ \textup{Ad}_{\Lambda_{2}}^{*}(\mu)\\
\\
&=\mu\circ \Psi_{\Lambda_{1}}\circ \Psi_{\Lambda_{2}}-\textup{Ad}_{\Lambda_{1}}^{*}(\mu\circ \Psi_{\Lambda_{2}})\\
\\
&\phantom{=}+\textup{Ad}_{\Lambda_{1}}^{*}(\mu\circ \Psi_{\Lambda_{2}})-\textup{Ad}_{\Lambda_{1}}^{*}\circ \textup{Ad}_{\Lambda_{2}}^{*}(\mu)\\
\\
&=(\mu\circ \Psi_{\Lambda_{1}}-\textup{Ad}_{\Lambda_{1}}^{*}(\mu))\circ \Psi_{\Lambda_{2}}\\
\\
&\phantom{=}+\textup{Ad}_{\Lambda_{1}}^{*}(\mu\circ \Psi_{\Lambda_{2}}-\textup{Ad}_{\Lambda_{2}}^{*}(\mu))\\
\\
&=\vartheta(\Lambda_{1})+\textup{Ad}_{\Lambda_{1}}^{*}(\vartheta(\Lambda_{2})).
\end{array}\]

A special case of momentum map is that where the 1-cocycle $\vartheta$ of Eq.(\ref{eq:equivariance}) is zero. Such an expression can be reinterpreted as
\begin{equation}
\label{eq:equivariance_2}
\mu_{\textup{Ad}_{\Lambda}(\chi)}\circ \Psi_{\Lambda}(x)=\mu_{\chi}(x).
\end{equation}
We call these momentum maps equivariant \cite{souriau1997structure,marsden1999,beckett2022}. Formally, an equivariant momentum map satisfies the commutation diagram
\begin{figure}[H]
	\begin{center}
		\tikzset{every picture/.style={line width=0.75pt}} 

\begin{tikzpicture}[x=0.75pt,y=0.75pt,yscale=-1,xscale=1]

\draw    (162,38) -- (310.83,38) ;
\draw [shift={(313.83,38)}, rotate = 180] [fill={rgb, 255:red, 0; green, 0; blue, 0 }  ][line width=0.08]  [draw opacity=0] (8.93,-4.29) -- (0,0) -- (8.93,4.29) -- cycle    ;
\draw    (160,166.28) -- (308.83,166.28) ;
\draw [shift={(311.83,166.28)}, rotate = 180] [fill={rgb, 255:red, 0; green, 0; blue, 0 }  ][line width=0.08]  [draw opacity=0] (8.93,-4.29) -- (0,0) -- (8.93,4.29) -- cycle    ;
\draw    (149,46) -- (149,154.28) ;
\draw [shift={(149,157.28)}, rotate = 270] [fill={rgb, 255:red, 0; green, 0; blue, 0 }  ][line width=0.08]  [draw opacity=0] (8.93,-4.29) -- (0,0) -- (8.93,4.29) -- cycle    ;
\draw    (324.83,46) -- (324.83,154.28) ;
\draw [shift={(324.83,157.28)}, rotate = 270] [fill={rgb, 255:red, 0; green, 0; blue, 0 }  ][line width=0.08]  [draw opacity=0] (8.93,-4.29) -- (0,0) -- (8.93,4.29) -- cycle    ;

\draw (141,30) node [anchor=north west][inner sep=0.75pt]    {$U$};
\draw (140,159) node [anchor=north west][inner sep=0.75pt]    {$U$};
\draw (318,30) node [anchor=north west][inner sep=0.75pt]    {$\mathfrak{g}^{*}$};
\draw (318,155) node [anchor=north west][inner sep=0.75pt]    {$\mathfrak{g}^{*}$};
\draw (227,20) node [anchor=north west][inner sep=0.75pt]    {$\mu $};
\draw (228,177) node [anchor=north west][inner sep=0.75pt]    {$\mu $};
\draw (115,90) node [anchor=north west][inner sep=0.75pt]    {$\Psi _{\Lambda }$};
\draw (330,88) node [anchor=north west][inner sep=0.75pt]    {$\textup{Ad}_{\Lambda}^{*}$};

\end{tikzpicture}
		\caption{Commutation diagram of an equivariant momentum map.}
		\label{fig:commutation_equivariant}
	\end{center}
\end{figure}
\noindent Geometrically, an equivariant momentum map sends the $G$-orbit $\{\Psi_{\Lambda}(x)\}_{\Lambda\in G}$ initiated at a point $x\in U$, to the $G$-orbit $\{\textup{Ad}^{*}(\mu(x))\}_{\Lambda\in G}$ initiated at the point $\mu(x)\in\mathfrak{g}^{*}$.

\section{Induced measures in $\mathcal{M}$ hypermanifolds}
\label{ap:induced_measures}
Let $\lambda_{\omega}|_{\mathcal{M}}$ be the induced measure defined in Eq.(\ref{eq:induced_measure}). As a consequence of the  definition of the submanifold $\mathcal{M}$ as a union of $\mathcal{P}_{x}^{-1}(0)\subset T_{x}^{*}M$ for all $x\in \Sigma_{0}$ (see Eq.(\ref{eq:end_support})), we can note that the integrals $\int_{\mathcal{M}}d\lambda_{\omega}|_{\mathcal{M}}(u)(...)$ can be expressed as
\[\displaystyle\int_{\mathcal{M}}d\lambda_{\omega}|_{\mathcal{M}}(u)(...)=\displaystyle\int_{\Sigma_{0}}d\gamma(x)\displaystyle\int_{\mathcal{P}_{x}^{-1}(0)}d\sigma_{x}(p)(...),\]
where, if $(x^{\mu},p_{\mu})$ is a coordinate system in which we have $p_{\mu}=i_{\partial/\partial x^{\mu}}p $, $\gamma(x)$ is the measure induced, over $\Sigma_{0}$, by the top-form on $M$
\[d^{D}x=dx^{0}\wedge dx^{1}\wedge...\wedge dx^{d},\]
while $\sigma_{x}$ is the measure, over $\mathcal{P}_{x}^{-1}(0)$, induced by the top-form on $T_{x}^{*}M$:
\[d^{D}p=dp_{0}\wedge dp_{1}\wedge...\wedge dp_{D}.\]
Indeed, this is true because the coordinates $(x^{\mu},p_{\mu})$ are Darboux coordinates \cite{da2008lectures}, in which $\Omega=dp_{\mu}\wedge dx^{\nu} $ and thus the Liouville measure locally looks like the Lebesgue measure on $\mathbb{R}^{2D}$ \cite{da2008lectures}:
\[\displaystyle\int_{A}d\lambda_{\omega}(u)(...)=\displaystyle\int_{\varphi(A)}d^{D}xd^{D}p(. ..).\]

Taking into account that at the point $x\in \Sigma_{0}$ the metric tensor returns the metric $g(x)$ on $T_{x} M$, by duality, it also induces the metric $g(x)^{-1}$ on the dual space $T_{x}^{*}M$. Thus, both $(T_{x}M,g(x))$ and $(T_{x}^{*}M\simeq \mathbb{R}^{D},g(x)^{-1})$ are metric vector spaces. Therefore, it is convenient to state that $\gamma$ is induced, instead of $d^{D}x$, by the volume-invariant top-form
\begin{equation}
\label{eq:volume_invariant}
\textup{vol}_{g}(x)=|g(x)|^{\frac{1}{2}}d^{D}x,
\end{equation}
while $\sigma_{x}$ is induced, instead of $d^{D}p$, by the volume-invariant top-form\footnote{An equivalent, but formally more developed mathematical approach, using Sasaskian structures on $T^{*}M$, can be found in reference \cite{sarbach2022}.}
\begin{equation}
\label{eq:volume_invariant_p}
\textup{vol}_{g(x)^{-1}}(p)=|g(x)^{-1}|^{\frac{1}{2}}d^{D}p.
\end{equation} 

Considering that we can calculate the gradient field\footnote{To simplify the notation, we drop the argument of $g^{-1}(x)$.} $\textup{grad}_{g^{-1}}\,\mathcal{P}_{x}$ of the function defined in (\ref{eq:mass_shell_constriction}), thanks to the metric $ g^{-1}$, and recognizing that this field is normal to $\mathcal{P}^{-1}_{x}(0)$ at every point\footnote{Let $X$ be a tangent vector field to the zero-level hypersurface $\mathcal{P}_{x}^{-1}(0)$, i. e., $i_{X}d\mathcal{P}_{x}(p)=0$. By definition of the gradient field, we have that $\textup{grad}_{g^{-1}}\mathcal{ P}_{x}$ is orthogonal to $X$: $g^{-1}(\textup{grad}_{g^{-1}}\mathcal{P}_{x},X)=i_ {X}d\mathcal{P}_{x}=0$. Since this is the case for all tangent fields to $\mathcal{P}_{x}^{-1}(0)$, $\textup{grad}_{g^{-1}}\mathcal{P}_ {x}$ is normal to $\mathcal{P}_{x}^{-1}(0)$.}, we can express $\sigma_{x}$ as \cite{lee2012}
\begin{equation}
\label{eq:sigmax}
\sigma_{x}(p)=\displaystyle\frac{i_{\textup{grad}_{g^{-1}}\mathcal{P}_{x}}\textup{vol}_{g^{-1}}(p)}{\Vert \textup{grad}_{g^{-1}}\mathcal{P}_{x}\Vert}\Bigr|_{\mathcal{P}_{x}^{-1}(0)}.
\end{equation}

Now, the easiest way to compute the gradient of $\mathcal{P}_{x}$ and perform the evaluation over $\textup{vol}_{g^{-1}}$ is to pass to an orthogonal basis  over $T_{x}^{*}M$, where the metric $g^{-1}$ becomes $\eta=\textup{diag}(-1,1,1,...,1)$, which is equivalent to using a Cartesian system $(P^{0},P^{1},...,P^{d})=(P^ {\boldsymbol{0}},\vec{P})$, where
\[\mathcal{P}_{x}(P)=-(P^{0})^{2}+\Vert \vec{P}\Vert^{2}+m^{2}.\]
Hence, the normalized gradient field (over $\mathcal{P}_{x}^{-1}(0)$)  is just
\begin{equation}
\label{eq:gradient_Px}
\displaystyle\frac{\textup{grad}_{g^{-1}}\,\mathcal{P}_{x}}{\Vert \textup{grad}_{g^{-1}}\,\mathcal{P}_{x}\Vert}\Bigr|_{\mathcal{P}_{x}^{-1}(0)}=\displaystyle\frac{P^{\mu}}{m}\displaystyle\frac{\partial}{\partial P^{\mu}},
\end{equation}
with
\begin{equation}
\label{eq:P0}
P^{0}=\left(m^{2}+\Vert\vec{P}\Vert^{2}\right)^{\frac{1}{2}},
\end{equation}
and the $D$-form $\textup{vol}_{g^{-1}}$ is given by
\begin{eqnarray}
\nonumber
\textup{vol}_{g^{-1}}\Bigr|_{\mathcal{P}_{x}^{-1}(0)}&=&dP^{0}\wedge dP^{1}\wedge...\wedge dP^{d}\Bigr|_{\mathcal{P}_{x}^{-1}(0)}\\
\label{eq:vol_g1}
&=&dP^{0}\wedge d^{d}P\Bigr|_{\mathcal{P}_{x}^{-1}(0)}.
\end{eqnarray}
Indeed, by definition, in the vector basis $(\frac{\partial}{\partial P^{0}},\frac{\partial}{\partial P^{1}},...,\frac{\partial}{\partial P^{d}})$ generated by the Cartesian coordinates $(P^{0},\vec{P})$, we have
\[\begin{array}{ll}
\textup{grad}_{g^{-1}}\mathcal{P}_{x}\!\!\!&=\eta^{\mu\nu}\displaystyle\frac{\partial \mathcal{P}_{x}}{\partial P^{\mu}}\displaystyle\frac{\partial}{\partial P^{\nu}}\\
\\
&=-2\eta^{00}P^{0}\displaystyle\frac{\partial}{\partial P^{0}}+2\displaystyle\sum\limits_{i=1}^{d}\eta^{ii}P^{i}\displaystyle\frac{\partial}{\partial P^{i}}\\
\\
&=2P^{0}\displaystyle\frac{\partial}{\partial P^{0}}+2P^{i}\displaystyle\frac{\partial}{\partial P^{i}}\\
\\
&= 2P^{\mu}\displaystyle\frac{\partial}{\partial P^{\mu}},
\end{array}\]
such that
\[\begin{array}{ll}
\Vert \textup{grad}_{g^{-1}}\mathcal{P}_{x}\Vert\Bigr|_{\mathcal{P}_{x}^{-1}(0)}\!\!\!\!\!\!\!&=|\eta(\textup{grad}_{g^{-1}}\mathcal{P}_{x},\textup{grad}_{g^{-1}}\mathcal{P}_{x})|^{\frac{1}{2}}\\
\\
&=2|(P^{0})^{2}-\Vert \vec{P}\Vert^{2}|^{\frac{1}{2}}=2m,
\end{array}\]
while the expression (\ref{eq:vol_g1}) for Eq.(\ref{eq:volume_invariant_p}) is the result of the invariance of the volume form with respect to diffeomorphisms $(p_{\mu})\mapsto (P^{\mu})$.

By substituting the expressions (\ref{eq:gradient_Px}) and (\ref{eq:vol_g1}) into Eq.(\ref{eq:sigmax}), we find
\[\sigma_{x}(P)=\displaystyle\frac{1}{m}\left(P^{0}d^{d}P+\displaystyle\sum\limits_{i=1}^{d}d_{i}^{D}P\right)_{\mathcal{P}_{x}^{-1}(0)},\]
where, just seeking to simplify the notation, we have defined
\[\begin{array}{ll}
d_{i}^{D}P\!\!\!&=(-1)^{i}P^{i}dP^{0}\wedge...\wedge dP^{i-1}\wedge dP^{i+1}\wedge....\wedge dP^{d}\\
\\
&=-dP^{1}\wedge...\wedge dP^{i-1}\wedge P^{i}dP^{0}\wedge dP^{i+1}\wedge...\wedge dP^{d}.
\end{array}\]
In this regard, we note that, as a consequence of $d\mathcal{P}_{x}|_{\mathcal{P}_{x}^{-1}(0)}=0$, this last $d$-form can be rewritten as
\[d_{i}^{D}P\Bigr|_{\mathcal{P}_{x}^{-1}(0)}=-\displaystyle\frac{(P^{i})^{2}}{P^{0}}d^{d}P\Bigr|_{\mathcal{P}_{x}^{-1}(0)},\]
in such a way that, going back to $\sigma_{x}$, we find
\begin{equation}
\label{eq:sigmax2}
\sigma_{x}(P)=\displaystyle\frac{m}{P^{0}}d^{d}P\Bigr|_{\mathcal{P}_{x}^{-1}(0)}=\displaystyle\frac{m d^{d}P}{(m^{2}+\Vert\vec{P}\Vert^{2})^{\frac{1}{2}}}.
\end{equation}
Indeed,
\[\begin{array}{ll}
\sigma_{x}(P)\!\!\!&=\displaystyle\frac{1}{m}\left(P^{0}d^{d}P-\displaystyle\sum\limits_{i=1}^{d}\displaystyle\frac{(P^{i})^{2}}{P^{0}}d^{d}P\right)_{\mathcal{P}_{x}^{-1}(0)}\\
\\
&=\displaystyle\frac{1}{m P^{0}}\left((P^{0})^{2}-\Vert\vec{P}\Vert^{2}\right)d^{d}P\Bigr|_{\mathcal{P}_{x}^{-1}(0)}\\
\\
&=\displaystyle\frac{m}{P^{0}}d^{d}P\Bigr|_{\mathcal{P}_{x}^{1-}(0)}.
\end{array}\]

The case of $\gamma$ is even simpler since, by definition of the slices $\Sigma_{0}$, the normal vector field is the Killing vector field $\xi_{0}$. Then, using a coordinate system $(t,\vec{x})$, where the metric $g$ splits into the sum given by Eq.(\ref{eq:static_g}), in which case $|g|^{\frac{1}{2}}=|g_{00}|^{\frac{1}{2}}|g_{S}|^{\frac{1}{2}}$, we find that 
\begin{equation}
\label{eq:gammax}
\gamma(x)=|g_{S}(\vec{x})|^{\frac{1}{2}}d^{d}x=\textup{vol}_{g_{S}}(\vec{x}).
\end{equation}
Indeed,
\[\begin{array}{ll}
\gamma(x)\!\!\!&=\displaystyle\frac{i_{\xi_{0}}\textup{vol}_{g}}{\Vert\xi_{0}\Vert}\\
\\
&=\displaystyle\frac{|g(\vec{x})|^{\frac{1}{2}}i_{\frac{\partial}{\partial t}}dt\wedge dx^{1}\wedge...\wedge dx^{d}}{|g_{00}(\vec{x})|^{\frac{1}{2}}}\\
\\
&=|g_{S}(\vec{x})|^{\frac{1}{2}}d^{d}x=\textup{vol}_{g_{S}}(\vec{x}).
\end{array}\]

Putting together the results (\ref{eq:sigmax2}) and (\ref{eq:gammax}), we effectively obtain the expression (\ref{eq:medida_inducida_local}):
\[\omega_{\Omega}|_{\mathcal{M}}(\vec{x},\vec{P})=\gamma(\vec{x})\wedge \sigma_{x}(P)=  \displaystyle\frac{m|g_{S}(\vec{x})|^{\frac{1}{2}}d^{d}x\wedge d^{d}P(\vec{x})}{(m^{2}+\Vert\vec{P}(\vec{x})\Vert^{2})^{\frac{1}{2}}}.\]
\color{black}
\section{One-particle partition function for time-like Killing fields}
\label{ap:partition_function}
It is interesting to note that we can perform the integration with respect to  the moments in the integral that defines the one-particle partition function $z(\chi)$ (see Eq.(\ref{eq:partition_function_one_particle})) for any parameter $\chi\in \mathfrak{g}_{\star}$ with an associated time-like Killing vector field $\xi_{\chi}$. Indeed, to show this, let us start by defining the integral of moments as
\begin{equation}
\label{eq:int_i}
\mathcal{I}(\chi)=\displaystyle\int_{\mathbb{R}^{d}}\!\!d^{d}P\,\displaystyle\frac{e^{ -P^{0}\xi_{\chi}^{0}}}{P^{0}}e^{-\vec{P}\cdot \vec{\xi}_{\chi}},
\end{equation}
where $P^{0}=(m^{2}+\Vert P\Vert^{2})^{\frac{1}{2}}$.

The trick\footnote{This trick is found in  reference \cite{chacon2009}, where it is used for a very similar calculation, but from the perspective of kinetic theory.} consists in changing the Cartesian coordinates $(P^{1},...,P^{d})$ by the hyperspherical coordinates $(P,\phi,\theta_{1},\theta_{2},...,\theta_{d-2})$ through the transformation \cite{blumenson1960}
\begin{equation}
\label{eq:cartesian_to_hyperspherical}
\begin{array}{rl}
P^{1}\!\!\!&=P\cos(\theta_{1}),\\
P^{2}\!\!\!&=P\sin(\theta_{1})\cos(\theta_{2}),\\
P^{3}\!\!\!&=P\sin(\theta_{1})\sin(\theta_{2})\cos(\phi_{3}),\\
\vdots\\
P^{d-1}\!\!\!&=P\sin(\theta_{1})...\sin(\theta_{d-2})\cos(\phi),\\
P^{d}\!\!\!&=P\sin(\theta_{1})...\sin(\theta_{d-1})\sin(\phi),\\
\end{array}
\end{equation}
where $P\in [0,\infty)$ symbolizes the radial coordinate, while $\phi\in [0,2\pi)$ is the polar angle and $\theta_{i}\in [0,\pi]$ are azimuth angles \cite{blumenson1960}. With this change of coordinates, the volume element $d^{d}P$ becomes $dP P^{d-1} d\Omega^{d}$, where $\Omega^{d}$ is the solid angle in $d$ dimensions, i. e., \cite{blumenson1960}
\begin{equation}
\label{eq:solid_angle}
d\Omega^{d}=d\phi \prod\limits_{i=1}^{d-2}d\theta_{i} \sin^{d-i-1}(\theta_{i}).
\end{equation}
Thus, choosing $\theta_{1}$ to be the angle between the vectors $\vec{\xi}_{\chi}$ and $\vec{P}$, $\vec{P}\cdot \vec{\xi}_{\chi}=P \Vert\vec{\xi}_{\chi}\Vert \cos(\theta_{1})$, we can express Eq.(\ref{eq:int_i}) as
\[\begin{array}{ll}
\mathcal{I}\!\!\!&=\!\!\displaystyle\frac{2\pi^{\frac{d-1}{2}}}{\Gamma(\frac{d-1}{2})}\!\displaystyle\int_{0}^{\infty}\!\!\!\!dP P^{d-1} \displaystyle\frac{e^{-P^{0}\xi_{\chi}^{0}}}{P^{0}}\!\!\displaystyle\int_{0}^{\pi}\!\!\!d\theta_{1}\displaystyle\frac{\sin^{d-2}(\theta_{1})}{e^{P\Vert\!\vec{\xi}_{\chi}\!\Vert\!\cos(\theta_{1})}}.
\end{array}\]

Expanding now the exponential $e^{-P\Vert\vec{\xi}_{\chi}\Vert\cos(\theta_{1})}$ in powers of $u=-P\Vert\vec{\xi}_{ \chi}\Vert\cos(\theta_{1})$ around $u=0$, we arrive at the result
\[\begin{array}{ll}
\mathcal{I}&=\displaystyle\frac{2\pi^{\frac{d-1}{2}}}{\Gamma(\frac{d-1}{2})}\displaystyle\sum\limits_{n=0}^{\infty} (-1)^{n}\displaystyle\frac{\Vert\vec{\xi}_{\chi}\Vert^{n}}{n!}\times\\
\\
&\phantom{=}\displaystyle\int_{0}^{\infty}\!\!dP P^{d+n-1}\,\displaystyle\frac{e^{-P^{0}\xi_{\chi}^{0}}}{P^{0}}\displaystyle\int_{0}^{\pi}d\theta_{1}\,\sin^{d-2}(\theta_{1})\cos^{n}(\theta_{1}),
\end{array}\]
where the integral in $\theta_{1}$ is nonzero only for even values of $n$, in which case it takes the value
\begin{equation}
\label{eq:int_theta_1}
\displaystyle\int_{0}^{\pi}d\theta_{1} \sin^{d-2}(\theta_{1})\cos^{2n}(\theta_{1})=\displaystyle\frac{\Gamma(\frac{d-1}{2})\Gamma(n+\frac{1}{2})}{\Gamma(n+\frac{d}{2})}.
\end{equation}
Plugging Eq.(\ref{eq:int_theta_1}) into the last expression for $\mathcal{I}$, we get
\[\mathcal{I}=2\pi^{\frac{d-1}{2}}\displaystyle\sum\limits_{n=0}^{\infty}\displaystyle\frac{|\vec{\xi}_{\chi}|^{2n}}{(2n)!}\displaystyle\frac{\Gamma(n+\frac{1}{2})}{\Gamma(n+\frac{d}{2 })}\displaystyle\int_{0}^{\infty}\!\!dP\,P^{d+2n-1} \displaystyle\frac{e ^{-P_{\boldsymbol{0}}(P)\xi_{\chi}^{\boldsymbol{0}}}}{P_{\boldsymbol{0}}(P)}.\]

To solve the remaining integral, now in $P$, it is worth considering the change of variable from $P$ to $\zeta=\frac{P_{\boldsymbol{0}}}{m}$, via the mass-shell condition, i. e.,
\[\begin{array}{ll}
\mathcal{I}\!\!\!&=2\pi^{\frac{d-1}{2}}\displaystyle\sum\limits_{n=0}^{\infty}\displaystyle\frac{|\vec{\xi}_{\chi}|^{2n}}{(2n)!}\displaystyle\frac{\Gamma(n+\frac{1}{2})}{\Gamma(n+\frac{d}{2})}m^{d+2n-1}\times\\
\\
&\phantom{=}\displaystyle\int_{1}^{\infty} \!\!d\zeta\,e^{-(m\xi_{\chi}^{\boldsymbol{0}})\zeta}(\zeta^{2}-1)^{(\frac{d-1}{2}+n)-\frac{1}{2}}.
\end{array}\]
Then, we use the definition of the modified Bessel functions of the second kind (see Eq. (\ref{eq:besse_second_kind})):
\[\begin{array}{ll}
\mathcal{I}\!\!\!&=2\Bigl(\displaystyle\frac{2\pi m}{\xi_{\chi}^{\boldsymbol{0}}}\Bigr)^{\frac{d-1}{2}}\displaystyle\sum\limits_{n=0}^{\infty}\displaystyle\frac{1}{n!} \Bigl(\displaystyle\frac{ m|\vec{\xi}_{\chi}|^{2}}{2\xi_{\chi}^{\boldsymbol{0}}}\Bigr)^{n} K_{\frac{d-1}{2}+n}(m\xi_{\chi}^{\boldsymbol{0}}).
\end{array}\]
It is at this point that the time-like nature of the Killing vector field $\xi_{\chi}$ becomes crucial, since then the parameter $\lambda=\Vert\xi_{\chi}\Vert/\xi_{\chi} ^{0}$ satisfies the inequality
\[|\lambda^{2}-1|=\displaystyle\frac{\Vert\vec{\xi}_{\chi}\Vert^{2}}{(\xi_{\chi}^{0})^{2}},\]
which allows us to use the so-called multiplication theorem given by \cite{abramowitz}
\begin{equation}
\label{eq:bessel_multiplication}
\displaystyle\frac{K_{\nu}(\lambda a)}{\lambda^{\nu}}=\displaystyle\sum\limits_{k=0}^{\infty}\displaystyle\frac{1}{ k!}\Bigl(\displaystyle\frac{a(1-\lambda^{2})}{2}\Bigr)^{k}K_{\nu+k}(a).
\end{equation}
Finally, we arrive at the result
\begin{equation}
\label{eq:int_I_final}
\mathcal{I}=2m^{d-1}(2\pi)^{\frac{d-1}{2}} k_{\frac{d-1}{2}}(m\Vert\xi_{\chi}\Vert),
\end{equation}
which is used  in the expression (\ref{eq:partition_function_one_particle}) to arrive at the  desired result of Eq.(\ref{eq:partition_function_time_like}):
\[z(\chi)=2m^{d}(2\pi)^{\frac{d-1}{2}}\displaystyle\int_{\Sigma_{0}}\textup{vol}_{g_{S}}(\vec{x}) k_{\frac{d-1}{2}}(m\Vert\xi_{\chi}(\vec{x})\Vert).\]

\section{Series expansion of the Schwarzschild one-particle partition function}
\label{ap:series}

Let us start by rewriting the integrand of Eq.(\ref{eq:partition_function_sch}) as
\begin{equation}
\label{eq:integrand_sch}
\Gamma_{\textup{Sch}}=\displaystyle\frac{r^{2}\sin(\theta)}{m\beta R}K_{1}\bigl(f_{\textup{Sch}}(\zeta)\bigr),
\end{equation}
where we have introduced the notation $\zeta=\frac{r_{s}}{r}$ and  have defined the function
\begin{equation}
\label{eq:f_def}
f_{\textup{Sch}}(\zeta)=m\beta R^{\frac{1}{2}}(\zeta)=m\beta (1-\zeta)^{\frac{1}{2}}.
\end{equation}

The idea is to expand Eq.(\ref{eq:integrand_sch}) into a power series of $\zeta$, for which we will focus exclusively on the case $r>r_{s}$, i. e., $\zeta<1$. Our plan is to calculate separately the expansions of $R^{-1}$ and $K_{1}(f_{\textup{Sch}}(\zeta))$, and then multiply them using the Cauchy product, which is defined for any pair of series $\Sigma_{n=0}^{\infty}a_{n}\zeta^{n}$ and $\Sigma_{n=0}^{\infty} b_{n}\zeta^{n}$ as the series $\Sigma_{n=0}^{\infty}c_{n}\zeta^{n}$ with coefficients given by the discrete convolution \cite{grads2014,apostol1997}
\begin{equation}
\label{eq:convolution}
c_{n}=\displaystyle\sum\limits_{k=0}^{n}a_{k}b_{n-k}=\displaystyle\sum\limits_{k=0}^{n}b_{k}a_{ n-k}.
\end{equation}

While the series expansion of $R^{-1}$ is simply the geometric series, that is, $R^{-1}=\sum_{n=0}^{\infty} \zeta^{n}$, the series expansion of $K_{1}( f_{\textup{Sch}}(\zeta))$ can be calculated by means of the Taylor-MacLaurin series
\begin{equation}
\label{eq:taylor_series_Kf}
K_{1}(f_{\textup{Sch}}(\zeta))=\displaystyle\sum\limits_{n=0}^{\infty} \displaystyle\frac{K_{1}\circ f_{\textup{Sch}}^{(n)}(0)}{n!}\zeta^{n},
\end{equation}
where the derivatives $K_{1}\circ f_{\textup{Sch}}^{(n)}(0)=\frac{d K_{1}\circ f_{\textup{Sch}}}{d\zeta}\bigr|_{\zeta=0}$ can be computed
using the Faá di Bruno's formula for the $n$-th derivative of a composition of functions \cite{johnson2002curious}, i. e.,
\begin{equation}
\label{eq:faa_di_bruno}
(K_{1}\circ f_{\textup{Sch}})^{(n)}(0)=\displaystyle\sum\limits_{k=0}^{n}K_{1}^{(k)}(f_{\textup{Sch}}(0)) B_{n,k}(\{f_{\textup{Sch}}^{(s)}(0)\}_{s\in I}),
\end{equation}
where $I=\{1,2,...,n-k+1\}$ and  $B_{n,k}$ are the incomplete Bell polynomials, formally defined as \cite{zhang2012}
\begin{equation}
\label{eq:bell_pol}
B_{n,k}(\{a_{s}\}_{s\in I})=\displaystyle\sum\limits_{j's}\displaystyle\frac{n \prod\limits_{s\in I}( a_{s}/s!)^{j_{s}}}{j_{1}!j_{2}!...j_{n-k+1}!},
\end{equation}
with $\sum_{i\in I}j_{i}=k$ and $\sum_{i\in I}i j_{i}=n$. Fortunately, when the series in the argument of Bell's incomplete polynomials is a series of derivatives evaluated to zero, as is the case with $a_{s}=f_{\textup{Sch}}^{ (s)}(0)$, then the polynomials can be expressed as \cite{zhang2012}
\begin{eqnarray}
\nonumber
B_{n,k}(\{f_{\textup{Sch}}^{(s)}(0)\}_{s\in I})&=&\displaystyle\sum\limits_{j= 0}^{k}\displaystyle\frac{(-1)^{k+j}}{k!}\binom{k}{j}f^{k-j}(0)(f^{j})^{(n)}(0)\\
\nonumber\\
\label{eq:bell_pol_property}
&=&(-1)^{k}\displaystyle\frac{ n!}{k!}(m\beta)^{k} b_{n,k},
\end{eqnarray}
where the quantities $b_{n,k}$ were already defined previously in Eq.(\ref{eq:b_n_k}) as
\[b_{n,k}=\displaystyle\sum\limits_{j=0}^{k}(-1)^{j+n}\binom{k}{j}\binom{\frac{j}{2 }}{n}.\]

Regarding the $n$-th derivative of the modified Bessel function $K_{1}$, some properties are useful to us. The first of these properties is the relationship between neighboring functions, i. e.,  \cite{abramowitz}
\begin{equation}
\label{eq:bessel_property_1}
K_{\nu-1}(\zeta)=K_{\nu+1}(\zeta)-\displaystyle\frac{2\nu}{\zeta}K_{\nu}(\zeta).
\end{equation}
The second property allows us to express the derivative $K_{\nu}^{(k)}$ in terms of $K_{\nu}$ and $K_{\nu-1}$ as \cite{wolfram}
\begin{equation}
\label{eq:bessel_property_2}
\begin{array}{ll}
\displaystyle\frac{d^{k}K_{\nu}}{d\zeta^{k}}\!\!\!&=\zeta^{-k}\displaystyle\sum\limits_{j_{2}=0}^{k}(-1)^{j_{2}+k}\binom{k}{j_{2}}(-\nu)_{k-j_{2}}\\
\\
&\phantom{=}\displaystyle\sum\limits_{j_{1}=0}^{j_{2}}\displaystyle\frac{(-1)^{j_{1}}2^{2j_{1}-j_{2}}(-j_{2})_{2(j_{2}-j_{1})}(\nu)_{j_{1}}}{(j_{2}-j_{1})!}\\
\\
&\phantom{=}\left(\displaystyle\sum\limits_{j_{0}=0}^{j_{1}}\displaystyle\frac{(j_{1}-j_{0})!(-\zeta^{2}/4)^{j_{0}}K_{\nu}(\zeta)}{j_{0}!(j_{1}-2j_{0})!(1-j_{1}-\nu)_{j_{0}}(\nu)_{j_{0}}}\right.\\
\\
&\phantom{=}\left.+\displaystyle\sum\limits_{j_{0}=0}^{j_{1}-1}\displaystyle\frac{(j_{1}-j_{0}-1)!(-\zeta^{2}/4)^{j_{0}}(\zeta K_{\nu-1}(\zeta)/2)}{j_{0}!(j_{1}-2j_{0}-1)!(1-j_{1}-\nu)_{j_{0}}(\nu)_{j_{0}+1}}\right).
\end{array}
\end{equation}
For the case $\nu=1$, this last property  can be rewritten in a more convenient notation as
\begin{equation}
\label{eq:bessel_property_3}
\displaystyle\frac{d^{k}K_{1}}{dx^{k}}=\displaystyle\frac{(-1)^{k}k!}{\zeta^{k}}\left[ P_{0}(k;\zeta)K_{1}(x)+P_{1}(k;\zeta)\displaystyle\frac{\zeta}{2}K_{\nu-1}(\zeta)\right],
\end{equation}
where $P_{\kappa}(k;\zeta)$ has been defined already in Eq.(\ref{eq:pol_P_coef}). Now, combining the properties (\ref{eq:bessel_property_1}) and (\ref{eq:bessel_property_3}), we get the expression we were looking for $K_{1}^{(k)}(m\beta )$, i. e.,
\begin{eqnarray}
\label{eq:der_k_1}
\displaystyle\frac{d^{k}K_{1}}{d\zeta^{k}}\Bigr|_{m\beta}&=&K_{1}(m\beta)\displaystyle\frac{(-1) ^{k}k!}{\zeta^{k}}\times\\
\nonumber\\
\nonumber
&&\left[P_{0}(k;m\beta)- P_{1}(k;m\beta)\left(1-\displaystyle \frac{\mathcal{E}(m\beta)}{2}\right)\right],
\end{eqnarray}
where the quantity $\mathcal{E}$ was already defined in Eq.(\ref{eq:mathcal_E}).

Using these results in the Taylor-MacLaurin series (\ref{eq:taylor_series_Kf}), we obtain, as an expansion of $K_{1}(f_{\textup{Sch}}(\zeta))$, the series
\begin{equation}
\label{eq:series_K1fsch}
K_{1}(f_{\textup{Sch}}(\zeta))=K_{1}(m\beta)\displaystyle\sum\limits_{n=0}^{\infty} \widetilde{A}_{n}(m\beta)\zeta^{n},
\end{equation}
where the coefficients $\widetilde{A}_{n}$ are defined as
\begin{equation}
\begin{array}{l}
\label{eq:A_tilde}
\widetilde{A}_{n}(1;m\beta)=\\
\\
\displaystyle\sum\limits_{k=0}^{n} b_{n,k}\left[P_{0}(k,1;m\beta)-P_{1}(k,1;m\beta)\left(1-\displaystyle\frac{\beta}{2}\langle E_{\eta}\rangle_{\eta}\right)\right].
\end{array}
\end{equation}
Finally, calculating the Cauchy product between the geometric series $R^{-1}=\sum_{n=0}^{\infty}z^{\eta}$ and the series Eq.(\ref{eq:series_K1fsch}) we obtain, after multiplying by $r^{2}\sin(\theta)$, the series Eq.(\ref{eq:integrand_series}) with the coefficients Eq.(\ref{eq:A_n}), formally defined by means of the discrete convolution Eq.(\ref{eq:convolution}) as
\[A_{n}(m\beta)=\displaystyle\sum\limits_{\ell=0}^{n}\widetilde{A}_{\ell}(m\beta).\]


As an example of the application of the above results, in the following tables, we present the explicit values of some coefficients.

\begin{table}[H]
	\begin{center}
		\setlength{\tabcolsep}{10pt} 
		\renewcommand{\arraystretch}{3} 
		\begin{tabular}{ |c|l| }
			\hline
			$n$ & $A_{n}(1;m\beta)$\\
			\hline
			0&1\\
			\hline
			1&$\displaystyle\frac{1}{2}(1+\beta \langle E_{\eta}\rangle_{\eta})$\\
			\hline
			2&$\displaystyle\frac{1}{8}\left(3+6 \beta \langle E_{\eta}\rangle_{\eta}+m^{2}\beta^{2}\right)$\\
			\hline
			3&$\displaystyle\frac{1}{48}\left(15+9m^{2}\beta^{2}+\beta \langle E_{\eta}\rangle_{\eta}(45+m^{2}\beta^{2})\right)$\\
			\hline
			4&$\displaystyle\frac{1}{384}(105+90 m^{2}\beta^{2}+m^{4}\beta^{4}+4\beta \langle E_{\eta}\rangle_{\eta}(105+4m^{2}\beta^{2}))$\\
			\hline
		\end{tabular}
	\end{center}
	\label{tab:coeff_A}
	\caption{First five coefficients $A_{n}(m\beta)$.}
\end{table}

\begin{table}[H]
	\begin{center}
		\setlength{\tabcolsep}{10pt} 
		\renewcommand{\arraystretch}{3} 
		\begin{tabular}{ |c|l| }
			\hline
			$n$ & $\mathcal{A}_{n}(m\beta,r)$\\
			\hline
			0&$1$\\
			\hline
			1&$\displaystyle\frac{\beta   E_{\eta}}{r}-\displaystyle\frac{1}{2}\langle r^{-1}\rangle_{\eta}(1+\beta \langle E_{\eta}\rangle_{\eta})$\\
			\hline
			2&$\displaystyle\frac{\beta^{2} E_{\eta}^{2}}{2 r^{2}}-\displaystyle\frac{1}{2} \displaystyle\frac{\langle r^{-1}\rangle_{\eta}}{r} \beta  E_{\eta}(1+\beta \langle E_{\eta}\rangle_{\eta})+\displaystyle\frac{1}{4}\times$\\
			&$\left(\langle r^{-1}\rangle_{\eta}^{2}(1+\beta \langle E_{\eta}\rangle_{\eta})^{2}-\langle r^{-2}\rangle_{\eta}(\frac{3}{2}+3\beta\langle E_{\eta}\rangle_{\eta}+\frac{m^{2}\beta^{2}}{2})\right)$\\
			\hline
		\end{tabular}
	\end{center}
	\label{tab:mathcal_A}
	\caption{First three coefficients $\mathcal{A}_{n}(m\beta,r)$.}
\end{table}



\begin{thebibliography}{1}
	
\bibitem{tolman1934} Tolman, RC. [1934]. Relativity, thermodynamics, and cosmology. Dove Publications, New York.

\bibitem{bergmann1951} Bergmann, P. [1951]. Generalized statistical mechanics. Physical Review (\textbf{5}), vol. 84, p.p. 10-26, APS.

\bibitem{vankampen1968} van Kampen, NG. [1968]. Relativistic thermodynamics of moving systems. Physical Review, 173 (\textbf{1}), p.p. 295-301.

\bibitem{horvath1968} Horv\'ath, JI. [1968]. Contributions to the Relativistic Generalization of the Kinetic Theory of Gases and Statistical Mechanics. Symposia on Theoretical Physics and Mathematics 8, p.p. 47-60, Springer.

\bibitem{werner1976} Israel, W. [1976]. Nonstationary irreversible thermodynamics: a causal relativistic theory. Annals of Physics, vol. 100, 1-2, p.p. 310-331. Elsevier.

\bibitem{lipparini2008}
Lipparini, E. [2008]. Modern many-particle physics: atomic gases, nanostructures and quantum liquids. World Scientific Publishing Company.

\bibitem{pomeau2007} Pomeau, Y. [2007]. Statistical Mechanics of a Gravitational Plasma. Advances in Chemical Physics, vol. 135, p.p. 153-172, Wiley Online Library.

\bibitem{livadiotis2018} Livadiotis, G. [2018]. Derivation of the entropic formula for the statistical mechanics of space plasmas. Nonlinear Processes in Geophysics, vol. 25 (\textbf{1}), p.p. 77-88, Copernicus GmbH.

\bibitem{mati2020} Mati, P. [2020]. Statistical theory of photon gas in plasma. Journal of Statistical Mechanics: Theory and Experiment, vol. 2020 (\textbf{2}). IOP Publishing

\bibitem{bauche2015}
Bauche, J., Bauche-Arnoult, C., and  Peyrusse, O. [2015]. Atomic properties in hot plasmas: From levels to superconfigurations. Springer.

\bibitem{uzdensky2008} Uzdensky, D. \& Goodman, J. [2008]. Statistical description of a magnetized corona above a turbulent accretion disk. The Astrophysical Journal, vol. 682 (\textbf{1}), p.p. 608, IOP Publishing.

\bibitem{bar2014} Bar-Or, B. \& Tal, A. [2014]. The statistical mechanics of relativistic orbits around a massive black hole. Classical and Quantum Gravity, vol. 31 (\textbf{24}), p.p. 244003, IOP Publishing.

\bibitem{herpich2017} Herpich, J. \& Tremaine, S. \& Rix, H. [207]. Galactic disc profiles and a universal angular momentum distribution from statistical physics. Monthly Notices of the Royal Astronomical Society, vol. 467 (\textbf{4}), p.p. 5022-5032, Oxford University Press.

\bibitem{ftq2022}
Faraji, S., Trova, A., and  Quevedo, H. [2022]. Relativistic equilibrium fluid configurations around rotating deformed compact objects. The European Physical Journal C, 82(12), 1149.

\bibitem{ogorodnikov1957} Ogorodnikov, KF. [1957]. Statistical Mechanics of the Simplest Types of Galaxies. Soviet Astronomy,
 vol. 1, p.p. 748.

\bibitem{binney1987} Binney, J. \& Tremaine, S. [1987]. Galactic Dynamics. Princeton, New Jersey: Princeton University Press, 1st ed.

\bibitem{pietronero2005} Pietronero, L. \& Labini, F. [2005]. Statistical physics for cosmic structures. Complexity, Metastability and Nonextensivity, p.p. 91-101, World Scientific.

\bibitem{hameeda2021} Hameeda, M. \& Plastino, A. \& Rocca, MC. [2021]. Galaxies clustering generalized theory. Physics of the Dark Universe, vol. 32, p.p. 100816, Elsevier.

\bibitem{ourabah2022} Ourabah, K. [2022]. Generalized statistical mechanics of stellar systems. Physical Review E, vol. 105 (\textbf{6}), p.p. 064108, APS.

\bibitem{pietronero2007} Pietronero, L. \& Sylos, F. [2007]. The problem of cosmological dark matter and statistical physics. The European Physical Journal Special Topics, vol. 143 (\textbf{1}), p.p. 223-230, Springer.

\bibitem{feron2008} F\'eron, C. \& Hjorth, J. [2008]. Simulated dark-matter halos as a test of nonextensive statistical mechanics. Physical Review E, vol. 77 (\textbf{2}), p.p. 022106, APS.

\bibitem{patwardhan2008} Patwardhan, A. [2008]. Statistical Physics of Dark and Normal Matter Distribution in Galaxy Formation: Dark Matter Lumps and Black Holes in Core and Halo of Galaxy. arXiv preprint arXiv:0805.2360.

\bibitem{chavanis2015} Chavanis, P. \& Lemou, M. \& M\'ehats, F. [2015]. Models of dark matter halos based on statistical mechanics: The fermionic King model. Physical Review D, vol. 92 (\textbf{12}), p.p. 123527, APS.

\bibitem{hkv2022}
Howard, M., Kosowsky, A., and Valogiannis, G. [2022]. Galaxy Cluster Statistics in Modified Gravity Cosmologies. arXiv preprint arXiv:2205.13015.

\bibitem{sato2021} Sato, N. [2021]. The effect of spacetime curvature on statistical distributions. Classical and Quantum Gravity, vol. 36 (\textbf{16}), p.p. 165003, IOP Publishing.

\bibitem{kolekar2011} Kolekar, S. \& Padmanabhan, T. [2011]. Ideal gas in a strong gravitational field: Area dependence of entropy. Physical Review D, vol. 83 (\textbf{6}), p.p. 064034, APS.

\bibitem{sarbach2022} C\'ardenas, A. \& Rub\'en, O. \& Gabarrete, C. \& Sarbach, O. [2022]. An introduction to the relativistic kinetic theory on curved spacetimes. General Relativity and Gravitation, vol. 54, 3, pp. 1-120, Springer.

\bibitem{souriau1997structure} Souriau, J. [1997]. Structure of dynamical systems: a symplectic view of physics. Vol. 149. Springer Science \& Business Media.

\bibitem{barbaresco2014} Barbaresco, F. [2014]. Koszul information geometry and Souriau geometric temperature/capacity of Lie group thermodynamics. Entropy, vol. 16 (\textbf{8}), p.p. 4521-4565, MDPI.

\bibitem{marle2016} Marle, C. [2016]. From tools in symplectic and poisson geometry to J.-M. Souriau’s theories of statistical mechanics and thermodynamics. Entropy, vol. 18 (\textbf{10}), p.p. 370, MDPI.

\bibitem{barbaresco2019} Barbaresco, F. [2019]. Lie groups thermodynamics \& Souriau-Fisher metric. SOURIAU 2019 conference, Institut Henri Poincar\'e, 31st May.

\bibitem{marle2020gibbs} Charles-Michel, M. [2020]. On Gibbs states of mechanical systems with symmetries. Journal of Geometry and Symmetry in Physics, vol. 58, p.p. 45-85, Bulgarian Academy of Sciences, Institute of Mechanics.

\bibitem{rovelli2011} Rovelli, C. \& Smerlak, M. [2011]. Thermal time and Tolman–Ehrenfest effect: 'temperature as the speed of time'. Classical and Quantum Gravity. 28 (7): 075007.

\bibitem{cubero2007} Cubero, D. \& Casado-Pascual J. \& Dunkel J. \& Talkner P. \& Hanggi P. [2007]. Phys. Rev. Lett. 99 170601.

\bibitem{chacon2009} Chac\'on-Acosta, G. \& Dagdug, L. \& Morales-T\'ecotl, H. [2009]. Manifestly covariant J\"{u}ttner distribution and equipartition theorem. Phys. Rev. E., Statistical, nonlinear, and soft matter physics.

\bibitem{aragon2018modified} Arag\'on-Mu\~noz, L. \& Chac\'on-Acosta, G. [2018]. Modified relativistic J{\"u}ttner-like distribution functions with $\eta$-parameter. Journal of Physics: Conference Series, Vol. 1030, No. 1, p.p. 012004, IOP Publishing.


\bibitem{aleksandr1949} Aleksandr, I. \& Khinchin, A. [1949]. Mathematical foundations of statistical mechanics. Courier Corporation.

\bibitem{callen1985} Callen, H. [1985]. Thermodynamics and an introduction to thermostatics. John Wiley \& Sons, Inc.,
New York.

\bibitem{tuckerman2010} Tuckerman, M. [2010]. Statistical mechanics: theory and molecular simulation. Oxford university press.

\bibitem{barbaresco2021geometric} Barbaresco, F. \& Nielsen, F. et al. [2021]. Geometric Structures of Statistical Physics, Information Geometry, and Learning. Springer.

\bibitem{marsden1999} Marsden, J. \& Ratiu, E. \& Tudor, S. [1999]. Introduction to Mechanics and Symmetry. Springer, p.p. 165-180.

\bibitem{da2008lectures} da Silva, A. [2008]. Lectures on symplectic geometry. Vol. 3575, Springer.

\bibitem{guilleminmeasure} Ritchie, M. \& Guillemin, V. [1996]. Measure theory and probability. Springer.

\bibitem{athreyameasure} Athreya, K. \& Soumendra, N. [2006]. Measure theory and probability theory. Springer, vol. 19.

\bibitem{jaynes1957information} Jaynes, E. [1957]. Information theory and statistical mechanics. Physical review,
vol. 106 (\textbf{4}).

\bibitem{callensym} Callen, H. [1974]. Thermodynamics as a science of symmetry. Foundations of Physics, vol. 4 (\textbf{4}), p.p. 423-443.


\bibitem{nakahara2018} Nakahara, M. [2018]. Geometry, topology and physics. CRC press.

\bibitem{izquierdogroups} Azc\'arraga, JA. \& Izquierdo, JM. [1998]. Lie groups, Lie algebras, chomology and some applications in physics. Cambridge University Press.


\bibitem{moore} Moore, T. [2013]. A general relativity workbook. University Science Books Mill Valley.

\bibitem{ehlers1973} Ehlers, J. [1973]. Survey of general relativity theory. Relativity, Astrophysics and Cosmology: Proceedings of the Summer School Held, 14-26 August, 1972 at the Banff Centre, Banff, Alberta. Springer, p.p. 1-125.

\bibitem{gerloch} Ehlers, J. \& Geroch, R. [2004]. Equation of motion of small bodies in relativity. Ann. Phys. 309 (2004) 232–236.

\bibitem{kriele1999} Kriele, M. [1999]. Spacetime: foundations of general relativity and differential geometry. Springer Science \& Business Media.

\bibitem{wald} Wald, R. [2010]. General relativity. University of Chicago Press.

\bibitem{jaynes1957} Jaynes, E. [1957]. Information theory and statistical mechanics: I. Physical Review 106,
620.

\bibitem{sachs2012general} Sachs, R. \& Wu, HH. [2012]. General relativity for mathematicians. Springer Science \& Bussines Media.

\bibitem{abramowitz} Abramowitz, M. \& Stegun, I. [1968]. Handbook of Mathematical Functions. Dover.

\bibitem{caroll2019} Caroll, S. [2019]. Spacetime and geometry. Cambridge University Press.

\bibitem{grads2014} Gradshteyn, I. \& Ryzhik, I. [2014]. Table of integrals, series, and products. Academic Press.

\bibitem{apostol1997} Apostol, T. [1997]. Modular Functions and Dirichlet Series in Number Theory. 2nd ed. New York: Springer-Verlag, p.p. 24.

\bibitem{beckett2022} Beckett, A. Homogeneous Symplectic Spaces and Central Extensions. Physical Sciences Forum, vol. 5 (\textbf{1}), p.p. 24, MDPI. 

\bibitem{lee2012} Lee, J. [2012]. Smooth manifolds. Springer.

\bibitem{blumenson1960} Blumenson, L. [1960]. A Derivation of n-Dimensional Spherical Coordinates. The American Mathematical Monthly. 67 (\textbf{1}), p.p. 63–66.

\bibitem{johnson2002curious}. Warren, J. [2002]. The curious history of Fa{\`a} di Bruno's formula. The American mathematical monthly, vol. 109 (\textbf{3}), p.p. 217-234.

\bibitem{zhang2012} Zhang, Z. \& Jizhen, Y. [2012]. Notes on some identities related to the partial Bell polynomials. Tamsui Oxf. J. Inf. Math Sci. vol. 28 (\textbf{1}), p.p. 39-48.

\bibitem{wolfram} Seen in \textup{http://functions.wolfram.com/03.04.20.0021.01}.

\end{thebibliography}
\end{document}